\documentclass[a4paper]{article}
\usepackage{etex}
\usepackage{a4wide}
\usepackage{graphicx}
\usepackage{epstopdf} %
\usepackage{amsmath,amsbsy,amssymb,amsgen,amsfonts}
\usepackage{color,latexsym}
\usepackage[breaklinks=true,bookmarks=true,colorlinks=true,linkcolor=blue,citecolor=blue,allcolors=blue]{hyperref}
\usepackage[toc,title]{appendix}
\usepackage{natbib}

\usepackage{doi}
\usepackage{tikz}
\usetikzlibrary{shapes,arrows,positioning,calc,through,backgrounds}
\tikzstyle{every picture}+=[remember picture]
\usetikzlibrary{patterns}
\usepackage{rotating}
\newcommand{\lc}[2][black]{\textcolor{#1}{#2}}
\newcommand{\bitem}{\begin{itemize} }
\newcommand{\eitem}{\end{itemize} }
\newcommand{\benum}{\begin{enumerate} }
\newcommand{\eenum}{\end{enumerate} }

\newcommand{\Vor} {Vorono\"i }
\newcommand{\caseA}{M121}
\newcommand{\caseB}{M178}
\newcommand{\caseAA}{S121}
\newcommand{\caseBB}{S178}
\newcommand{\ben} {\begin{equation}}
\newcommand{\een} {\end{equation}}
\newcommand{\be} [1] {\begin{equation} \label{#1}}
\newcommand{\ee} {\end{equation}}
\newcommand{\bse} [1] {\begin{subequations} \label{#1}}
\newcommand{\ese} {\end{subequations}}
\newcommand{\ban} {\begin{eqnarray*} }
\newcommand{\ean} {\end{eqnarray*} }
\newcommand{\bea} {\begin{eqnarray}}
\newcommand{\eea} {\end{eqnarray}}

\newcommand{\solid}{---$\!$---$\!$--}

\newcommand{\dashed}{\hbox{{--}\,{--}\,{--}\,{--}}}

\newcommand{\chndot}{---\,$\cdot$\,---}

\newsavebox{\mybox}
\sbox{\mybox}{\dashed}
\usepackage{sectsty}
\allsectionsfont{\sffamily}
\let\LaTeXmaketitle\maketitle
\renewcommand{\maketitle}{{\sf\LaTeXmaketitle}}

\begin{document}
\epstopdfsetup{suffix=} %
\title{%
  Sedimentation of a dilute suspension of rigid spheres at
  intermediate Galileo numbers: the effect of clustering upon the
  particle motion}
\author{Markus
  Uhlmann\footnote{\href{mailto:markus.uhlmann@kit.edu}{markus.uhlmann@kit.edu}} 
  \hspace*{1ex}and 
  Todor
  Doychev\footnote{\href{mailto:todor.doychev@kit.edu}{todor.doychev@kit.edu}}
  \\[1ex]
  {\small 
    Institute for Hydromechanics, Karlsruhe Institute of
    Technology}\\ 
  {\small 
    76131 Karlsruhe, Germany
  }
}
\date{} 
\maketitle
\begin{abstract}
  Direct numerical simulation of the gravity-induced settling of
  finite-size particles in triply-periodic domains has been performed
  under dilute conditions. For a single solid-to-fluid density ratio
  of $1.5$ we have considered two values of the Galileo number
  corresponding to steady vertical motion ($Ga=121$) and to steady
  oblique motion ($Ga=178$) in the case of one isolated sphere. 
  For the multi-particle system we observe strong particle clustering
  only in the latter case. 
  The geometry and time scales related to clustering are determined
  from \Vor tesselation and particle-conditioned averaging. 
  As a consequence of clustering, the average particle settling
  velocity is increased by 12\% 
  as compared to the value of an isolated sphere; 
  such a collective effect is not observed in the non-clustering case.  
  By defining a local (instantaneous) fluid velocity average in the
  vicinity of the finite-size particles it is shown that the
  observed enhancement of the settling velocity is due to 
  the fact that the downward fluid motion (with respect to the global
  average) which is induced in the cluster regions is preferentially
  sampled by the particles. 
  It is further observed that the variance of the particle velocity is
  strongly enhanced in the clustering case. 
    With the aid of a decomposition
    of the particle velocity it is shown that 
    this increase is due to enhanced fluid velocity fluctuations (due
    to clustering) in the vicinity of the particles. 
  Finally, we discuss a possible explanation for the observation of a
  critical Galileo number marking the onset of clustering under dilute
  conditions. 
\end{abstract}
\section{Introduction}
\label{sec-intro}
The gravity-driven motion of heavy particles in a viscous fluid is a
process which is relevant to many applications, such as  
precipitation in meteorology, 
waste water treatment, 
fluvial geomorphology %
and chemical engineering systems. 
In the general case the interaction between the fluid phase and the
suspended inclusions is highly complex, involving a large range of
spatial and temporal scales. 
Correspondingly, sedimenting suspensions exhibit a rich set of
dynamical features which can significantly affect their global 
properties. 
Questions of practical interest with respect to such systems include
the following:
what is the average settling velocity of the particles?
What are the particularities of their spatial distribution? 
What are the characteristics of the flow field induced by the particle
motion? 

Compared to the very idealized situation of a single fixed sphere in a 
uniform flow at low Reynolds number,
the additional complexity when treating sedimenting suspensions  stems
from the combined effects of 
larger Reynolds numbers, 
particle mobility 
as well as from the multi-particle collective. 
The Reynolds number effect by itself has been extensively studied in
the past 
\citep[e.g.][]{johnson:99,ghidersa:00,schouveiler:02,bouchet:06}. 
The picture that has emerged from these and other investigations of
single fixed spheres in uniform, unbounded flow is as follows. 
With increasing Reynolds number the wake -- which is axi-symmetric and
steady at low Reynolds number -- first undergoes a bifurcation which
breaks the axi-symmetry while steadiness is maintained 
(above $Re_{c1}\approx210$, where $Re$ is based upon the magnitude of the
unperturbed flow velocity $u_\infty$, the sphere diameter $D$ and the 
kinematic fluid viscosity $\nu$), leading to a wake structure in the
shape of double-threaded vortices exhibiting a planar symmetry. 
Above a second critical value ($Re_{c2}\approx275$) the flow becomes
unsteady and time-periodic, with vortex shedding taking place. 
Further increasing the Reynolds number leads to a loss of planar
symmetry and of temporal periodicity (above $Re_{c3}\approx360$). 

When allowing the sphere to move freely under the effect of gravity in
an unbounded fluid initially at rest, new features emerge due
to the additional degrees of freedom. 
  In this case path instabilities
  arise in the coupled problem which is now described by two parameters,
  i.e.\ the solid to fluid density ratio $\rho_p/\rho_f$ 
  and the Galileo number $Ga$ which is a measure for the ratio between
  gravitational and viscous forces acting on a submerged particle 
  ($Ga=u_gD/\nu$, where $u_g=(|\rho_p/\rho_f-1|Dg)^{1/2}$ with $g$
  being the magnitude of the gravitational acceleration).
In the two-parameter space a variety of particle path regimes are
encountered,  
involving vertical, oblique, time-periodic oscillating, zig-zagging,
helical and chaotic motion, as shown numerically \citep{jenny:04} and
experimentally \citep{veldhuis:07,horowitz:10}. 
When focusing upon heavy particles (for which $\rho_p/\rho_f>1$) the
bounds between the different regimes do not depend strongly upon the
value of the density ratio. The sequence of regimes which is
encountered as the Galileo number is increased from an initially small
value (where steady motion along a vertical path is observed) leads to
steady oblique motion, then to planar motion with periodic
oscillations around an average oblique path, and finally to a
breakdown of planar symmetry and time periodicity. 

Much less is known about the behavior of multi-particle
configurations.
As a first step it is instructive to consider the case of several  
fixed spheres in certain spatial arrangements, thereby separating the
pure collective effect from the pure effect of particle mobility. 
In this spirit, \cite{tsuji:82} and \cite{tsuji:03} have investigated
a pair of spheres either aligned along the direction of the incoming
flow or arranged transversally; 
\cite{kim:93b} and \cite{schouveiler:04} have focused upon side-by-side
configurations. In all these studies of fixed pairs of spheres the
Reynolds number values were of the order of several hundred. 
It has been observed that the presence of a second sphere has a
significant effect upon the wake structure and its stability
properties, with correspondingly non-negligible influence on the drag
and lift forces; this effect depends evidently upon the inter-particle
distance and in a strong manner also upon the relative orientation of
the particle pair. 

The flow through random arrays of fixed spheres at Reynolds number
values up to several hundreds has
been investigated numerically by \cite{hill:01}, \cite{beetstra:07}
and \cite{tennetti:11}. The results of these studies have led to
the establishment of different drag formulas given as a function of
the solid volume fraction $\Phi_s$ and of the Reynolds number. It
should be noted that the range of solid volume fractions in the
aforementioned studies was relatively large ($\Phi_s\geq0.1$). 
Contrarily, systematic direct numerical simulations or experiments
considering the flow through randomly positioned (fixed) spheres at
low solid volume fractions are not available to our knowledge. 
\cite{riboux:13} have performed simulations of the flow through a
random array of fixed spheres at low solid volume
fractions (down to $0.006$), however, deliberately under-resolving the
flow in the vicinity of the spheres and explicitly imposing a
time-dependency of the hydrodynamic particle force. They showed that
their model nevertheless reproduces the large-scale features of the
flow induced by laboratory bubble swarms relatively well. They were
further able to separate the individual contributions to the amplitude
of fluid velocity fluctuations stemming from the random wake
positioning and from the temporal fluctuations of individual wakes.   

Turning now to the case of gravity-induced sedimentation of many
freely-mobile particles (i.e.\ where the effects of mobility and of the
particle collective are present), a series of experiments have been
carried out for solid volume fractions of $\Phi_s\leq0.0001$ and for
Galileo numbers in the range of 40 to 340 
by \cite{parthasarathy:90a,parthasarathy:90b} and \cite{mizukami:92}. 
Under these very dilute conditions, reasonable predictions of the flow
field induced by the particle ensemble can be obtained when assuming a
linear superposition of isolated particle wakes (with a suitable
choice of the wake description). 

To our knowledge \cite{kajishima:02} were the
first to perform direct numerical simulations of dilute sedimenting
suspensions at particle Reynolds numbers of ${O}(100)$ in
periodic boxes. 
  The numerical method was based upon fourth-order
  finite-differences 
  on a non-conforming grid, representing the fluid-solid interface
  with the aid of a local marker function. 
  For the sake of computational efficiency, however, the angular
  particle motion was initially not accounted for. 
They observed strong accumulation of 
particles (with a solid-to-fluid density ratio of $8.8$) in columnar 
structures when the particle Reynolds number exceeded a value of
$300$. 
The appearance of clusters was shown to strongly enhance the average
settling velocity of the particles. 
This result was explained by a wake attraction effect, where
trailing particles experience lower drag, therefore accelerating and 
approaching the corresponding leading particles, resulting in locally
enhanced concentrations of the solid phase. 
\cite{kajishima:04b} later revisited the same configuration at a
Reynolds number value of $300$ (corresponding to $Ga\approx210$), now
properly accounting for rotational particle motion. No clustering was
observed for solid volume fraction values below $0.001$.   
Furthermore, he found that the rotational motion of the particles
plays a significant role in the regeneration cycle of particle
clusters. 

The problem of a large number of bubbles %
rising in an initially ambient fluid has received comparably more
attention in the past. A few of the results from the literature on
bubbles with relevance to solid particle sedimentation shall be
mentioned in the following. 
Focusing on bubble Reynolds numbers of a few hundreds and volume
fractions of the dispersed phase in the dilute regime, experiments
have shown that the bubble wakes in a swarm are considerably
attenuated as compared to isolated wakes, with an exponential decay of
the velocity deficit reported by \cite{risso:08}. 
The energy spectrum of the bubble-induced liquid 
velocity fluctuations contains a significant range with a decay
proportional to the inverse of the third power of the wavenumber
\citep{lance:91,riboux:10}. 
It was subsequently shown by \cite{risso:11} that such a spectral
range may be explained by a superposition of localized random
disturbances under certain conditions.  
Concerning the amplitude of the velocity fluctuations of the liquid
carrier phase, it has been shown by \cite{risso:02} and subsequent
experiments \citep{martinez-mercado:07,riboux:10} that it varies
with a power of $0.4$ of the disperse phase volume fraction.

This short summary shows that our understanding of dilute
sedimenting suspensions at Reynolds numbers of the order of
hundreds is still incomplete. 
Apart from the pioneering work of Kajishima and collaborators
\citep{kajishima:02,kajishima:04b} 
the phenomenon of wake-induced clustering in this setting has received
little attention. 
In particular, it is not known how particle clusters affect the
scaling of the gravity-induced motion of both phases, i.e.\ the
particle motion itself and the induced ``pseudo-turbulence''. 
Furthermore, it is not clear whether (and to what extend) the larger
set of results already obtained in the context of homogeneous bubbly
flows apply to the case of heavy solid particles.  
  The objective of the present work is to provide a comprehensive
  description of the particle motion in
  dilute settling suspensions. 
  The principal question which we attempt to solve is how the spatial
  distribution of the particulate phase affects its motion. 
We have simulated this configuration with full resolution of the
fluid-solid interface by means of an immersed boundary method. 
Care has been taken to choose a small-scale resolution which is
sufficient to resolve the details of the particle wakes at the
respective Galileo number value. 
Large computational domain sizes were employed such as to allow for
the characterization of large-scale features of the flow.
In this study we focus upon a single value of the solid-to-fluid
density ratio of $1.5$ and upon a solid volume fraction of $0.005$,
while two values of the Galileo number are investigated, measuring
$121$ and $178$. 
These two values correspond to the regimes where an isolated particle
in an unbounded fluid exhibits a straight vertical path and to a
steady oblique path, respectively. 
Here we find no significant particle clustering in the former case
while massive clustering occurs in the latter one. 
In the present paper the scope is limited to the analysis of the
spatial distribution of the particles and of their motion. 
Detailed results relating to the 
flow field induced by the particle motion 
will be presented in a future publication. 

The computational setup is described in the following section. 
The results of our simulations are then presented in
\S~\ref{sec-results}, where we first consider the temporal evolution
of global quantities, before we turn to a \Vor tesselation based
analysis of the disperse phase distribution and finally to the
particle motion.  
The paper closes with a summary and discussion in
\S~\ref{sec-conclusion}. 
\section{Computational setup}\label{sec-numa}
\subsection{Numerical method}
\label{sec-numa-meth}
The numerical method employed in the present simulations 
has been described in detail by \cite{uhlmann:04}.
The incompressible Navier-Stokes equations are solved by a fractional
step approach with an implicit treatment of the viscous terms
(Crank-Nicolson) and a three-step Runge-Kutta scheme for the
non-linear terms. The spatial discretization employs second-order
central finite-differences on a staggered mesh which is uniform and
isotropic.  
The no-slip condition at the surface of moving solid particles is
imposed by means of a specifically designed immersed boundary technique
\citep{uhlmann:04}. The motion of the particles is computed from the
Newton equations for linear and angular motion of rigid bodies, driven
by buoyancy, hydrodynamic force/torque and contact forces (in case of
collisions). 
Since the suspension under consideration is dilute, collisions are
treated by a simple repulsive force mechanism \citep{glowinski:99}
formulated such as to keep colliding particles from overlapping
non-physically. 
In the multi-particle simulations presented herein it turns out that
the average time interval between two collision events experienced by
a particle measures $52$ ($39$) times the gravitational time scale (in
simulations M121 and M178, respectively, cf.\
table~\ref{tab:phys_params}) showing that indeed collisions are
relatively infrequent.  

The computational code has been previously employed for the direct
numerical simulation of various flow configurations  
\citep[][]{uhlmann:08a,chan-braun:11a,villalba:12,kidanemariam:13,chan-braun:13}.
\subsection{Flow configuration and physical parameters}
\label{sec:sec-numa-config}
We consider the gravity-induced settling of multiple spherical
particles with identical diameter in an otherwise ambient, 
incompressible, unbounded fluid. 
Under such conditions, the system is characterized by 
the following physical variables: 
the fluid density $\rho_f$, 
the kinematic fluid viscosity $\nu$, 
the magnitude of the vector of gravitational acceleration
${g}=|\mathbf{g}|$,  
the particle diameter $D$, 
the particle density $\rho_p$ 
and the solid volume fraction $\Phi_s$.
Dimensional analysis then shows that the problem is determined by
three non-dimensional parameters: 
the aforementioned solid volume fraction $\Phi_s$, 
the density ratio $\rho_p/\rho_f$ and
the Galileo number $Ga$. 
The latter quantity can be written in the form of a Reynolds number,
viz.\ $Ga=u_g D/\nu$, with the gravitational velocity scale given by
$u_g=(|\rho_p/\rho_f-1|Dg)^{1/2}$. 
Incidentally, a gravitational time scale can be defined as 
$\tau_g=D/u_g$.   

In the present study we have performed direct numerical simulations of
two main parameter points (cf.\ table~\ref{tab:phys_params} for the
physical parameter values). 
In both cases the value of $\Phi_s=0.005$
was chosen for the global solid volume fraction, i.e.\ we consider a
rather dilute suspension for which, however, the indirect interaction
between the particles via the carrier phase can be significant as will
be shown. 
Furthermore, a solid-to-fluid density ratio of $\rho_p/\rho_f=1.5$ was
selected for both cases; this value corresponds to the value of some
plastic materials (such as PVC) in water. 
Finally, two values of the Galileo number have been simulated, namely
$Ga=121$ and $178$, which is the only physical parameter that has been
varied in the present study. 
These two main simulations are denoted as M121 and M178. 

Let us briefly recall the regimes of motion a single isolated particle
experiences at the chosen density ratio $\rho_p/\rho_f=1.5$, 
which are attained asymptotically in time when starting from an initially
ambient fluid and a particle at rest, as mapped out by \cite{jenny:04}
and \cite{uhlmann:13a}.  
For values of the Galileo number below $Ga_{cr1}\approx155$ the wake
is axi-symmetric and consists of a 
single toroidal vortex structure attached to the particle, the
particle settling steadily on a straight vertical path. 
When increasing the Galileo number beyond this value $Ga_{cr1}$, the
wake becomes oblique (with planar symmetry) and features a pair of
thread-like quasi-axial vortices; the motion in this regime is steady
and follows a path with a small inclination with respect to the
vertical direction, i.e.\ the particle experiences a finite horizontal
drift with a random direction in the horizontal plane. 
Above $Ga_{cr2}\approx185$ the wake becomes time-dependent and
hairpin-shaped, exhibiting periodic oscillations around the above
mentioned oblique state; 
the previous symmetry is still maintained and, therefore, the
fluctuating particle motion is confined to a plane in space.   
For values larger than $Ga_{cr3}\approx215$ the periodicity and the
planar symmetry are lost, and eventually chaotic particle motion with
zero average drift in the horizontal plane sets in. 
The sequence of flow regimes and the corresponding wake
structures for isolated particles are illustrated in
figure~\ref{fig-wake-regimes-isolated}.
Therefore, the two parameter values (regarding $Ga$ and
$\rho_p/\rho_f$) chosen for the present multi-particle sedimentation
simulations correspond to the axi-symmetric wake regime ($Ga=121$) and
to the regime of steady oblique motion ($Ga=178$) when considering an
isolated particle.
The simulations were performed in triply-periodic, cuboidal
computational boxes, one set of edges being aligned with the direction
of gravity.
The Cartesian coordinate system is arranged such that the
$z$-coordinate points opposite to the direction of gravity, while $x$
and $y$ are the remaining coordinates in the horizontal plane.
Since such a periodic system does not attain a steady state in
the presence of settling spheres, a constant vertical pressure
gradient (i.e.\ a homogeneous body force) was imposed whose integral
(over the domain) equals the total submerged weight of the suspended
particles.  
In both cases the domain is elongated in the direction of
gravity in order to account for the anisotropic length scales of the 
particle wakes. The employed domain sizes are shown in
table~\ref{tab:phys_params}. 
The vertical domain length $L_z$ is two orders of magnitude
larger than the particle diameter, corresponding to $341D$ and $170D$
in case M121 and M178, respectively. 
In each horizontal direction, a linear dimension of $68D$ (case
M121) and $85D$ (case M178), respectively, was chosen.  
The adequacy of these box sizes for the representation of the current
sedimentation problem will be discussed in \S~\ref{sec:sec-numa-transient}.

In order to represent the flow field in the vicinity of the particles
faithfully, we employ a (uniform) grid width $\Delta x=\Delta y=\Delta
z$ such that a particle diameter is
resolved by $15$ ($24$) grid points in case M121 (M178), as shown in
table~\ref{tab:num_params}. 
This choice takes into account the increase in Galileo number between
these cases, as reflected by the different basic wake regimes of the
corresponding isolated particle motion (axi-symmetric toroidal wake
versus double-threaded oblique wake).  
Further discussion of the grid convergence of our results is
presented in \S~\ref{sec-numa-valid}. 
From the domain size and the grid resolution normalized by the
particle diameter, the number of grid nodes of the computational
mesh is readily obtained (table~\ref{tab:num_params}), amounting to a
total of $1024^2\times5120$ ($2048^2\times4096$) in case M121 (M178).  
In our configuration the solid volume fraction is given by the
relation $\Phi_s=N_p(\pi/6)D^3/(L_xL_zL_y)$, where $N_p$ is the number
of particles in the periodic cell. With the above numbers it turns out
that the motion of $N_p=15190$ and $11867$ particles, respectively, is
tracked in the two cases (cf.\ table~\ref{tab:num_params}). 
A visualization of the computational domain and the initial particle
positions is shown in figure~\ref{fig:box_part_pos}. 
Additionally we have performed a set of complementary 
simulations of the settling of a single particle (denoted as ``S121''
and ``S178'' in table~\ref{tab:phys_params}) with identical density
ratio and Galileo number as in the multi-particle cases M121 and
M178. These additional simulations were performed in smaller boxes
than the main cases, yet at very low solid volume fraction (two orders
of magnitude smaller than M121 and M178). Consequently, they will be
used as reference data devoid of multi-particle effects in the
following discussion. 

\subsection{Validation}
\label{sec-numa-valid}
The present numerical method and its computer implementation have
undergone a wide range of validation tests and benchmarks 
\citep[cf.][]{uhlmann:04c,uhlmann:04,uhlmann:05a,uhlmann:06c,uhlmann:08a}. 

In order to determine the required small-scale resolution for the
computationally demanding simulations presented herein, we have
performed some additional tests. 
They consisted in grid convergence simulations of the settling of a
single mobile particle in a uniform flow (featuring a uniform inflow
plane and convective outflow instead of periodicity in the vertical
direction).    
Reliable reference data from spectral/spectral-element computations
with a boundary conforming grid (using a moving grid attached to the
particle) is 
available for this case. This set-up allowed us to benchmark our
method with regards to the subtle effects arising when the
coupled fluid-particle system undergoes a bifurcation at increasing
Galileo number. 
The proposed benchmark is, therefore, a very stringent test for a
non-boundary-conforming numerical method. 
We have found that a resolution of $D/\Delta x=15$ is capable of
reproducing the axi-symmetric wake regime (at $Ga=144$) correctly with
a maximum error on the particle velocity of approximately 6\%. 
Turning to the double-threaded wake regime ($Ga=178$), an increased
resolution of $D/\Delta x=24$ captures the oblique particle motion
faithfully with a comparable quality. 
Please note that although the (asymptotic) particle motion is steady
in the Lagrangian frame of reference in these two regimes, the
simulations on a fixed computational mesh are still time-dependent. 
The details of these validation tests as well as data at supplementary
parameter points can be found in \cite{uhlmann:13a}. 
\subsection{Simulation start-up procedure}
\label{sec:sec-numa-transient}
The initial particle positions were set randomly. As the \Vor 
analysis discussed in \S~\ref{sec-results-spatial-struct-voronoi}
shows, the initial state determined by the pseudo-random number
generator indeed reflects the expected statistics of a random Poisson
process.  

In order to speed up the computation of the initial transients when 
the particles accelerate from rest up to values comparable to their 
terminal velocity, we have combined two strategies: 
a)~simulating first the flow around fixed particles, then releasing
them;  
b)~\mbox{using} first a coarser grid, then successively refining it. 
In particular, we have started with a coarse-grid simulation (using
$D/\Delta x=7.5$ in case M121 and $D/\Delta x=6$ in case M178) of the
flow around fixed particles and evolved the simulation in time until a 
statistically stationary state was reached (covering an interval of
$531\tau_g$ and $746\tau_g$ in cases M121 and M178, respectively). 
The relative flow velocity (in the positive $z$-direction) was chosen
such that the expected average particle drag force approximately
balances the submerged weight of the particles, i.e.\ anticipating the
Reynolds number of the freely mobile particles to be equal to the one
of the flow around the fixed ones ($Re=142$ and $231$ in cases M121
and M178, respectively). 
In order to maintain the relative motion, a homogeneous body force
in the vertical component of the fluid momentum equation was added,
similar to the case of settling spheres.

After the initial coarse simulation, the final flow field was
interpolated linearly upon a twice finer grid ($D/\Delta x=15$ in case
M121 and $D/\Delta x=12$ in case M178) and the simulation was
continued with the particles fixed for another $142\tau_g$
($283\tau_g$) in case M121 (M178). 
In the lower Galileo number simulation M121 the particles were then
released, and the simulation was run for $3567\tau_g$. 
In the higher Galileo number case M178 the final flow field was once
more refined to the final resolution $D/\Delta x=24$ and the
simulation was run for another $54\tau_g$ with the particles fixed in
space. Subsequently the particles in case M178 were released and an
interval of 
$1672\tau_g$ 
at the final resolution was simulated considering mobile particles. 
Henceforth, the instant in time when particles were released is
arbitrarily set to zero. 
At this point let us quickly mention the computational resources spent
for the present computations. 
The simulation M121 was run on $640$ processor cores on the 
IBM System~x iDataPlex system at LRZ M\"unchen (requiring
$2.3\cdot10^6$~CPU core hours)   
and for case M178 we used $8192$ processor cores of the Blue Gene/p
system at JSC J\"ulich ($2.5\cdot10^7$~CPU core hours). 
\subsection{Evaluation of the box-size}
\label{sec:sec-numa-box-size}
\begin{figure}
  \begin{minipage}{2ex}
    \rotatebox{90}{
      $R_{\alpha\alpha}(r_x)$
    }
  \end{minipage}
  \begin{minipage}{0.45\textwidth}
    \centerline{$(a)$}
    \includegraphics[width=\textwidth]
    {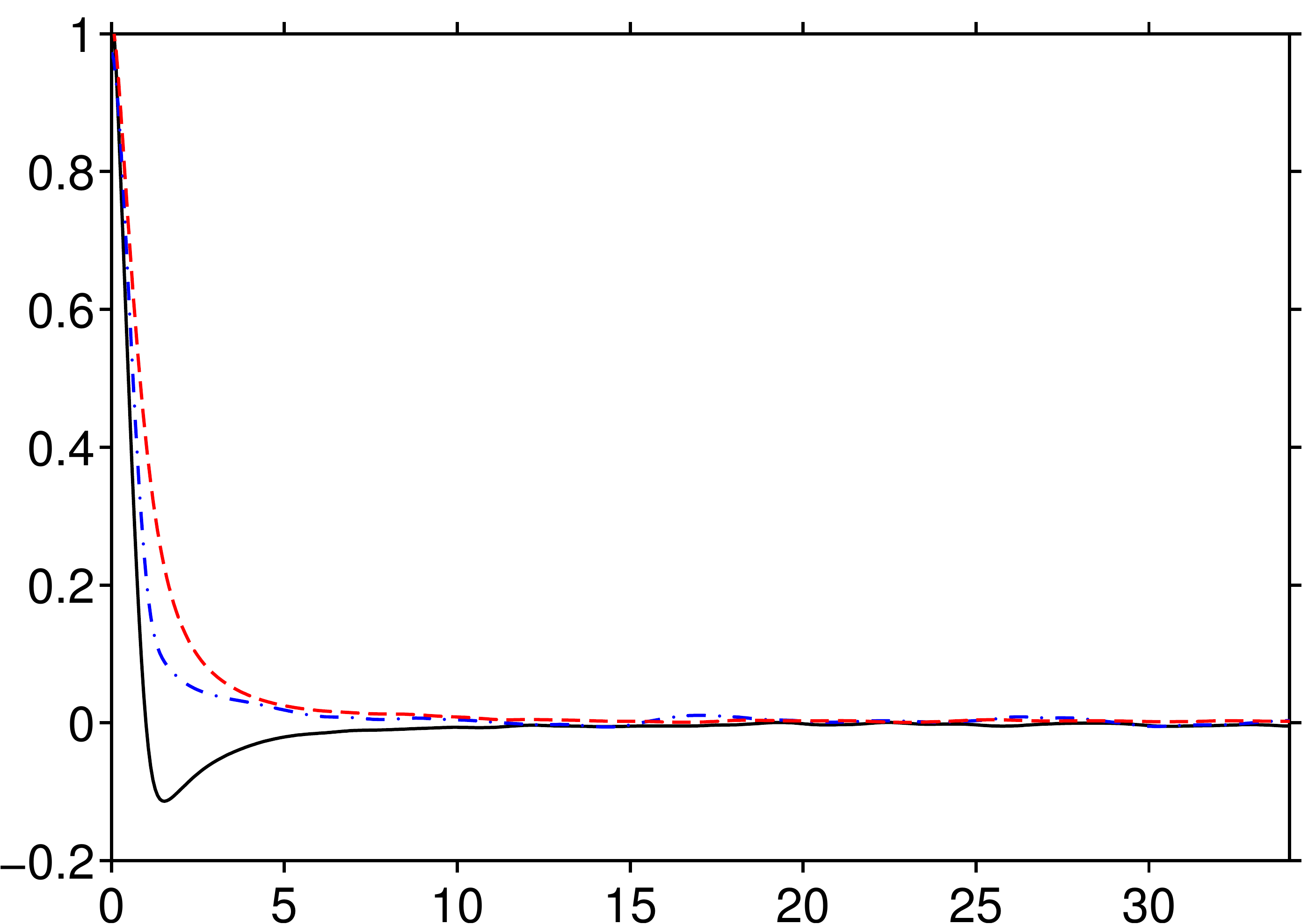}
    \centerline{$r_x/D$}
  \end{minipage}
  \hfill
  \begin{minipage}{2ex}
    \rotatebox{90}{
      $R_{\alpha\alpha}(r_z)$
    }
  \end{minipage}
  \begin{minipage}{0.45\textwidth}
    \centerline{$(c)$}
    \includegraphics[width=\textwidth]
    {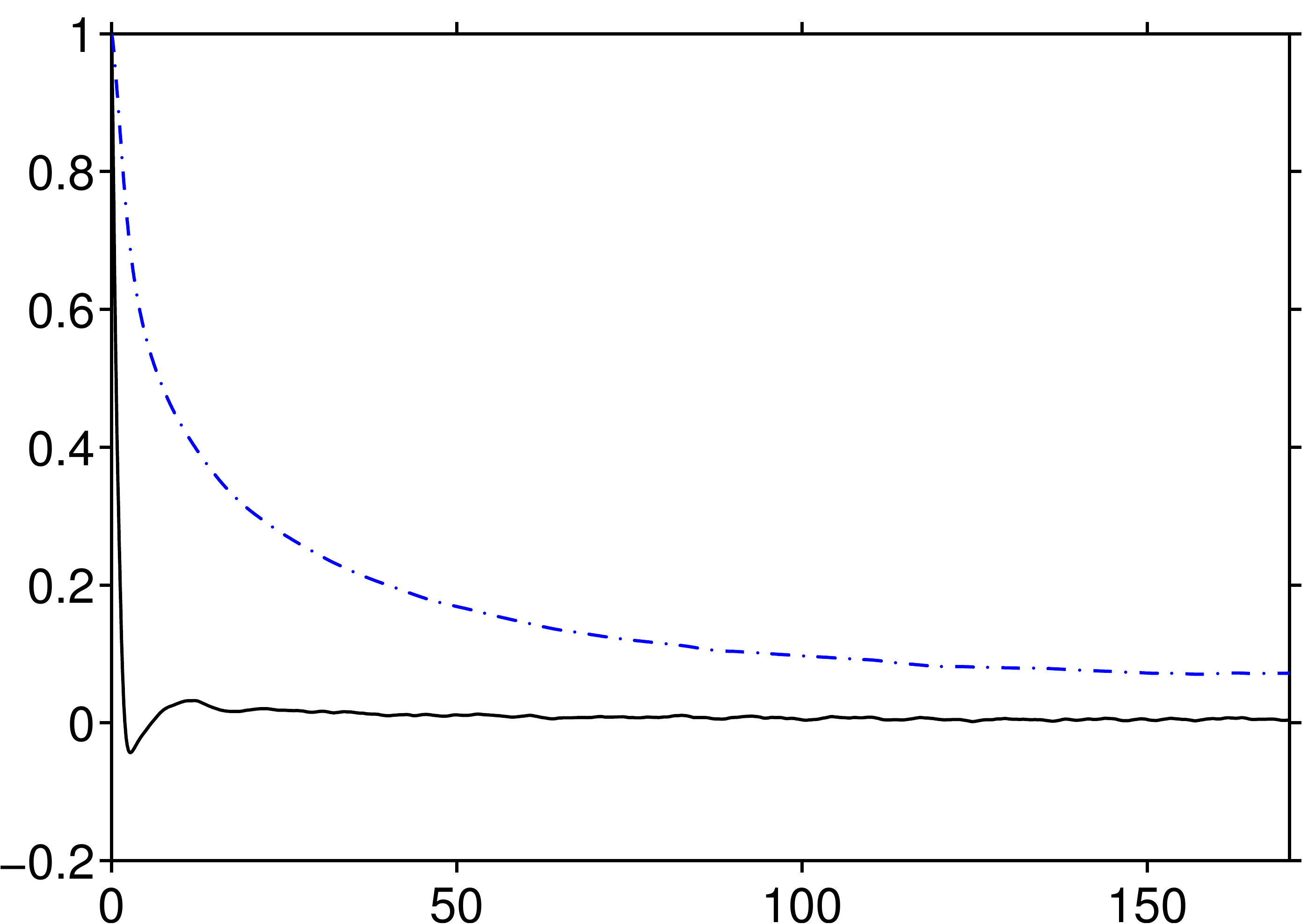}
    \centerline{$r_z/D$}
  \end{minipage}
  \caption{
    Two-point auto-correlation function of fluid velocity
    fluctuations in case \caseA{} prior to release (i.e.\ at
    $t=0$), for separations in: 
    $(a)$ the x-direction, and 
    $b$) the z-direction. 
    Lines show velocity components in the horizontal
    directions ($u$: \lc{\solid}, $v$: \lc[red]{\dashed}), and in
    the vertical direction ($w$: \lc[blue]{\chndot}).
    Note that in the graph in $(b)$ the 
    statistically equivalent component $R_{22}(r_z)$ has been omitted. 
  } 
  \label{fig:fluid-vel-corr-M121}
\end{figure}

\begin{figure}
  \begin{minipage}{2ex}
    \rotatebox{90}{
      $R_{\alpha\alpha}(r_x)$
    }
  \end{minipage}
  \begin{minipage}{0.45\textwidth}
    \centerline{$(a)$}
    \includegraphics[width=\textwidth]
    {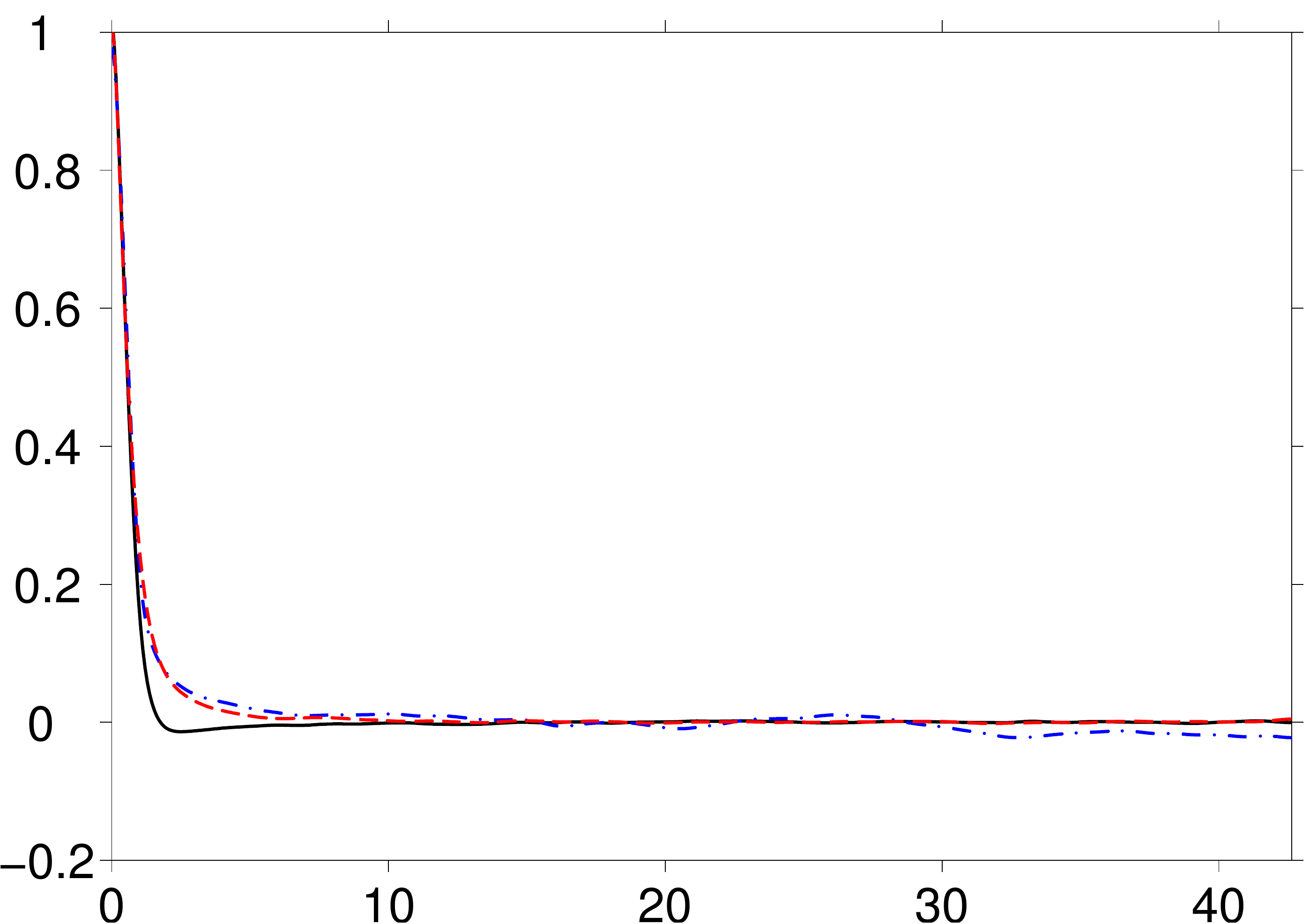}
   \centerline{$r_x/D$}
  \end{minipage}
  \hfill
  \begin{minipage}{2ex}
    \rotatebox{90}{
      $R_{\alpha\alpha}(r_z)$
    }
  \end{minipage}
  \begin{minipage}{0.45\textwidth}
    \centerline{$(c)$}
    \includegraphics[width=\textwidth]
    {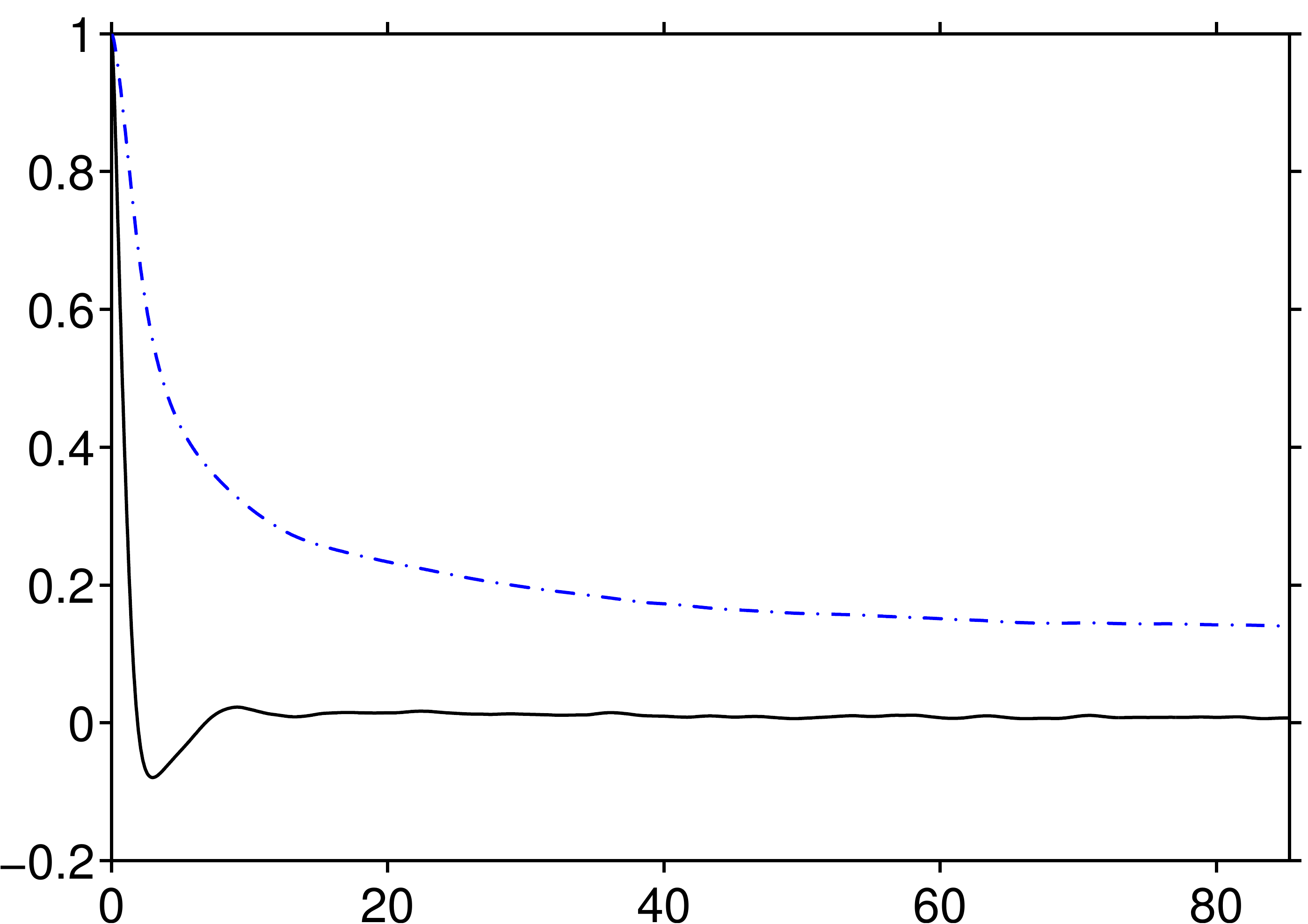}
    \centerline{$r_z/D$}
  \end{minipage}
  \caption{%
    As in figure~\ref{fig:fluid-vel-corr-M121}, but for
    case \caseB{}. 
  }
  \label{fig:fluid-vel-corr-M178}
\end{figure}

Figures~\ref{fig:fluid-vel-corr-M121} and
\ref{fig:fluid-vel-corr-M178} show the two-point 
correlation function 
of fluid velocity fluctuations for both cases 
just prior to the release of the particles. 
Recall that this corresponds to the flow at sphere Reynolds numbers
$Re=142$ and $231$, respectively, around an array of randomly
positioned, fixed spheres. 
Therefore, the particle wakes are already fully developed and
in a statistically stationary state at this point in time. 
It can be seen from the figures that in the horizontal directions all
velocity signals are essentially decorrelated for separations larger
than 10 particle diameters, a distance which is much smaller than the
horizontal domain size. 
Concerning vertical separations, the graphs show that the vertical
velocity component is not fully decorrelated at separations of half
the box height (while the horizontal velocity component does
decorrelate).  
The correlation coefficient $R_{ww}$ measures $0.07$ ($0.14$) in case 
M121 (M178) at the largest separation along the $z$-coordinate. 
This is obviously owed to the strong anisotropy of the particle
wakes. 
It should be kept in mind, that the velocity field of the
flow around the fixed particles contains a three-dimensional
time-average contribution which affects the correlation lengths. 
We can conclude this discussion by stating that even larger (axial)
domain extensions are necessary if full decorrelation of the flow
around dilute random arrays of fixed spheres is desired. 
However, preliminary coarse grid simulations (at $Re=230$) have 
shown that even an extended axial domain size of $L_z=680D$ does not 
warrant full decorrelation. 
\subsection{Notation}\label{sec-numa-notation}
Before turning to the results, let us fix the basic notation followed
throughout the present text. 
The spatial domain occupied by the two phases is given by $\Omega = \Omega_f
\cup \Omega_p$, where $\Omega_f$ and $\Omega_p$ are the domains
occupied by the fluid and the particles, respectively. 
Velocity vectors and their components corresponding to the fluid and
the particle phases are distinguished by subscripts ``f'' and ``p''
respectively, as in $\mathbf{u}_f = (u_f,v_f,w_f)^T$ and $\mathbf{u}_p
= (u_p,v_p,w_p)^T$. 
Similarly, the $i$th particle's position vector is denoted as
$\mathbf{x}_p^{(i)} = 
(x_p^{(i)},y_p^{(i)},z_p^{(i)})^T$, the vector of angular particle velocity as 
$\boldsymbol\omega_p^{(i)} =
(\omega_{p,x}^{(i)},\omega_{p,y}^{(i)},\omega_{p,z}^{(i)})^T$ and the
hydrodynamic force exerted by the fluid on the $i$th particle as
$\mathbf{f}_p^{(i)} = (f_{p,x}^{(i)},f_{p,y}^{(i)},f_{p,z}^{(i)})^T$.   

The fluctuations of the velocities of each phase are defined with
respect to the instantaneous phase average, i.e.\ 
$\mathbf{u}_{f}^\prime(\mathbf{x},t)=
\mathbf{u}_{f}(\mathbf{x},t)-\langle\mathbf{u}_{f}\rangle_{\Omega_f}(t)$,  
and 
$\mathbf{u}_{p}^{(i)\,\prime}(t)=
\mathbf{u}_{p}^{(i)}(t)-\langle\mathbf{u}_{p}^{(i)}\rangle_{p}(t)$,  
where the precise definitions of the averaging operators over the
particle ensemble, $\langle\,\rangle_p$, and over the volume occupied
by the fluid, $\langle\,\rangle_{\Omega_f}$, are given in
appendix~\ref{sec:app-averaging}.  
The fluctuations of the angular particle velocity are analogously
defined.  
\section{Results}  
\label{sec-results}
\subsection{Temporal evolution of the system}
\label{sec-results-transient}
Here we 
are concerned with the general evolution of the particulate flow
system starting with the moment when the particles are released.
Note that the flow field around the fixed particles is already
fully-developed at this time (as explained in
\S~\ref{sec:sec-numa-transient}), featuring an average particle
Reynolds number commensurate with the terminal settling velocity of
isolated spheres at the same Galileo number. 
Therefore, in the absence of multi-particle effects, 
the released spheres are expected to settle at the same Reynolds
number as was set for the flow around the corresponding fixed spheres.  

\begin{figure}
  \centering
  \begin{minipage}{2ex}
    \rotatebox{90}{$w_{s}D / \nu$}
  \end{minipage}
  \begin{minipage}{0.47\textwidth}
    \includegraphics[width=\textwidth]
    {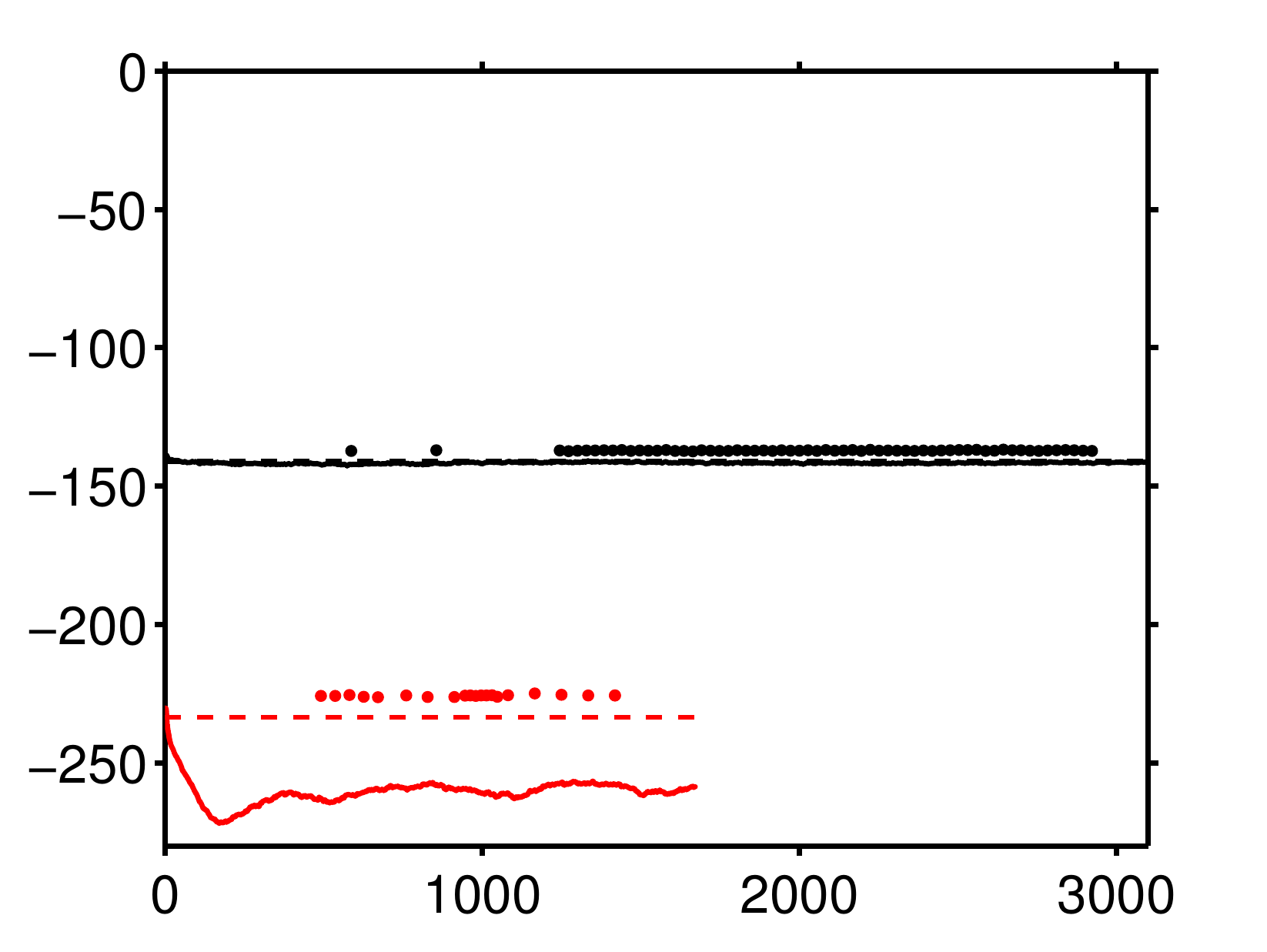}
    \hspace*{-.35\linewidth}\raisebox{.43\linewidth}{$Ga=121$}
    \hspace*{-14ex}\raisebox{.18\linewidth}{$Ga=178$}
    \\
    \centerline{$t/\tau_g$}
  \end{minipage}
  \caption{%
    Average particle settling velocity  as a  
    function of time for both cases M121 and M178 (solid lines),
    normalized with the particle diameter and kinematic viscosity. 
    The terminal velocity of a corresponding single particle (cases
    S121, S178) is indicated by dashed lines. 
    The settling velocity with respect to the fluid velocity in the
    particles' vicinity (cf.\ discussion in
    \S~\ref{sec-part-uf-seen-by-partices} and definition in 
    equation~\ref{eq:mean-app-vel-lag-shell}) is indicated by 
    filled circular symbols.
  } 
  \label{fig:part-wmean}
\end{figure}

Let us first consider the average settling velocity. For this purpose
we define (at time $t$) the $i$th particle's velocity relative to the 
fluid-phase-averaged velocity as follows
\begin{equation}\label{equ-res-trasient-def-uprel}
  \mathbf{u}_{pr}(i,t)=\mathbf{u}_{p}^{(i)}(t)-\langle\mathbf{u}_f\rangle_{\Omega_f}(t)
  \,.
\end{equation}
The standard definition of the average settling velocity then
corresponds to the average over all particles (at a given time) of the
vertical component of (\ref{equ-res-trasient-def-uprel}), viz.
\begin{equation}\label{equ-res-trasient-def-w-settling}
  w_{s}(t)=\langle {w}_{pr}(i,t)\rangle_{p}
  \,.
\end{equation}
Figure~\ref{fig:part-wmean} shows the temporal evolution of the
average settling velocity according to
relation~(\ref{equ-res-trasient-def-w-settling}), normalized 
with a viscous velocity scale, 
$w_sD/\nu$.   
It can be seen that the multi-particle ensemble in case M121 settles
at approximately the same rate as the corresponding isolated particle,
the curve being essentially flat with $Re=|\langle
w_s\rangle_t|D/\nu=141.6$ (cf.\ table~\ref{tab:phys_params}). 
The collective effect of many particles at solid volume fraction
$\Phi_s=0.005$ apparently does not lead to a modification of the
average settling velocity at $Ga=121$ (at a density ratio of $1.5$). 
Conversely, for case M178 figure~\ref{fig:part-wmean} shows an initial
increase of the magnitude of the average settling velocity; the curve
then levels off and fluctuates around a value of $Re=260.6$. 
Different time scales can be distinguished in the signal of the
average settling velocity in case M178 in the developed regime, the
largest of which being of the order of several hundred $\tau_g$.  
The collective enhancement of
the average settling velocity (as defined in
\ref{equ-res-trasient-def-w-settling})  
in case M178 measures approximately 12\% when compared to the value of
an isolated particle in case S178 (cf.\ table~\ref{tab:phys_params}). 
A difference of this amplitude is obviously significant in many
applications and merits a detailed investigation. 
Note that the only physical difference between cases M121 and M178 is
the value of the Galileo number. As we will see in the following, 
the particle phase is forming agglomerations in case M178 
(cf.\ \S~\ref{sec-results-spatial-struct}, where the properties of the
spatial particle distribution will be discussed), which lead to an 
increase in $|w_s|$ (cf.\ \S~\ref{sec-part-motion-settl-vel}). 
In figure~\ref{fig:fluid-part-vel-rms} the time evolution of the
intensities of velocity fluctuations of both phases are shown 
(cf.\ definition given in \S~\ref{sec-numa-notation}).  
In case M121 a short initial transient is visible after which the
vertical components of both phases exhibit slow fluctuations with a
time scale of the order of one thousand gravitational time units. 
On the other hand, the curves of case M178 again settle into an
asymptotic behavior after approximately $200\tau_g$ after which
oscillations with comparably shorter period ensue. 
While the scaling of the amplitude of these and other (globally
averaged) signals will be discussed in more detail in
\S~\ref{sec:sec-part-motion}, 
at present we are only concerned with the definition of 
the time interval over which temporal statistics are computed. 
For consistency between the two flow cases, we have
chosen to consider the flow as fully-developed for $t\geq200\tau_g$ in
both instances. 
Therefore, the time interval over which the
statistically stationary state is observed corresponds to 
$T_{obs}=2967\tau_g$ ($1472\tau_g$) in case M121 (M178).
\%
\subsection{Spatial structure of the dispersed phase} 
\label{sec-results-spatial-struct}
\begin{figure}
  \centering
  \begin{minipage}{2ex}
    \rotatebox{90}{$y/D$}
  \end{minipage}
  \begin{minipage}{0.35\textwidth}
    \centerline{$(a)$}
    \includegraphics[width=1.0\textwidth]
    {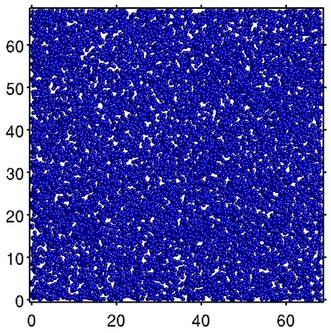}
    \\
    \centerline{$x/D$}
  \end{minipage}
  \hspace*{.1\linewidth}
  \begin{minipage}{2ex}
    \rotatebox{90}{$y/D$}
  \end{minipage}
  \begin{minipage}{0.35\textwidth}
    \centerline{$(b)$}
    \includegraphics[width=1.0\textwidth]
    {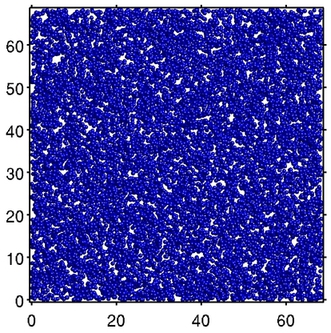}
    \\
    \centerline{$x/D$}
  \end{minipage}
  \caption{%
    Particle positions in case \caseA{} viewed from the top: 
    (a) initial state at $t=0$; 
    (b) in the statistically stationary regime at
    $t=1200\tau_g$.  
  }
  \label{fig:part-pos-top-M121}
\end{figure}

\begin{figure}
  \centering
  \begin{minipage}{2ex}
    \rotatebox{90}{$y/D$}
  \end{minipage}
  \begin{minipage}{0.35\textwidth}
    \centerline{$(a)$}
    \includegraphics[width=1.0\textwidth]
    {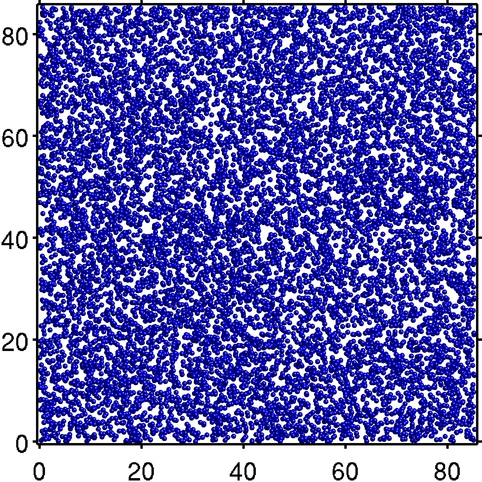}
    \\
    \centerline{$x/D$}
  \end{minipage}
  \hspace*{.1\linewidth}
  \begin{minipage}{2ex}
    \rotatebox{90}{$y/D$}
  \end{minipage}
  \begin{minipage}{0.35\textwidth}
    \centerline{$(b)$}
    \includegraphics[width=1.0\textwidth]
    {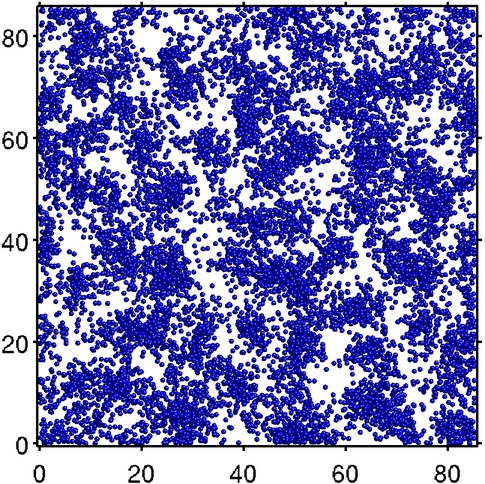}
    \\
    \centerline{$x/D$}
  \end{minipage}
 \caption{%
    The same as figure~\ref{fig:part-pos-top-M121}, but for case
    \caseB{}: 
    (a) $t=0$; 
    (b) $t=820\tau_g$.  
  }
  \label{fig:part-pos-top-M178}
\end{figure}

Figure~\ref{fig:part-pos-top-M121} shows a visualization of
instantaneous particle positions in case \caseA{} with the view
directed along the vertical axis. 
Comparing the random initial positions
(figure~\ref{fig:part-pos-top-M121}a) to those at some instant when
the system is in a statistically stationary state 
(figure~\ref{fig:part-pos-top-M121}b) reveals no significant
difference.  
The corresponding visualization for case \caseB{} is shown in
figure~\ref{fig:part-pos-top-M178}. It is clearly observable that the
particle distribution after the initial transient 
deviates significantly from the initial distribution. 
Localized regions in the projection upon the horizontal plane with a
high number density of particles alternate with regions practically
devoid of particles.  
It should be noted that the box size (in multiples of the particle
diameter) and, consequently, the number of particles in case M121
is significantly larger than in case M178. Therefore, the
illustrations of the initial random state
(figure~\ref{fig:part-pos-top-M121}a vs.\
\ref{fig:part-pos-top-M178}a) differ somewhat.  

Hereafter, regions with high particle concentration are loosely
referred to as ``clusters'' and regions with low particle
concentration as ``voids'', where the definition of high- and low-
particle concentration 
will be made more precise later in this section.
Seen from this vertical view the particles in case M178 appear to form
structures which are elongated in the vertical direction. 
Close inspection reveals that the clusters and the voids extend
throughout the entire height of the computational domain. 
The characteristic size of the clusters in the horizontal direction,
however, is much smaller than the computational domain.
The present visual detection of clusters is in line with previous 
findings 
where a similar ``columnar particle accumulation'' 
was reported at similar parameter values 
\citep{kajishima:04b}. 
Time sequences of particle position graphs show that these structures
persist over long time intervals. We will return to this point
shortly.  
\subsubsection{\Vor analysis} 
\label{sec-results-spatial-struct-voronoi}
In the following we quantify the tendency of the particles
to form clusters  
by performing a \Vor analysis of the particle positions.  
A number of techniques have been established for the characterization
of the spatial structure of the dispersed phase: 
box counting \citep{fessler:94}, 
pair correlation function estimation \citep{sundaram:97},
genuine clustering detection algorithms
\citep{wylie:00,melheim:05}, 
nearest-neighbor statistics \citep{kajishima:04b} 
and \Vor tessellation \citep{monchaux:10b,monchaux:12}. 
The latter technique has the advantage of providing a rich set of
geometrical information without the need for choosing any length scale
a priori, while at the same time being computationally efficient.  

\begin{figure}
  \centering
  \begin{minipage}{3ex}
    $(a)$
  \end{minipage}
  \begin{minipage}{0.25\linewidth}
      \includegraphics[width=\textwidth]
      {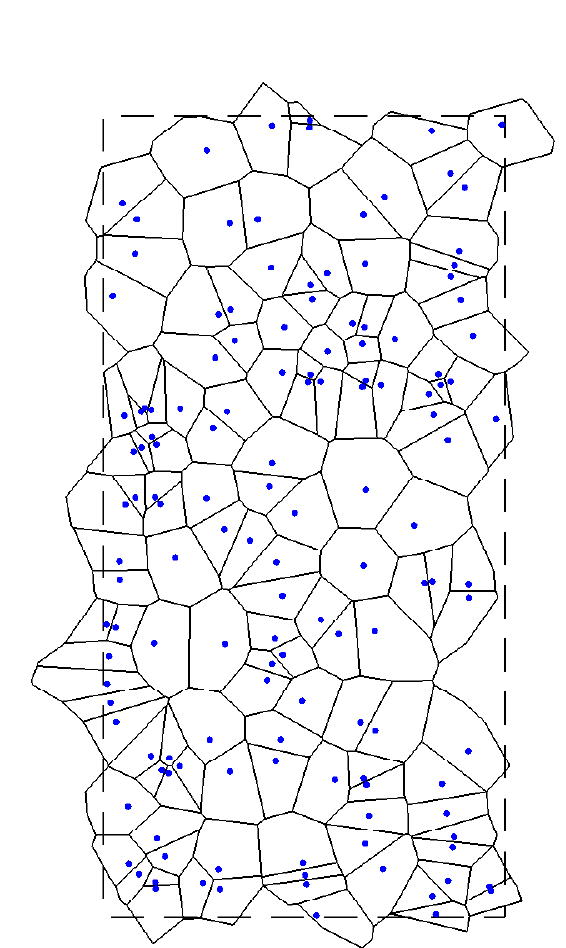}
  \end{minipage}
  \hspace{.1\linewidth}
  \begin{minipage}{3ex}
    $(b)$
  \end{minipage}
  \begin{minipage}{.3\linewidth}
    \includegraphics[width=\linewidth]
    {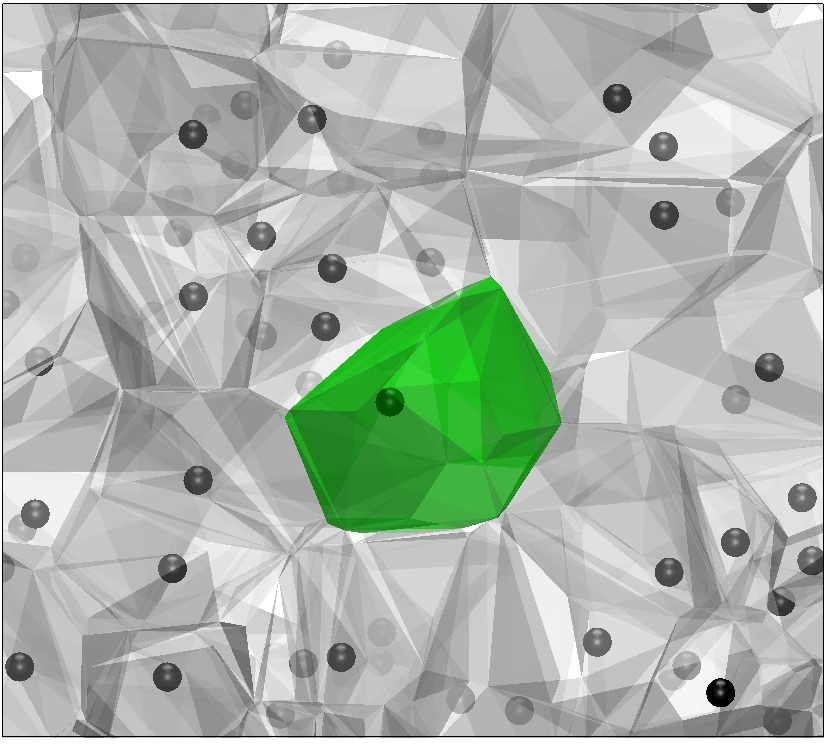}
  \end{minipage}
  \caption{%
      (a) A two dimensional Vorono\"i diagram in a vertical plane with 
      double-periodic boundaries (indicated with dashed lines). 
      The projected center locations of all particles intersecting the
      plane are taken as the Vorono\"i cell sites. Note that this
      graph is purely intended to illustrate the treatment of periodic
      directions; the actual analysis performed in this work considers 
      the three-dimensional case. 
      (b) Close-up of a three dimensional Vorono\"i diagram.
      Data is from case \caseB. 
  } 
  \label{fig:voronoi-3D-2D}
\end{figure}

The term \Vor tessellation 
corresponds to a decomposition of space into 
an ensemble
of cells ${\cal V}_i$, each of which being associated to one particle, 
i.e.\ with $1\leq i\leq N_p$.
The \Vor cells possess the property that all points
inside the $i$th cell are located closer to the centroid of the $i$th
particle than to any other particle \citep{okabe:92}. 
Examples of \Vor diagrams in two- and three space dimensions are shown
in figure~\ref{fig:voronoi-3D-2D}. 
The cells are periodically extended such as to
fill the space (figure~\ref{fig:voronoi-3D-2D}a).  
Note that the \Vor analysis performed here is three-dimensional.
From the tesselation one can determine the volume of each cell,
henceforth denoted as $V_i$, which can be related to the inverse of
the local particle concentration, i.e.\ \Vor cells with small (large) 
volume indicate regions with high (low) particle concentration. 
The \Vor cell volumes analyzed in the following are normalized such
that the mean value measures unity, which removes the dependence upon
the particle number density \citep{ferenc:07}.  

\begin{figure}
  \begin{minipage}{2ex}
    \rotatebox{90}{pdf}
  \end{minipage}
  \begin{minipage}{0.45\textwidth}
    \centerline{ $(a)$ }
    \includegraphics[width=\textwidth]
    {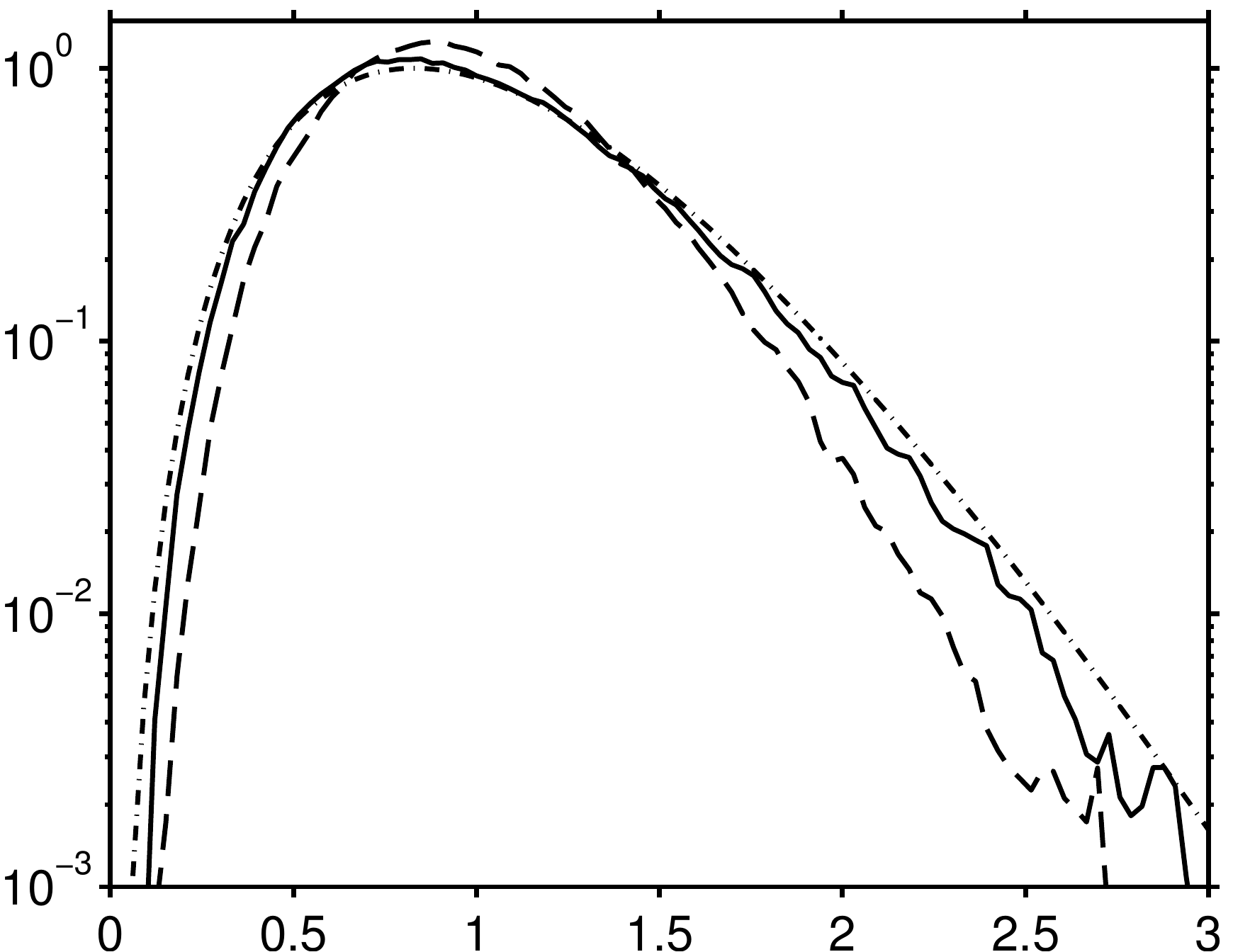}
    \centerline{ $V_i/\langle V \rangle_{p,t}$ }
  \end{minipage}
  \hfill
  \begin{minipage}{2ex}
    \rotatebox{90}{pdf}
  \end{minipage}
  \begin{minipage}{0.45\textwidth}
    \centerline{ $(b)$ }
    \includegraphics[width=\textwidth]
    {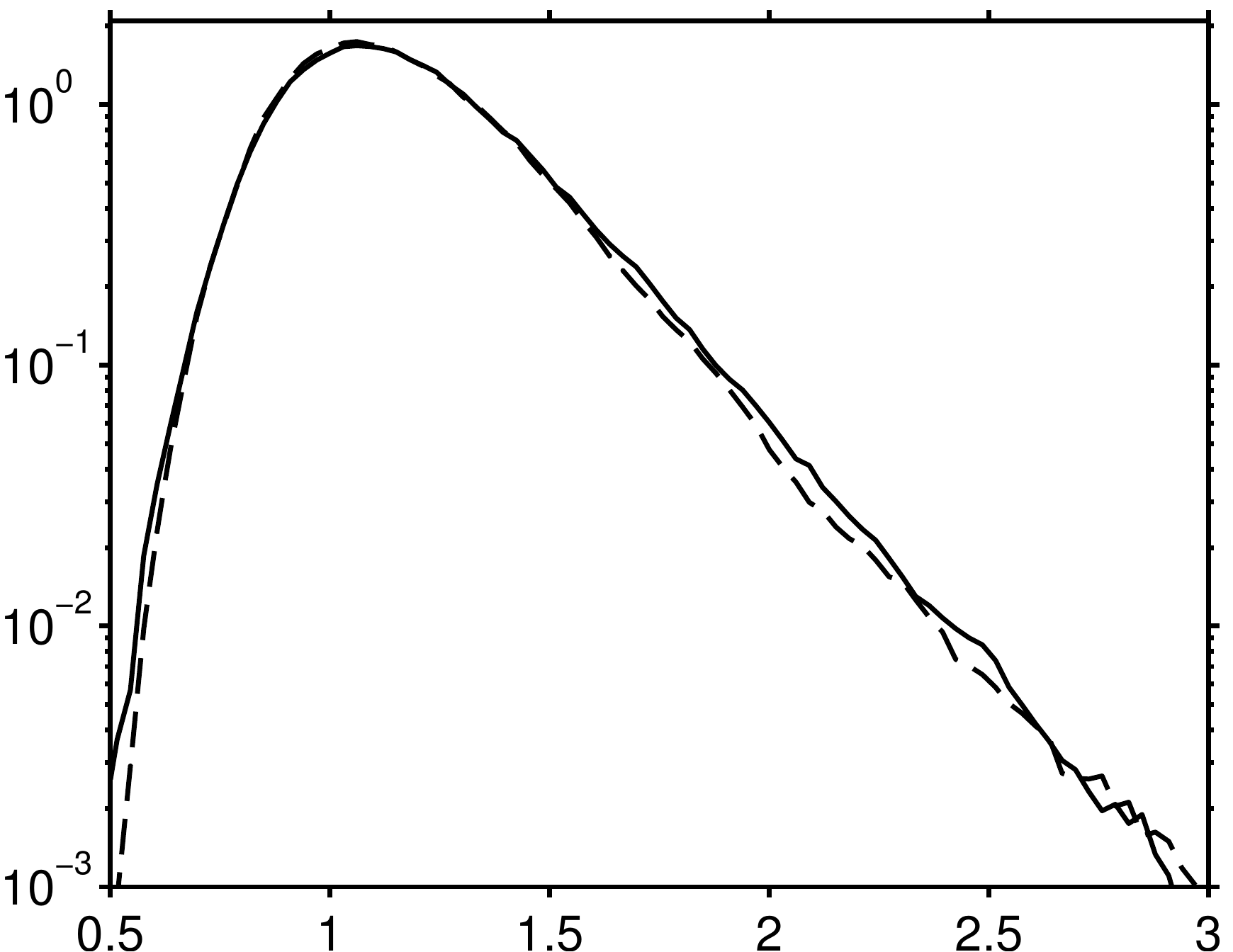}
    \centerline{ $A_i$ }
  \end{minipage}
  \caption{%
    Case M121: probability density function of (a) the
    normalized \Vor cell volumes $V_i$ and 
    (b) the \Vor cell aspect ratios $A_i$ defined in equation
    \ref{eq:VorAR}. 
    Line styles correspond to different instants in time: 
    the initial (random) particle distribution (solid line); 
    the statistically stationary state (dashed line). 
    In $(a)$ the chain-dotted line 
    represents the Gammma distribution corresponding to a random
    Poisson process \citep{ferenc:07}. 
  }
  \label{fig:voronoi-VOL-AR-PDF-M121}
\end{figure}

\begin{figure}
  \begin{minipage}{2ex}
    \rotatebox{90}{pdf}
  \end{minipage}
  \begin{minipage}{0.45\textwidth}
    \centerline{ $(a)$ }
    \includegraphics[width=\textwidth]
    {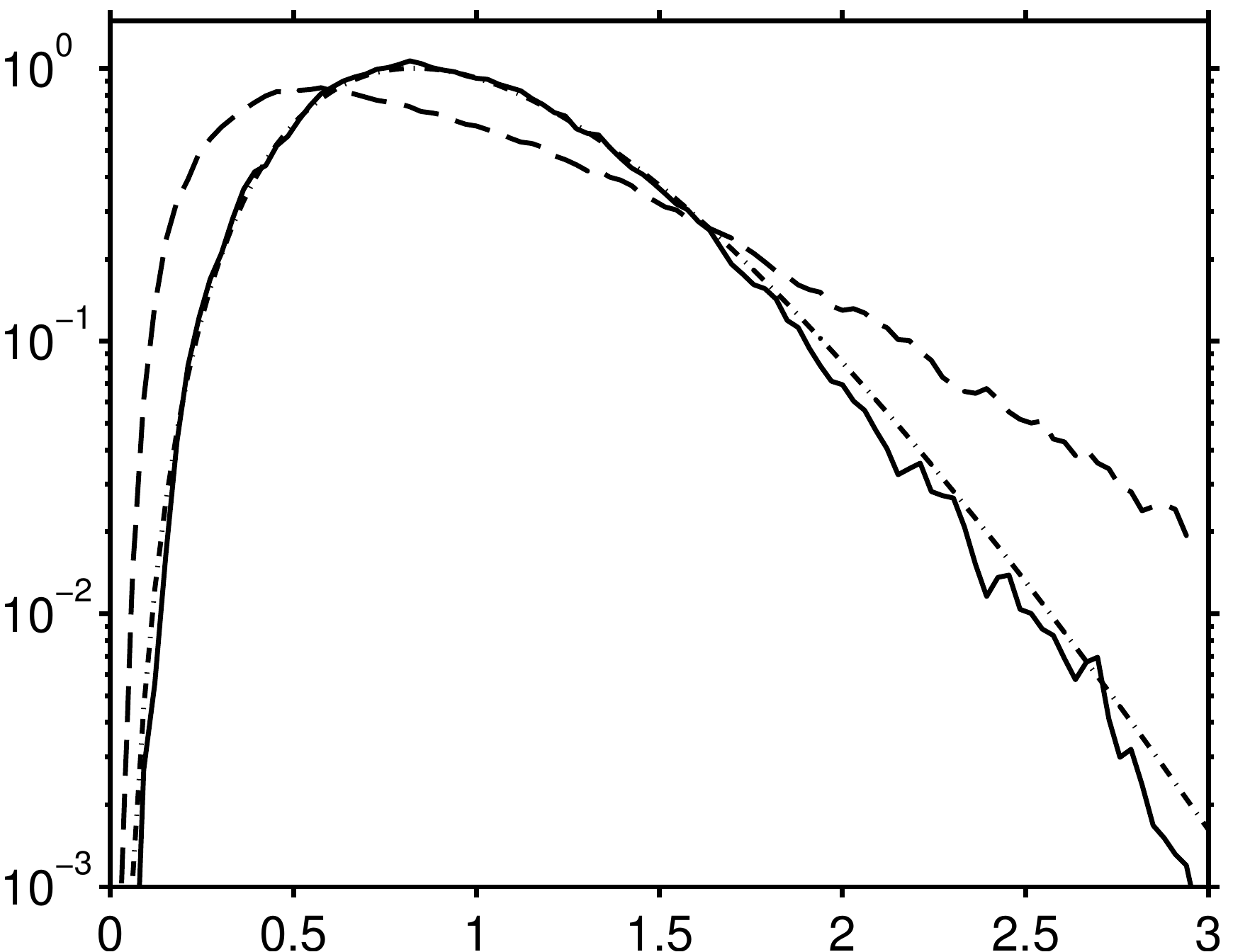}
    \centerline{ $V_i/\langle V \rangle_{p,t}$ }
  \end{minipage}
  \hfill
  \begin{minipage}{2ex}
    \rotatebox{90}{pdf}
  \end{minipage}
  \begin{minipage}{0.45\textwidth}
    \centerline{ $(b)$ }
    \includegraphics[width=\textwidth]
    {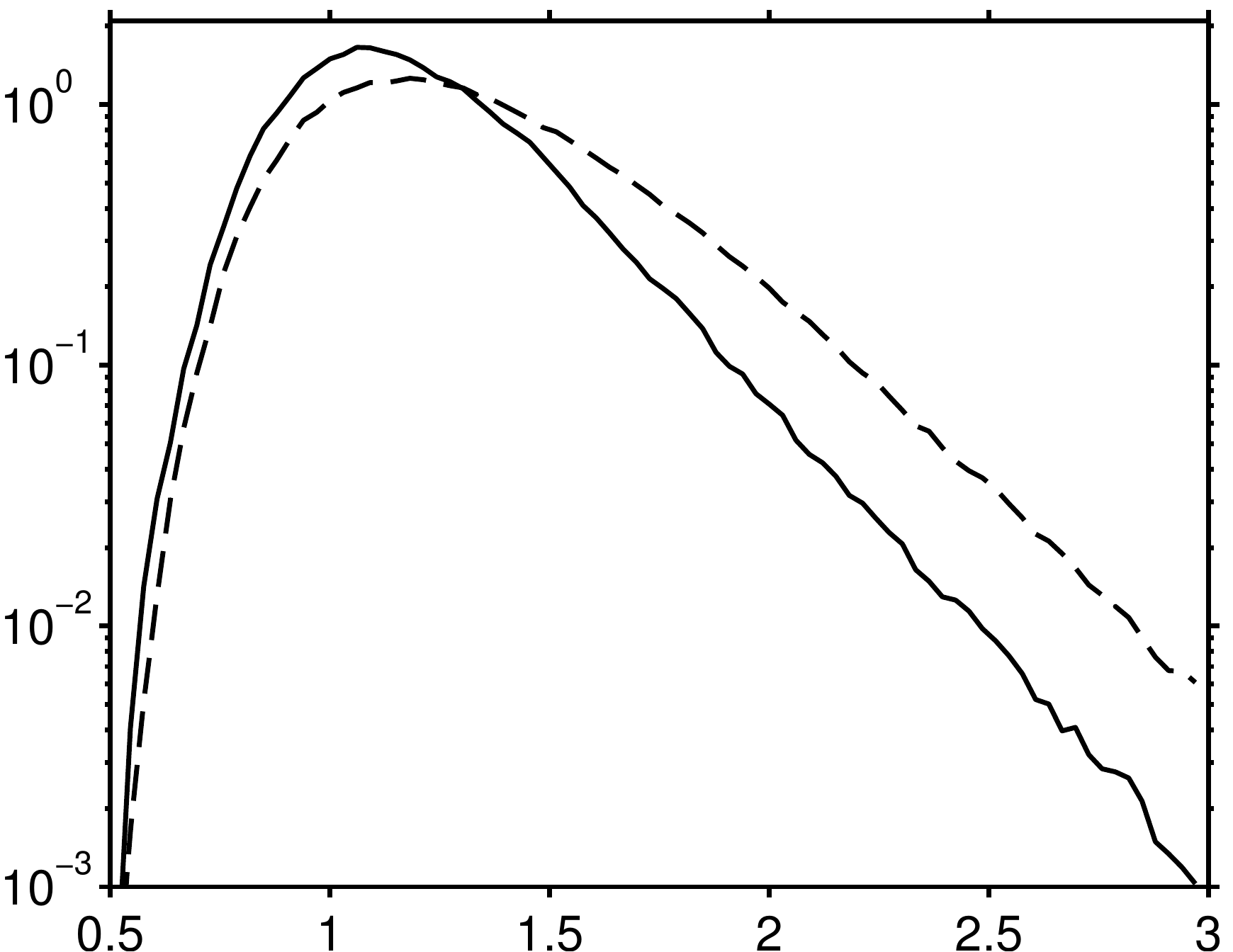}
    \centerline{ $A_i$ }
  \end{minipage}
  \caption{%
    As figure~\ref{fig:voronoi-VOL-AR-PDF-M121}, but for case
    M178.
  }
  \label{fig:voronoi-VOL-AR-PDF-M178}
\end{figure}

Figures~\ref{fig:voronoi-VOL-AR-PDF-M121}a and
\ref{fig:voronoi-VOL-AR-PDF-M178}a depict the p.d.f.\ of the
normalized \Vor volumes for both flow cases at different times during
the course of the simulations. Except for the initial state, the
tesselations are performed at several instants covering short time
intervals of length of less than $30\tau_g$ and then averaged to
increase the number of samples. 
It can be seen that the initial state (which was determined
pseudo-randomly) is well represented by a Gamma distribution which
corresponds to a random Poisson process \citep{ferenc:07}. 
In case \caseA{} (figure~\ref{fig:voronoi-VOL-AR-PDF-M121}a) the
p.d.f.\ of the \Vor cell volumes tends to a shape 
which is narrower than the initial random particle distribution. This
indicates that the particle distribution in case \caseA{} becomes more
ordered than randomly distributed particles.  
Conversely, in case \caseB{} (figure~\ref{fig:voronoi-VOL-AR-PDF-M178}a),
the curves broaden once particles are released. In the statistically
stationary regime they exhibit tails with significantly higher
probability of finding cells with very large or very small volumes
than in the random case, indicating that cluster (and void) formation
takes place.

\begin{figure}
  \centering
  \begin{minipage}{2ex}
    \rotatebox{90}{$\sigma(V_i)$ }
  \end{minipage}
  \begin{minipage}{0.45\textwidth}
    \includegraphics[width=\textwidth]
    {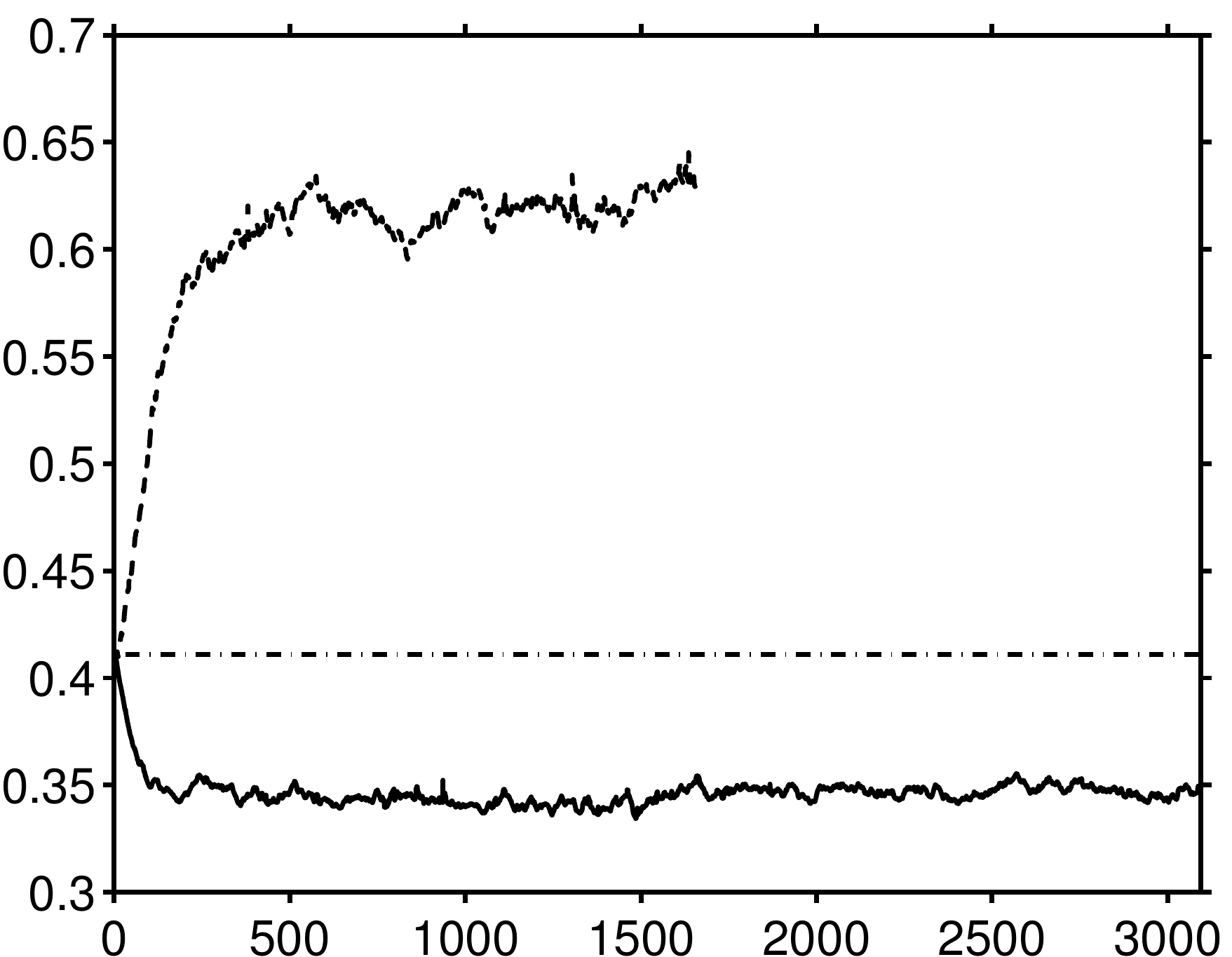}
    \hspace*{-.43\linewidth}\raisebox{.62\linewidth}{$Ga=178$}
    \hspace*{-5ex}\raisebox{.16\linewidth}{$Ga=121$}
    \\
    \centerline{$t/\tau_g$}
  \end{minipage}
  \caption{%
    Standard deviation of \Vor cell volumes as function of time: 
    case \caseA{} (solid line), case \caseB{} (dashed line), 
    randomly
    distributed particles (chain-dotted line). 
  }
  \label{fig:voronoi-VOL-STD}
\end{figure}

\cite{monchaux:10} have used the standard deviation of the p.d.f.\
of the normalized \Vor cell volumes in order to quantify the
particles' tendency to cluster. 
Figure~\ref{fig:voronoi-VOL-STD} depicts this quantity as a 
function of time for our two cases. 
As can be seen, the standard
deviation in case \caseA{} decreases after particles have been
released to settle freely and quickly reaches a state where it 
fluctuates lightly around a value of $0.345$. 
  By way of comparison, the value corresponding to the random
  arrangement measures $0.411$ \citep[note that this value is slightly
  smaller than the value of $0.447$ computed by][due to the finite
  size of the present particles]{ferenc:07}.
Contrarily, the standard deviation in case \caseB{} increases with
time and reaches a statistically steady value of $0.616$,
exhibiting somewhat larger temporal fluctuations.  
From the figure it is clearly observable that
the transient time for the particles to reach a
statistically steady state is considerably longer in case \caseB{} 
than in case \caseA{}, in line with the observed transients 
of particle and fluid velocity fluctuations discussed in
\S~\ref{sec-results-transient}. 

We have further analyzed the particle distribution by
computing the aspect ratio of each \Vor cell, defined as the
ratio of its largest horizontal extension $l_{H}({\cal V}_i)$ to its largest
vertical extension $l_{V}({\cal V}_i)$, viz.
\begin{equation}
  \label{eq:VorAR}
  A_{i} = l_{H}({\cal V}_i) / l_{V}({\cal V}_i)
  \quad \quad \forall\;i=1...N_p.
\end{equation}
The so-defined aspect ratio provides a measure of the anisotropy of
the \Vor cells. 
Figures~\ref{fig:voronoi-VOL-AR-PDF-M121}b and
\ref{fig:voronoi-VOL-AR-PDF-M178}b show the data for both flow cases
again at different times during the simulation. 
In case \caseA{} no significant deviation from the p.d.f.\ of the
randomly distributed particles is observed.  
This means that the tendency towards a more ordered state (observed in
conjunction with the data for the \Vor cell volume, cf.\
figure~\ref{fig:voronoi-VOL-AR-PDF-M121}a) is not accompanied by 
a change in the geometrical aspect ratio at the lower Galileo number. 
In contrast, in case \caseB{} an appreciable
difference is observed in the statistically stationary state. 
Finding \Vor cells with large values of $A_i$ is more probable than in
the case of randomly distributed particles. 
This is an indication that the majority of the \Vor cells are
squeezed (stretched) in the vertical (horizontal) direction. 
The result is consistent with the fact that the particles in case
\caseB{} have a tendency to be aligned in the vertical direction, as
already visually noted for one snapshot in
figure~\ref{fig:part-pos-top-M178}b.    

The \Vor cell volume data can be used to define an intrinsic definition of
cluster regions and voids by determining the 
intersection points between the p.d.f.\ of the DNS data and that of
the random distribution: particles whose \Vor cell
volume is smaller than the lower intersection point $V^c$ are
considered to be located in a cluster and those whose \Vor cell volume is
larger than the upper intersection point $V^v$ are attributed to void
regions \citep{monchaux:10}. 
By tracking the particle positions in time, 
we can compute statistics on the temporal behavior with respect to
clustering. 
It turns out that in case M178 the mean residence time of the particles in 
clusters (voids) measures $24.9 \tau_g$ ($87.7 \tau_g$). 
This indicates, that once the particles enter clusters or void regions
they remain trapped therein for relatively long times, with a 
bias towards voids. 
It should be noted, however, that the \Vor tesselation attributes
rather large cell volumes to particles located on the periphery of
clusters. 
We have checked that with the above definition, at any given time in
the statistically stationary regime of case M178 
approximately $30\%$ of the particles reside in clusters and $15\%$ in
void regions, with little temporal fluctuations. 

Since each \Vor cell can be uniquely assigned to a single particle,
the derived quantities can be tracked in time. 
This allows us to study the behavior of these geometrical quantities
in a Lagrangian manner. 
The auto-correlation function related to a quantity $Q_{\cal V}(t)$
is defined as 
\begin{equation}
  \label{equ-res-spatial-def-lag-corr-qv}
  R_{Q_{\cal V} Q_{\cal V}} (\tau_{sep}) 
  = 
  \frac{\langle Q_{\cal V}^\prime(t) Q_{\cal V}^\prime(t + \tau_{sep})
    \rangle_{p,t}}
  {\langle Q_{\cal V}^\prime(t) Q_{\cal V}^\prime(t) \rangle_{p,t}},
\end{equation}
where $Q_{\cal V}^\prime(t) = Q_{\cal V}(t) - \langle Q_{\cal V}
\rangle_{p,t}$ is the 
fluctuation with respect to the mean over the particle ensemble and
time (cf.\ appendix~\ref{sec:app-averaging}). 

\begin{figure}
  \begin{minipage}{2ex}
    \rotatebox{90}{$R_{Q_{\cal V} Q_{\cal V}}$}
  \end{minipage}
  \begin{minipage}{0.45\textwidth}
    \centerline{$(a)$}
    \includegraphics[width=\textwidth]
    {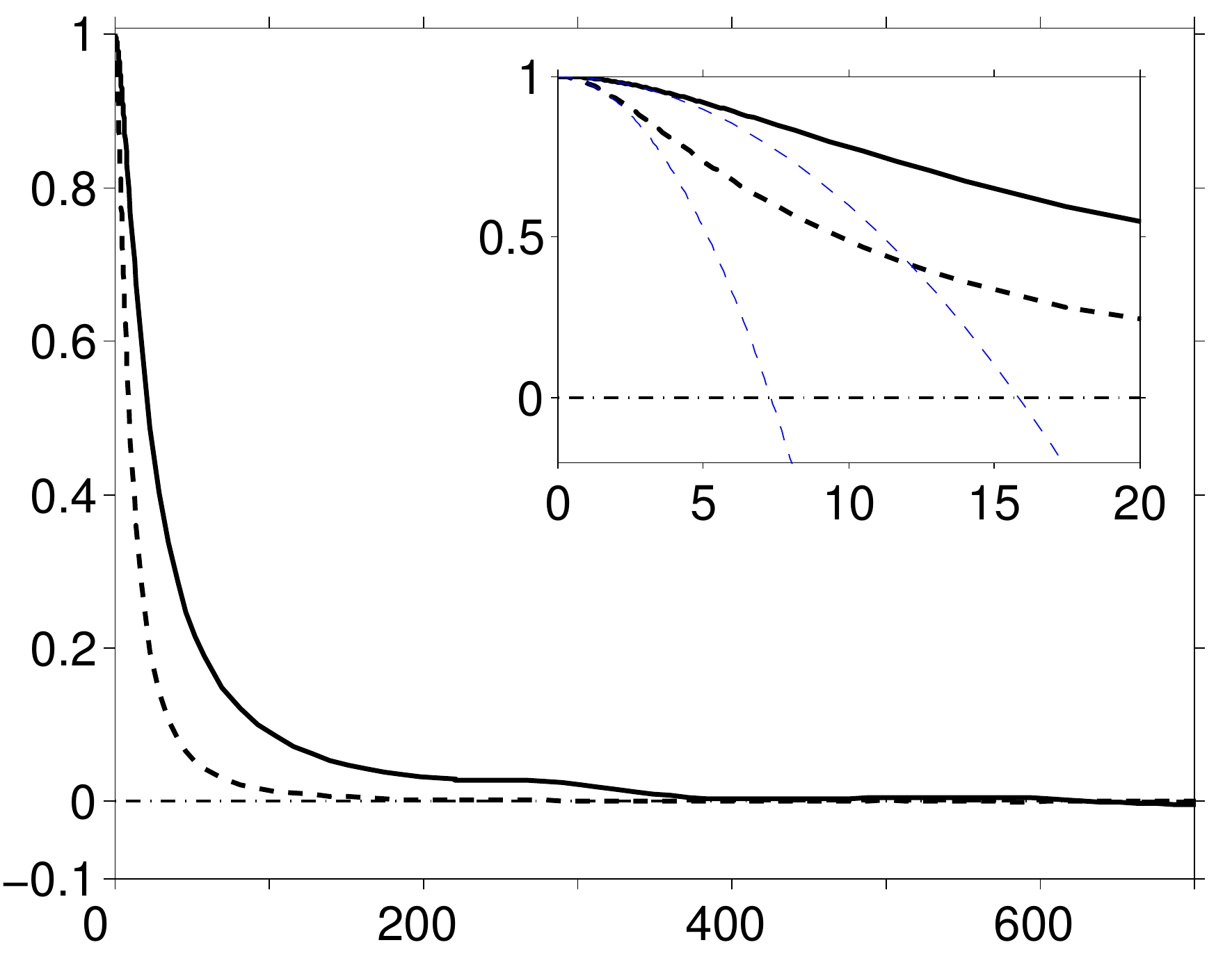}
    \centerline{$\tau_{sep}/\tau_g$}
  \end{minipage}
  \hfill
  \begin{minipage}{2ex}
    \rotatebox{90}{$R_{Q_{\cal V} Q_{\cal V}}$}
  \end{minipage}
  \begin{minipage}{0.45\textwidth}
    \centerline{$(b)$}
    \includegraphics[width=\textwidth]
    {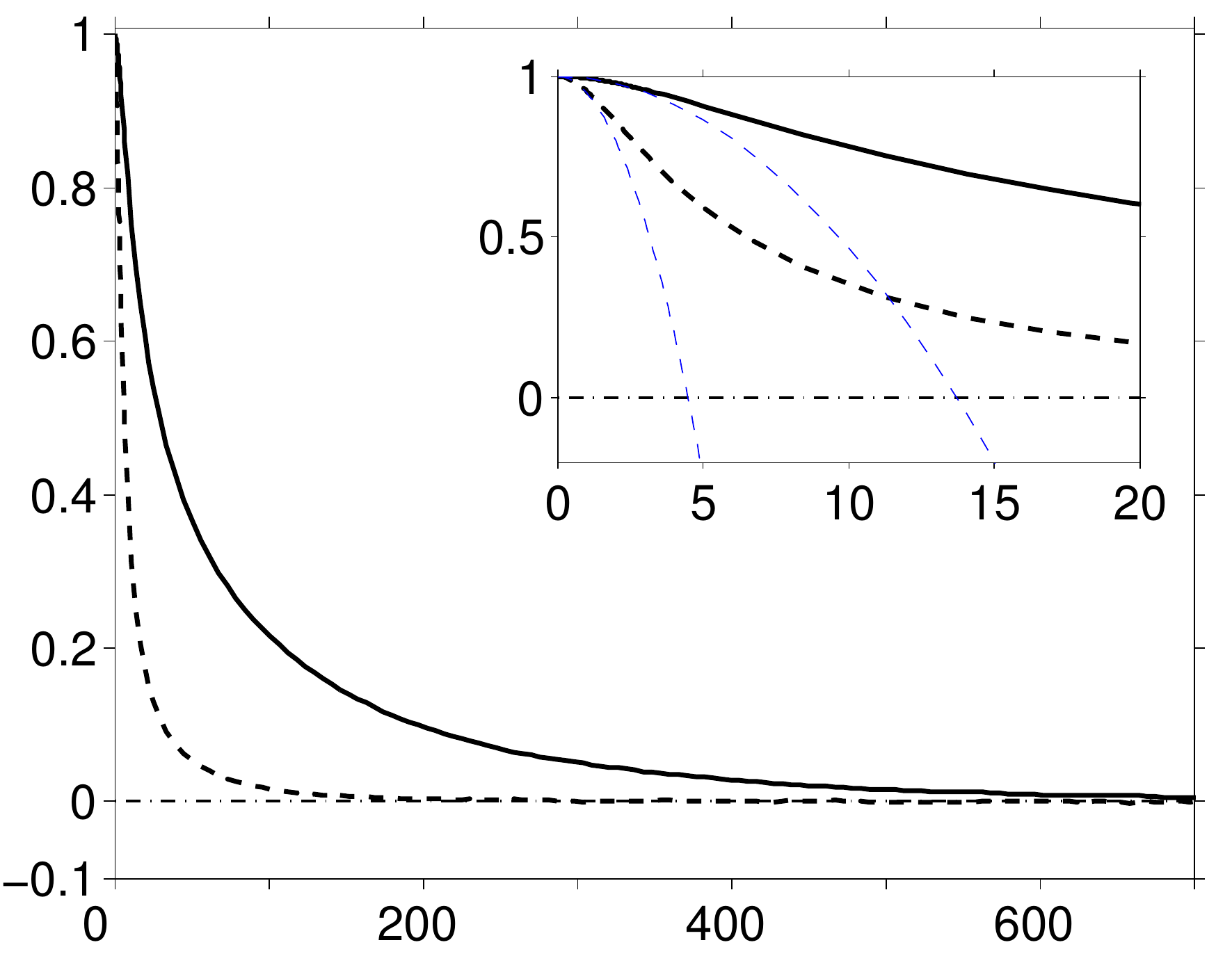}
    \centerline{$\tau_{sep}/\tau_g$}
  \end{minipage}
  \caption{%
    Lagrangian autocorrelation of \Vor volumes $R_{V_iV_i}$ (solid 
    line) and 
    of the \Vor cell aspect ratio $R_{A_i A_i}$ (dashed line)
    as function of the separation time $\tau_{sep}$ (cf.\ definition in
    \ref{equ-res-spatial-def-lag-corr-qv}) for  
    (a) case \caseA{} and
    (b) case \caseB{}. 
    The insets show closeups of the same data 
    for small separations, including the osculating parabolas  
    at $\tau_{sep}= 0$.  
  }
  \label{fig:voronoi-VOL-CORR}
\end{figure}

The auto-correlation corresponding to the \Vor cell volume is shown in
figure~\ref{fig:voronoi-VOL-CORR}.  
It can be observed that the correlation time is somewhat larger in the
clustering case M178 than in the counterpart M121 where no clustering
takes place. The corresponding integral time scale measures  
$43.7\tau_g$ ($73.7 \tau_g$) in case M121 (M178). 
The Taylor micro-scale, however, is similar in both cases:
$15.8\tau_g$ (case M121), $13.7 \tau_g$ (case M178).  
Regarding the \Vor cell aspect ratio 
(included in the same figure)  
it exhibits a similar behavior in both flow cases, featuring a 
considerably faster decorrelation than the cell volume. 
The corresponding integral time scale measures $17.2 \tau_g$ ($14.2
\tau_g$) in case M121 (M178). 
Note that the \Vor cell aspect ratio is a geometrical property of the
shape of the cells; therefore, it may be significantly modified while
the particle positions dynamically rearrange without necessarily changing
the volume of the corresponding cells.
\subsubsection{Conditionally averaged particle concentration} 
\label{sec-results-spatial-struct-partconc}
From the \Vor analysis of \S~\ref{sec-results-spatial-struct-voronoi} 
we are able to determine whether the particles tend to form
agglomerations as a whole -- which was seen to be the case in
simulation M178.  
In this section we study the {\it local} particle distribution 
in the vicinity of a test particle. 
This kind of analysis will provide more insight into the
micro-structure of the dispersed phase. 
For this purpose we have performed particle-conditioned averaging of
the locations of the remaining particles. 
Since the conditioned field is axi-symmetric (the axis coinciding with
the vertical direction), the significant data is defined in a plane
including the vertical direction.
Using the computational grid of the simulations we first define for
each grid cell a solid phase indicator function. The origin of this
field is then shifted to each particle location (taking into account
periodic wrap-around) and added up; this process is then repeated
(additively) for a number of $16816$ ($26180$) particle snapshots in
case M121 (M178) and dividing by the number of samples, yielding a
field with values in the interval [0,1] \citep{kidanemariam:13}. 
The result, which will be denoted by
$\phi_s^{cond}(\tilde{r},\tilde{z})$, can be interpreted as the
probability that a point with coordinates $(\tilde{r},\tilde{z})$ --
with respect to the center of a test particle -- is located inside the 
solid phase domain $\Omega_p$ (cf.\
appendix~\ref{sec:app-cond-averaging} for a precise definition). 
When normalizing this quantity with the global solid volume fraction
$\Phi_s$, the result 
is closely related to the pairwise distribution function 
\citep{sundaram:97,shotorban:06,sardina:12}. 
For visual clarity the following visualizations are shown
(redundantly) for the half plane instead of the fundamental quadrant.  

\begin{figure}
  \begin{minipage}{3ex}
    \rotatebox{90}{$\tilde{z}/D$}
  \end{minipage}
  \begin{minipage}{0.4\textwidth}
    \centerline{$(a)$}
    \includegraphics[height=.87\textwidth,clip=true,
    viewport=190 40 1005 850]
    {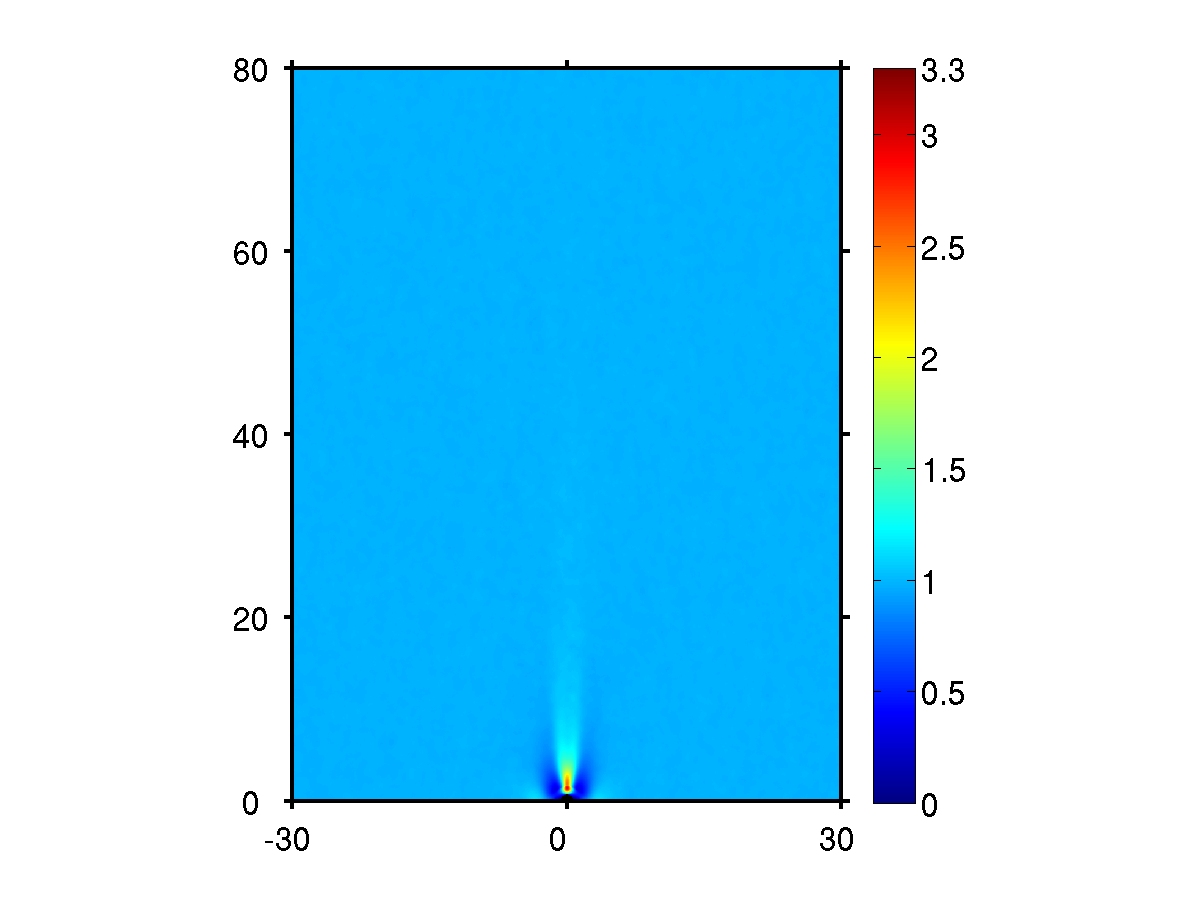}
    \\
    \centerline{$\tilde{r}/D$}
  \end{minipage}
  \hspace*{-5ex}
  \begin{minipage}{3ex}
    $\frac{\phi_s^{cond}}{\Phi_s}$ 
  \end{minipage}
  \hfill
  \begin{minipage}{3ex}
    \rotatebox{90}{$\tilde{z}/D$}
  \end{minipage}
  \begin{minipage}{0.348\textwidth}
    \centerline{$(b)$}
    \includegraphics[height=\textwidth,clip=true,
    viewport=260 40 940 850]
    {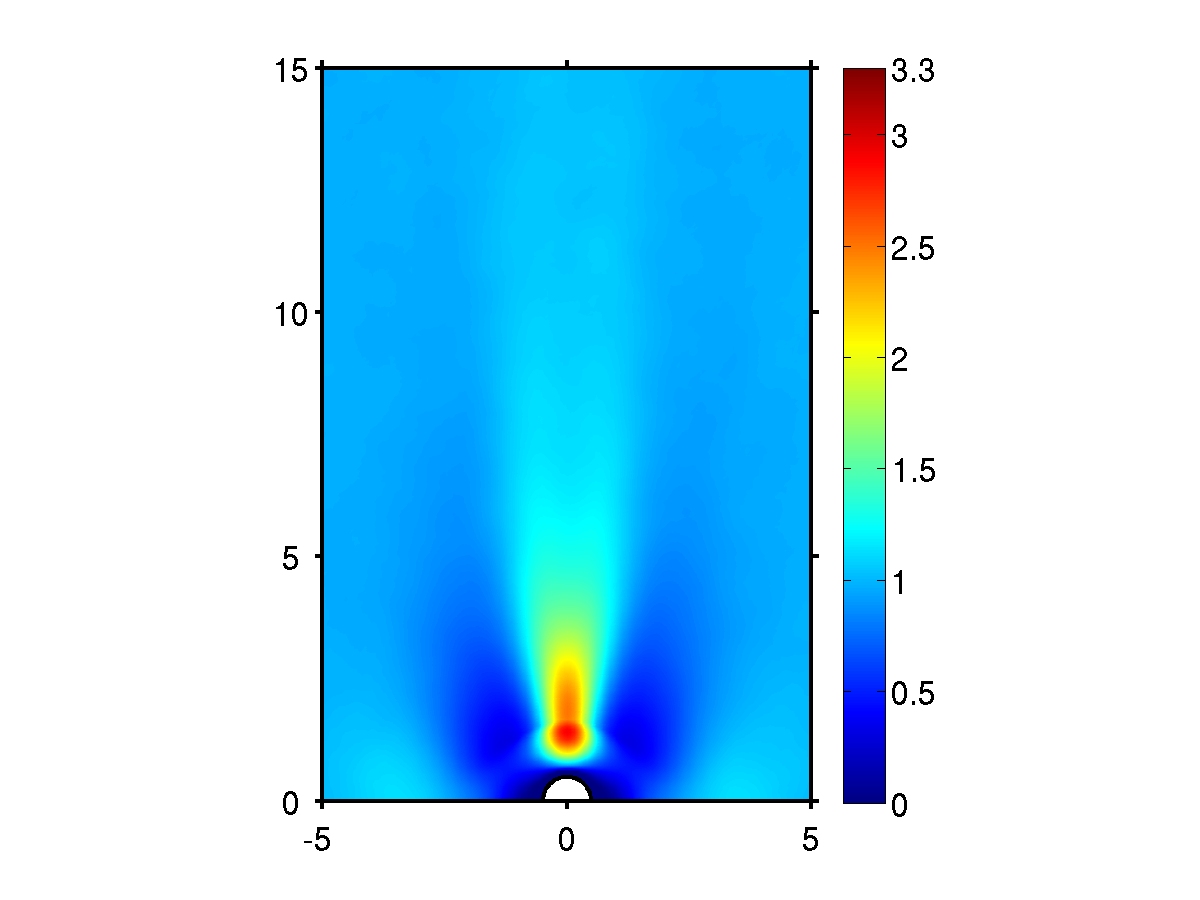}
    \\
    \centerline{$\tilde{r}/D$}
  \end{minipage}
  \hspace*{-5ex}
  \begin{minipage}{3ex}
    $\frac{\phi_s^{cond}}{\Phi_s}$ 
  \end{minipage}
  \caption{%
    (a) Averaged solid volume fraction 
    conditioned on particle positions for case \caseA{},
    normalized by the global solid volume fraction.  
    (b) Close-up of the same data. 
    The coordinates $\tilde{r}$ and $\tilde{z}$ are relative to
    the test particle. Since the data is axi-symmetric, $\tilde{r}$
    denotes the distance in the horizontal plane. 
  }
  \label{fig:part-2D-num-den-ver-M121}
\end{figure}

\begin{figure}
  \begin{minipage}{3ex}
    \rotatebox{90}{$\tilde{z}/D$}
  \end{minipage}
  \begin{minipage}{0.4\textwidth}
    \centerline{$(a)$}
    \includegraphics[height=.87\textwidth,clip=true,
    viewport=190 40 1005 850]
    {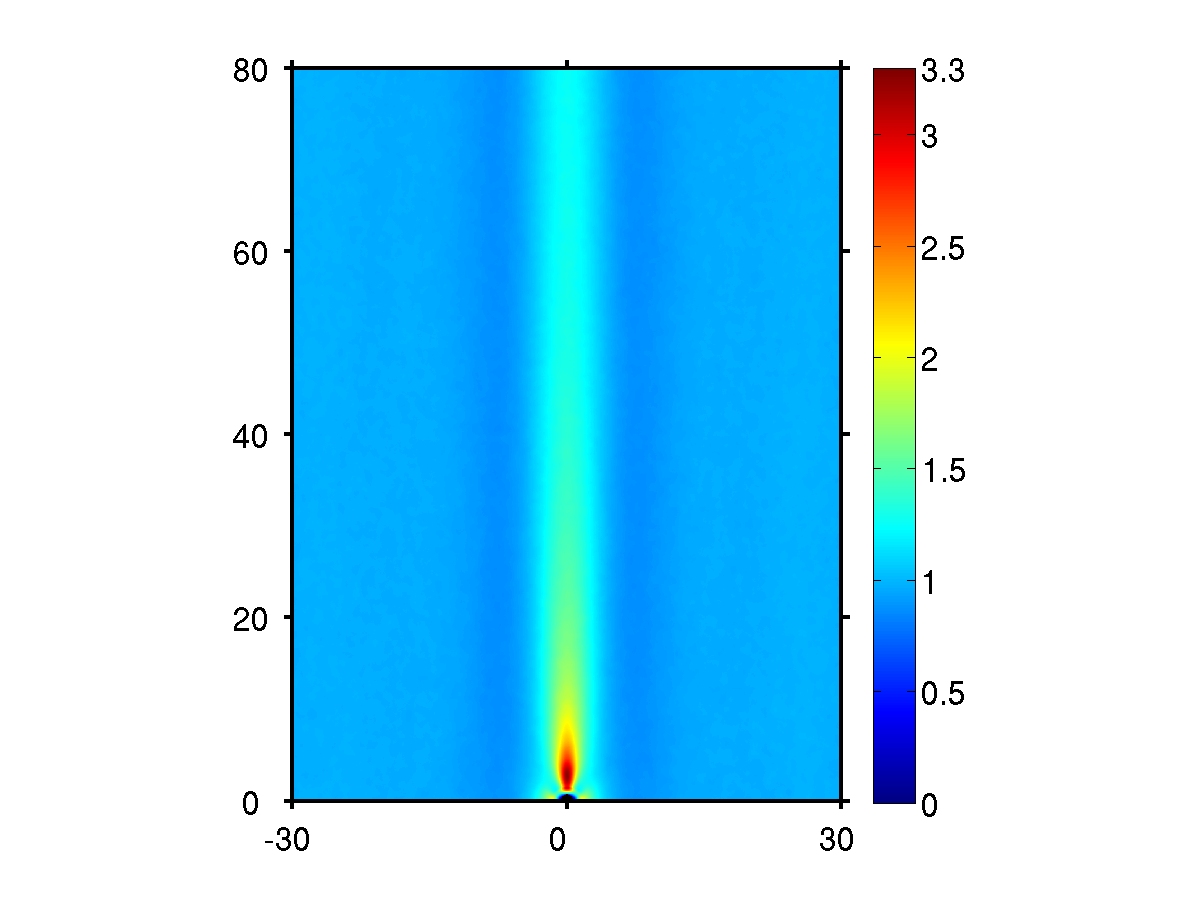}
    \\
    \centerline{$\tilde{r}/D$}
  \end{minipage}
  \hspace*{-5ex}
  \begin{minipage}{3ex}
    $\frac{\phi_s^{cond}}{\Phi_s}$ 
  \end{minipage}
  \hfill
  \begin{minipage}{3ex}
    \rotatebox{90}{$\tilde{z}/D$}
  \end{minipage}
  \begin{minipage}{0.348\textwidth}
    \centerline{$(b)$}
    \includegraphics[height=\textwidth,clip=true,
    viewport=260 40 940 850]
    {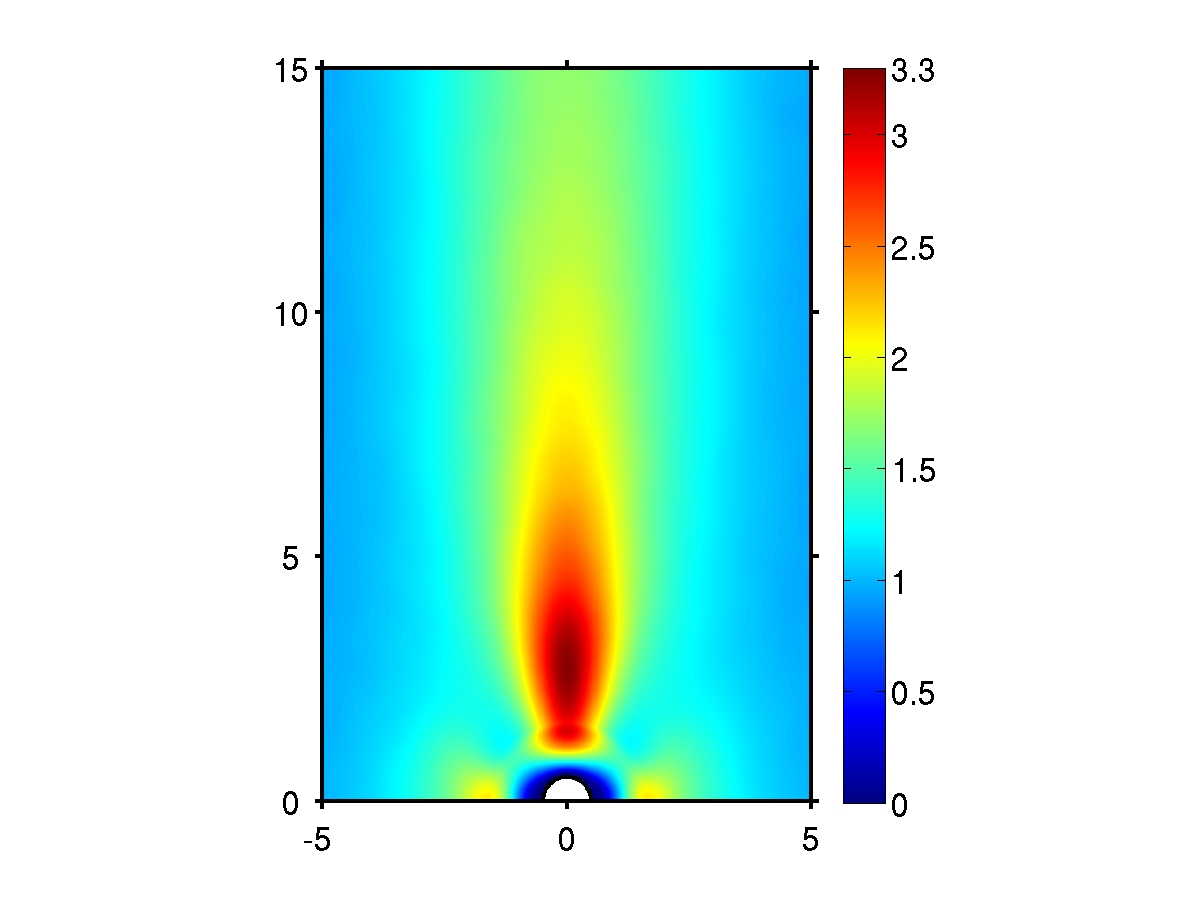}
    \\
    \centerline{$\tilde{r}/D$}
  \end{minipage}
  \hspace*{-5ex}
  \begin{minipage}{3ex}
    $\frac{\phi_s^{cond}}{\Phi_s}$ 
  \end{minipage}
  \caption{%
    As figure~\ref{fig:part-2D-num-den-ver-M121}, but for case M178. 
  }
  \label{fig:part-2D-num-den-ver-M178}
\end{figure}

Figures~\ref{fig:part-2D-num-den-ver-M121} and
\ref{fig:part-2D-num-den-ver-M178}  
show the two dimensional maps of 
$\phi_s^{cond}/\Phi_s$ 
for both flow cases.
As can be clearly seen, on a 
macroscopic level the dispersed phase in case \caseA{}
(figure~\ref{fig:part-2D-num-den-ver-M121}$a$) does not show any signs 
of accumulation, and particles appear to be randomly distributed. 
Only at close proximity of the reference particle can we observe a
non-trivial structure (figure~\ref{fig:part-2D-num-den-ver-M121}b): 
the probability of finding another particle is significantly larger
than the global average on the vertical axis at a distance around one
particle diameter; 
it is below average principally at small horizontal separations from
the test particle and over a range of angles with respect to the
horizontal. 

In case \caseB{} (figure~\ref{fig:part-2D-num-den-ver-M178}a), the
distribution of  
the conditioned solid volume fraction clearly exhibits a strongly 
anisotropic spatial structure even at the macroscopic level, featuring
high values of the local concentration in form of a streamwise
elongated stripe around the vertical axis. 
As can be seen, the region of high local particle density extends
throughout the entire height of the computational domain.  
This is in line with the observed clustering in columnar structures. 
The micro-structure in case M178
(figure~\ref{fig:part-2D-num-den-ver-M178}b) also differs from the
lower Galileo case M121: in particular, a small region with larger
than average solid volume fraction is found at horizontal distances
around $1.5D$. 

\begin{figure}
  \begin{minipage}{2ex}
    \rotatebox{90}{${\phi_s^{cond}}/{\Phi_s}$}
  \end{minipage}
  \begin{minipage}{0.45\textwidth}
    \centerline{$(a)$}
    \includegraphics[width=\textwidth]
    {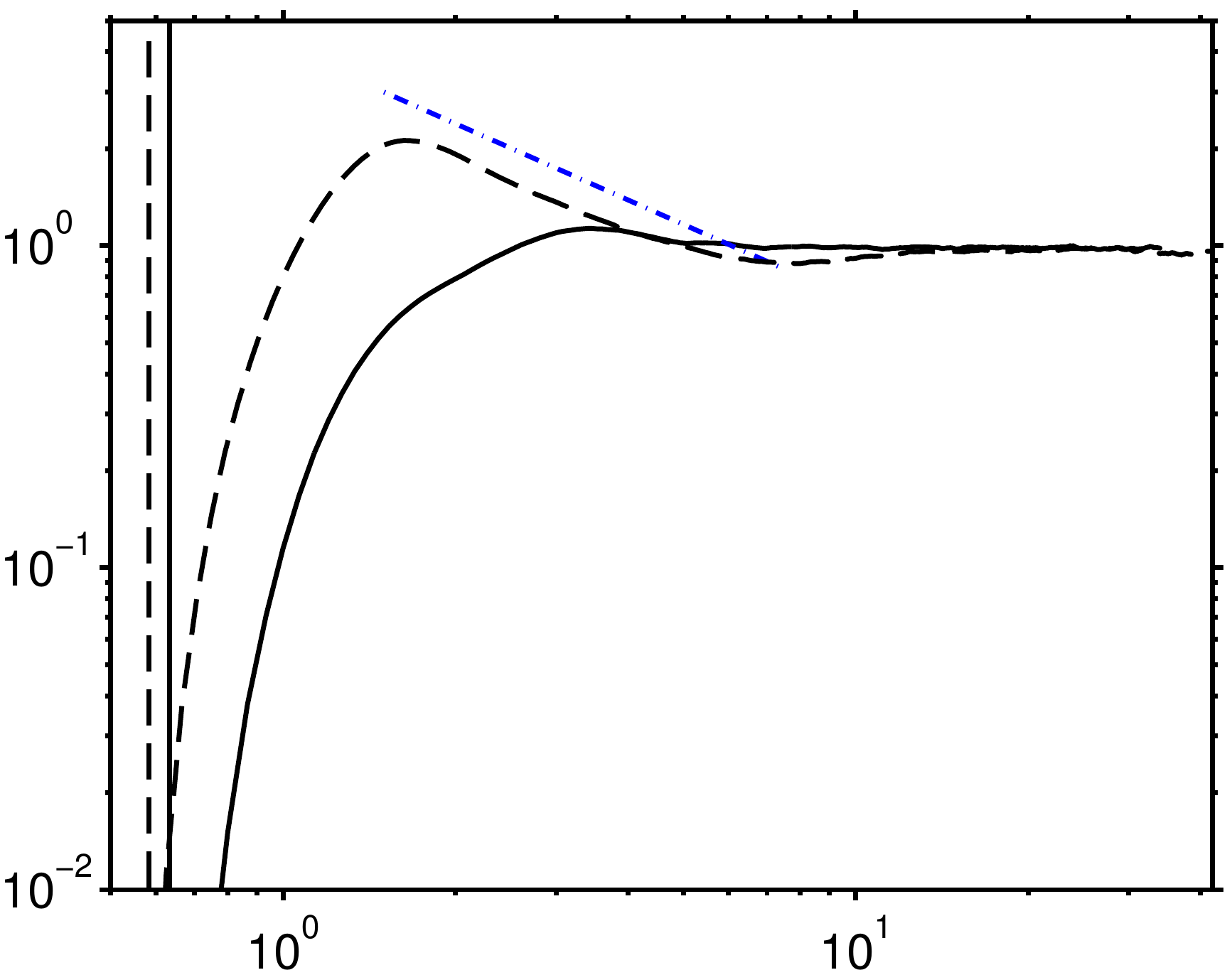}
    \centerline{$\tilde{r}/D$  }
  \end{minipage}
  \hfill
  \begin{minipage}{2ex}
    \rotatebox{90}{${\phi_s^{cond}}/{\Phi_s}$}
  \end{minipage}
  \begin{minipage}{0.45\textwidth}
    \centerline{$(b)$}
    \includegraphics[width=\textwidth]
    {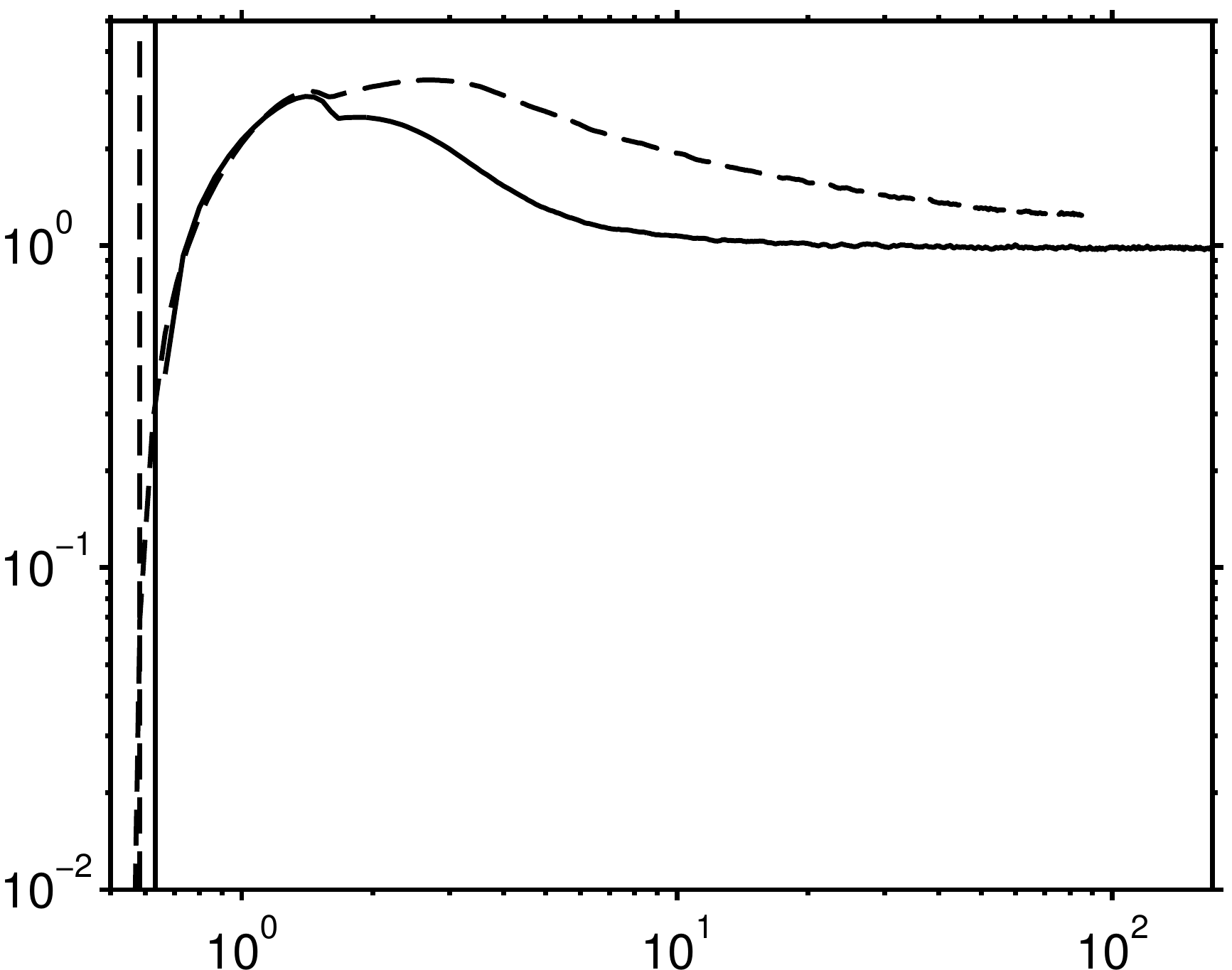}
    \centerline{$\tilde{z}/D$}
  \end{minipage}
  \caption{%
    Variation of the particle-conditioned solid volume fraction
    $\phi_s^{cond}$ along two perpendicular axes through the test
    particle:  
    (a) for $\tilde{z}=0$; 
    (b) $\tilde{r}=0$. 
    Shown are case \caseA{} (solid line) and case
    \caseB{} (dashed line). 
    The vertical lines
    represent for each case the distance
    $\vert \tilde{\mathbf{x}}_{ijk} \vert = D/2 + 2\Delta x $,
    which marks the limit of the range of action of the
    inter-particle repulsion force. 
    In ($a$) the chain-dotted line 
    is proportional to 
    $\tilde{r}^{-0.65}$.
  }
  \label{fig:part-2D-num-den-centerline}
\end{figure}

The spatial variation of the conditioned solid volume fraction 
can be more conveniently examined in a quantitative fashion by
considering vertical and horizontal profiles  
through the center of the reference particle,
as shown in figure~\ref{fig:part-2D-num-den-centerline}. 
Let us first consider the horizontal direction 
(figure~\ref{fig:part-2D-num-den-centerline}a).
In case \caseA{} the 
normalized solid volume fraction 
increases slowly and reaches unity at a horizontal distance of
approximately $\tilde{r}= 2.5D$ from the surface
of the reference particle; beyond that distance the solid volume
fraction remains close to unity. 
In case \caseB{}, on the other hand, a faster increase with
horizontal distance is observed, crossing unity at 
$\tilde{r}\approx D$, 
leading up to a maximum value $\phi_s^{cond}/\Phi_s\approx2$ located
at $\tilde{r}\approx1.5D$, then levelling back off 
towards unity at a rate approximately proportional to
$\tilde{r}^{-0.65}$. 
The lateral region which is depleted of particles is therefore much
smaller in the clustering case M178 than in case M121.  

Turning to the vertical direction
(figure~\ref{fig:part-2D-num-den-centerline}b), 
we observe nearly identical conditioned solid volume fraction values
in both cases for small distances from the test particle (up to
$1.5D$). Beyond that point, the curve corresponding to case M121
decreases and reaches unity from above at a distance of roughly $8D$. 
In case M178, however, the probability to find a particle keeps
increasing up to a vertical distance of approximately $3D$, beyond
which point it slowly
decays, not reaching unity within the present domain size.
Thus, the macroscopic behavior of the particle-conditioned solid
volume fraction along the vertical axis directly reflects the observed 
clustering of particles in case M178.  
The micro-structure, however, is remarkably similar in both cases. 
It appears that a ``robust'' mechanism is at play, which does not
strongly depend upon the Galileo number over the presently
investigated range. 

At lower Reynolds number (up to a value of 10) information on the
micro-structure of the dispersed phase is available for rising bubbles
from the experimental measurements of \cite{cartellier:09} and for
sedimenting suspensions from the numerical simulations of
\cite{yin:08}.  
In the former study, a deficit of close vertically-aligned bubble
pairs was observed with a decreasing extent when the bubble
concentration (void fraction) was increased.  
In the case of \cite{yin:08} the solid volume fraction was kept
constant at a value of $0.01$, while the particle Reynolds number was
varied (in the range of 1 to 10). The authors found that the extent of
the deficit of vertical pairs grows with the Reynolds number and that
an excess of horizontal pairs occurs at $Re=10$; both observations 
were attributed to wake-induced ``drafting-kissing-tumbling'' of
particle pairs \citep[cf.][]{fortes:87,wu:98}. 
The micro-structure observed by \cite{yin:08} is therefore
qualitatively different from the present cases. 
At the time being, however, the gap in parameter space with respect to
those lower Reynolds number suspensions appears too large to be
bridged by straightforward arguments. 
\subsection{Particle motion} 
\label{sec:sec-part-motion}
In order to analyze the motion of finite-size particles in detail, it
is useful to work with relative velocities between the two phases. 
This immediately leads to the question of the most adequate definition
of a fluid velocity with respect to which the relative particle motion
should be computed. 
Therefore, before turning to the actual statistical results, we
first provide a definition of the fluid velocity in the
vicinity of a finite-size particle. We believe that this step is vital
to an understanding of the dynamics of finite-size particles where no
straightforward measure of the fluid velocity relevant to a given
particle exists.  
\subsubsection{Fuid velocity seen by the particles} 
\label{sec-part-uf-seen-by-partices}
There have been various approaches to defining a fluid velocity ``seen''
by the particles. 
\cite{bagchi:03} investigated the effect of turbulence on the drag
and lift of a single particle, performing DNS of homogeneous-isotropic
turbulence swept past a fixed sphere. The fluid velocity
``seen'' by the particle was alternatively defined either
(i)~as the fluid velocity at the position of the particle in an
(otherwise identical) particle-free simulation, or
(ii)~as the fluid velocity averaged over a spherical volume centered
at the particle centroid in the corresponding particle-free
simulation. 
\cite{merle:05} investigated the forces acting on a bubble in a
turbulent pipe flow. Taylor's hypothesis was invoked in order to
define the fluid velocity ``seen'' by the bubble without recurring to
a particle-free companion simulation. 
\cite{lucci:10} investigated the modification of 
homogeneous-isotropic turbulence by particles of size comparable to the
Taylor micro scale (in the absence of gravity). They proposed two
definitions of the fluid velocity ``seen'' by the particles: 
(i) the fluid velocity at a distance (comparable to the Taylor
micro-scale) away from the particle surface, taken in the direction of
the instantaneous particle velocity vector, and 
(ii) essentially the same definition, but averaging over a small
spherical cap.
In both cases, the use of the directional information from the
instantaneous particle velocity (in an inertial frame) is
questionable, as noted by \cite{kidanemariam:13}.  

Here, we use the definition proposed by \cite{kidanemariam:13}, where the
relevant fluid velocity in the vicinity of a particle is approximated
by the average of the fluid velocity over the surface
$\mathcal{S}$ of a sphere with radius ${R}_S$ centered at the
particle's center location; this velocity corresponding to the $i$th
particle at time $t$ will henceforth be denoted by  
$\mathbf{u}_f^\mathcal{S}(i,t,{R}_S)$.
Note that \cite{cisse:13} have recently proposed a similar 
  definition based upon the mass flux entering a spherical shell.
We presently choose the same averaging radius as
\cite{kidanemariam:13}, i.e.\ we set ${R}_S=1.5D$. 
This choice essentially assures on the one hand that the particle's
own boundary layer is not significantly affecting the resulting fluid
velocity measure while at the same time $R_S$ is sufficiently small
for the result to be directly relevant to the generation of the
hydrodynamic force (and torque) acting on the respective particle.  
The reader is referred to that reference for a discussion of the
chosen value of the averaging radius. 
From now on the symbol ``${R}_S$'' will be
dropped from the list of arguments of the quantity
$\mathbf{u}_f^\mathcal{S}$ for the sake of conciseness.   
In order to characterize the effect of particle-conditioned
averaging over a spherical surface, we have compared the statistics of
the resulting velocity values $\mathbf{u}_f^\mathcal{S}$ to those of
the entire (unconditioned) fluid velocity field $\mathbf{u}_f$. 
Information related to the first and second moments of the vertical
component are listed in table~\ref{tab-res-w-shell-moments}.  
It can be seen that in both flow cases the mean value of the
particle-conditioned vertical fluid velocity is smaller than the
unconditioned one. However, in case M121 the difference 
$\langle w_f^{\cal S}\rangle_{p,t}-\langle w_f\rangle_{\Omega_f,t}$
only amounts
to approximately 3\% of the absolute value of the conventional
settling velocity $|\langle w_s\rangle_t|$ (cf.\ equation
\ref{equ-res-trasient-def-w-settling}), 
while the (absolute value of the) difference is larger than 11\% in
case M178.  
Consequently, it can be said that the particles in the higher Galileo
number case M178 
are surrounded (on average) 
by fluid with a significantly smaller
vertical velocity than the box-averaged value, i.e.\ they are located
inside regions with a downward fluid motion (as compared to the
globally-averaged fluid velocity).   
Incidentally, let us mention that in both flow cases the standard
deviation of the vertical component of the particle-conditioned fluid
velocity is significantly smaller than the unconditioned
counterpart. This result is believed to reflect the fact that
averaging over the surface of a sphere with diameter $3D$ acts like a
spatial filter which tends to suppress small-scale fluctuations. 
However, the shapes of the normalized p.d.f.s of the
particle-conditioned fluid velocity field are very similar to those of
the unconditioned field (figures omitted).  
With the chosen definition of the relevant fluid velocity in the
vicinity of the particles, we can now define an instantaneous relative
velocity based upon a local reference value, viz.
\ben
\label{equ-res-part-motion-def-uprel-shell}
\mathbf{u}_{pr}^\mathcal{S}(i,t) 
=
\mathbf{u}_p^{(i)}(t)
-
\mathbf{u}_f^\mathcal{S}(i,t) 
\qquad \forall\; i=1 \ldots N_p
\,.
\een
The difference between this definition and the standard one given in
(\ref{equ-res-trasient-def-uprel}) is that in
(\ref{equ-res-part-motion-def-uprel-shell}) both 
values -- the particle and the fluid velocity -- are evaluated
locally. 
It can therefore be expected that the definition given in
(\ref{equ-res-part-motion-def-uprel-shell}) provides a better
representation of the local flow conditions which affect the particle
motion.  
The mean settling velocity with respect to each particle's surrounding
fluid region is then simply defined by applying an average over the
particle ensemble to the vertical component of
(\ref{equ-res-part-motion-def-uprel-shell}), viz.  
\ben
\label{eq:mean-app-vel-lag-shell}
w_{s}^{\mathcal{S}}(t) 
= 
\langle w_{pr}^\mathcal{S}(i,t) \rangle_p
\,.
\een 
In the limit of the averaging radius ${R}_S$ tending to infinity the 
particle-conditioned settling velocity according to definition
(\ref{eq:mean-app-vel-lag-shell}) tends to the conventional settling
velocity based on the globally averaged fluid velocity
(\ref{equ-res-trasient-def-w-settling}). 
For finite values of the shell radius $R_S$, the difference between
the present definition of the average settling velocity
(\ref{eq:mean-app-vel-lag-shell}) and the conventional definition
(\ref{equ-res-trasient-def-w-settling}) is, therefore, entirely due to 
the two different definitions of the fluid velocity, reflecting the
fact that particles do not sample the flow field uniformly. 
\subsubsection{Average settling velocity} 
\label{sec-part-motion-settl-vel}
Let us return to 
figure~\ref{fig:part-wmean}
which shows the average settling velocity (according to the two
alternative definitions) as a function of time.
Recall that in case M121 (where no clustering occurs) the collective
settling velocity defined with respect to the box-averaged fluid
velocity $w_s$ (cf.\ equation~\ref{equ-res-trasient-def-w-settling}) is
practically identical to the value of an isolated particle, whereas in
case M178 (in the presence of clustering) the collective effect
manifests itself through a strongly enhanced absolute value of $w_s$. 
When using the average settling velocity with respect to the local fluid
velocity $w_{s}^{\mathcal{S}}$ (defined in
\ref{eq:mean-app-vel-lag-shell}),  
the absolute value of the settling velocity is correspondingly reduced.
More specifically, in case M121 the average settling velocity with
respect to the fluid in each particle's vicinity remains close to the
settling velocity of an isolated particle (since $\langle w_f^{\cal
  S}\rangle_{p,t}$ is not too different from $\langle
w_f\rangle_{\Omega_f,t}$, as seen in
table~\ref{tab-res-w-shell-moments}).  
In case M178, however, the significant difference in the definition of
the fluid velocity brings the settling velocity 
$w_{s}^{\mathcal{S}}$ close to the single particle reference value 
(the absolute value of the former is upon time-average approximately
$3.3\%$ smaller than the latter). 
\%
\%
It appears that the particles in case M178 are (on average) 
locally moving at a vertical relative velocity 
which is similar to the value for 
a single particle, albeit slightly smaller. 
Our results therefore suggest that the observed enhanced global
settling rate in case M178 is a direct consequence of particles being
located preferentially in regions where the fluid velocity
fluctuation (with respect to the box-average) is negative. 

\begin{figure}
  \begin{minipage}{2ex}
    \rotatebox{90}{%
      $w_{pr}^\prime/
      \langle w_{pr}\rangle_{p,t}$
    }
  \end{minipage}
  \begin{minipage}{0.45\textwidth}
    \centerline{$(a)$}
    \includegraphics[width=\linewidth]
    {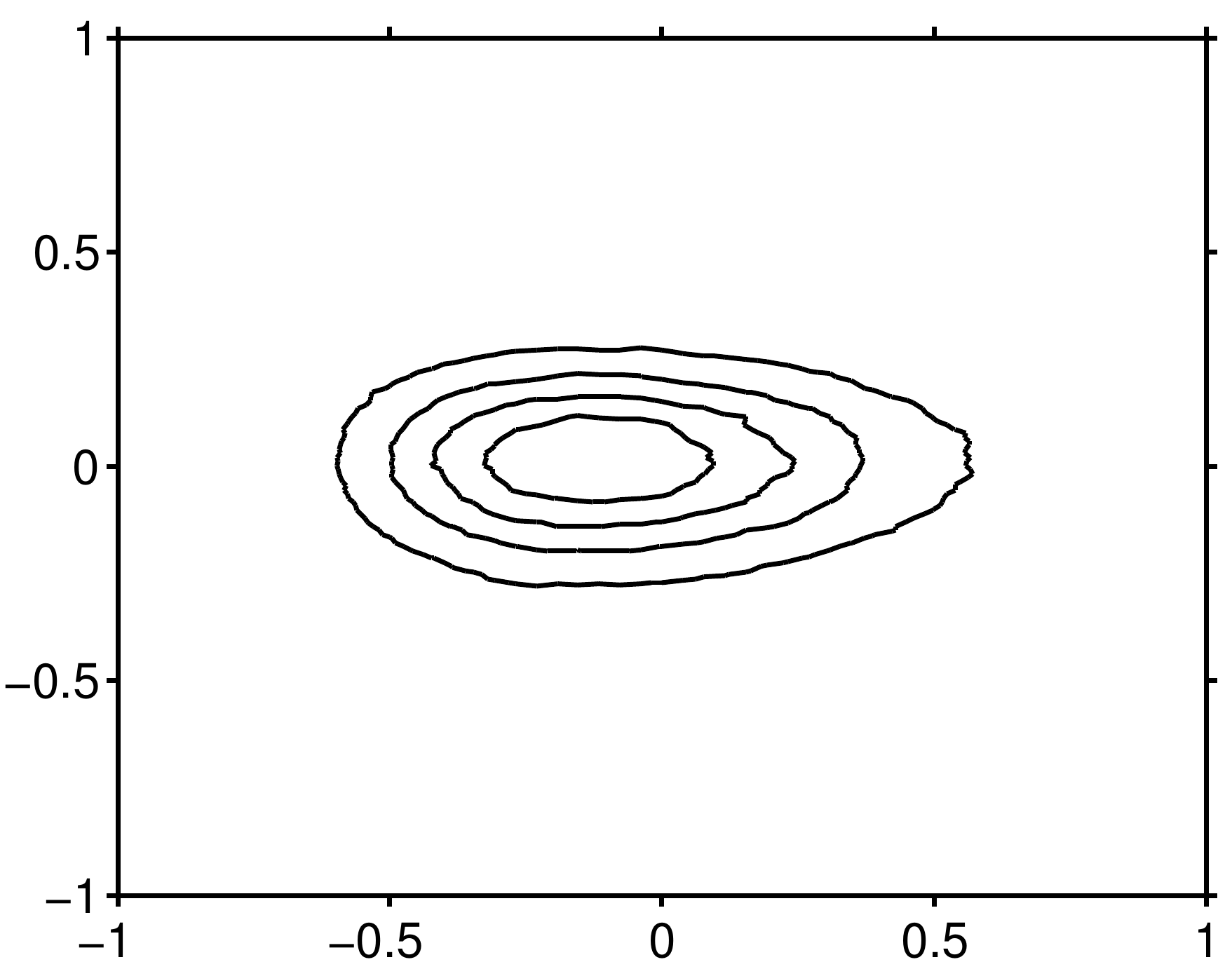}
    \\
    \centerline{
      $V_i^\prime/\langle V\rangle_{p,t}$
    }
  \end{minipage}
  \hfill
  \begin{minipage}{2ex}
    \rotatebox{90}{%
      $w_{pr}^\prime/
      \langle w_{pr}\rangle_{p,t}$
    }
  \end{minipage}
  \begin{minipage}{0.45\textwidth}
    \centerline{$(b)$}
    \includegraphics[width=\linewidth]
    {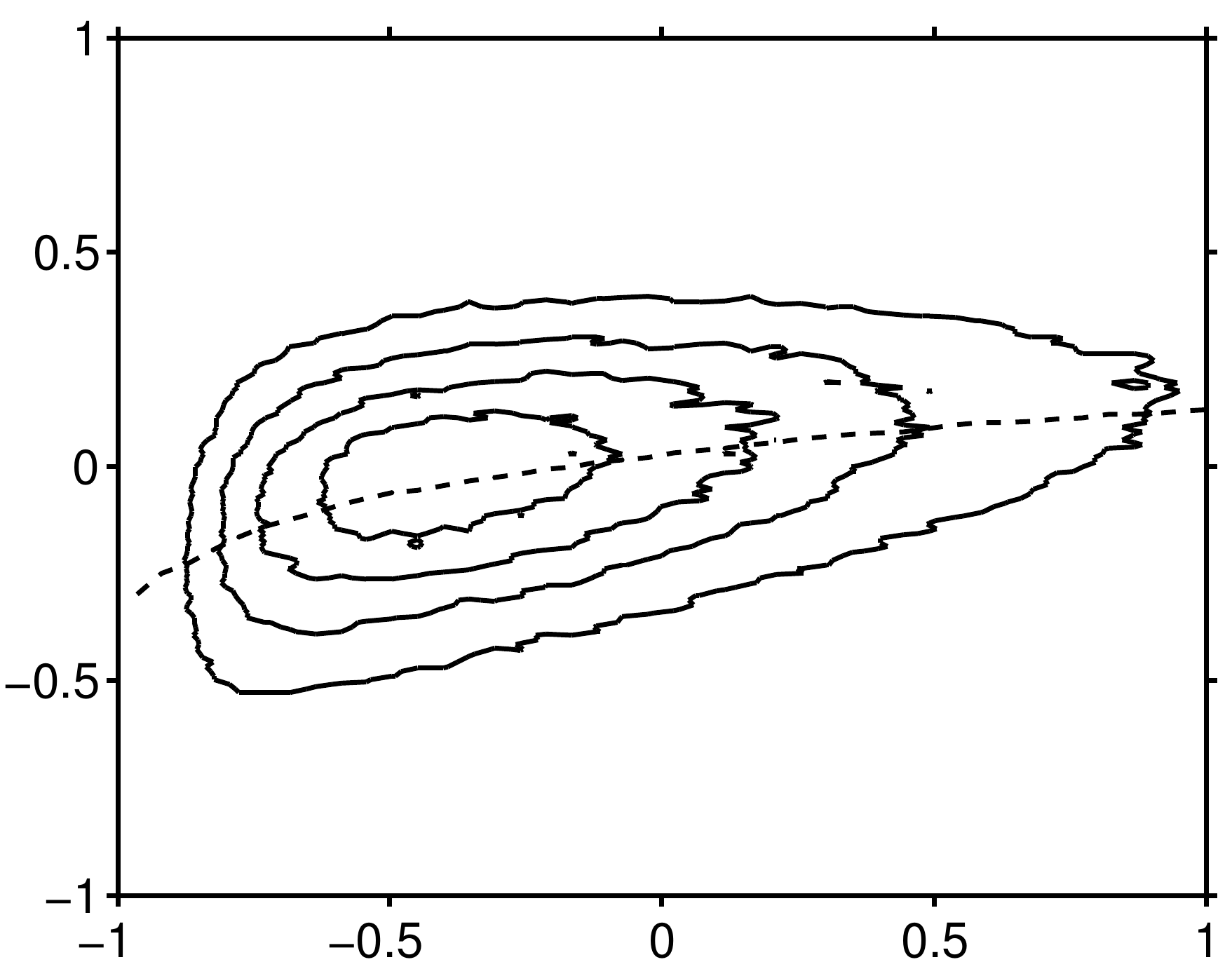}
    \\
    \centerline{
      $V_i^\prime/\langle V\rangle_{p,t}$
    }
  \end{minipage}
  \caption{%
    Joint p.d.f.s of the vertical component of the relative particle
    velocity fluctuations $w_{pr}^\prime=w_{pr}^\prime-\langle
    w_{pr}\rangle_{p,t}$ and the fluctuations of the Vorono\"i 
    cell volumes $V_i^\prime=V_i-\langle V\rangle_{p,t}$ for: 
    $(a)$ case M121, 
    $(b)$ case M178.
    In both graphs the lines indicate contours at
    $\{0.2,0.4,0.6,0.8\}$ 
    times the maximum number of occurrences. 
      The dashed line in $(b)$ indicates the conditional mean of 
      $w_{pr}^\prime$ (conditional on $V_i^\prime$) under the same
      normalization, i.e.\ 
      $\langle w_{pr}^\prime|V_i^\prime\rangle$. 
  }
  \label{fig-particles-joint-pdf-settling-velocity-voronoi-cell-volume}
\end{figure}

\begin{figure}
  \begin{minipage}{2ex}
    $\displaystyle\frac{y}{D}$
  \end{minipage}
  \begin{minipage}{0.45\textwidth}
    \centerline{$(a)$}
    \includegraphics[width=\textwidth]%
    {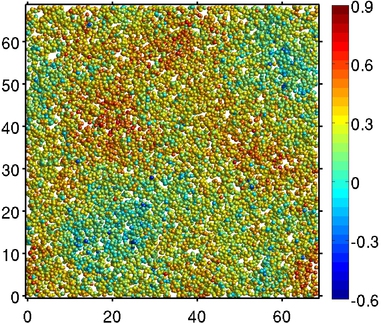}
    \\
    \centerline{$x/D$ }
  \end{minipage}
  \hfill
  \begin{minipage}{2ex}
    $\displaystyle\frac{y}{D}$
  \end{minipage}
  \begin{minipage}{0.45\textwidth}
    \centerline{$(b)$}
    \includegraphics[width=\textwidth]%
    {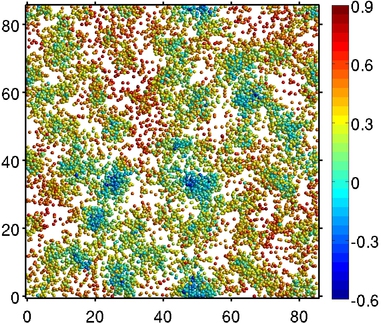}
    \\
    \centerline{$x/D$ }
  \end{minipage}
  \caption{%
    Top view of the particle locations at one instant in time. 
    The particles are colored according to the fluctuation of their 
    settling velocity  
    $w_{pr} ^\prime/ \langle w_{pr} \rangle_{p,t}$, 
    where $w_{pr} ^\prime=w_{pr} -\langle w_{pr} \rangle_{p,t}$.
    (a) Case \caseA{}, (b) case \caseB{}. 
  }
  \label{fig:part_pos_col_vel}
\end{figure}

\begin{figure}
  \begin{minipage}{2ex}
    $\displaystyle\frac{y}{D}$
  \end{minipage}
  \begin{minipage}{0.45\textwidth}
    \centerline{$(a)$}
    \includegraphics[width=\textwidth]%
    {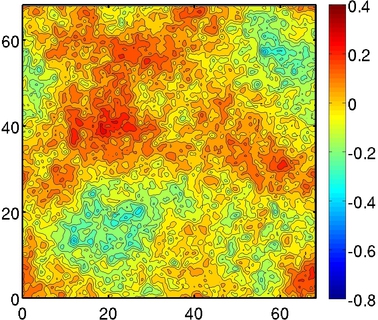}
    \\
    \centerline{$x/D$ }
  \end{minipage}
  \hfill
  \begin{minipage}{2ex}
    $\displaystyle\frac{y}{D}$
  \end{minipage}
  \begin{minipage}{0.45\textwidth}
    \centerline{$(b)$}
    \includegraphics[width=\textwidth]%
    {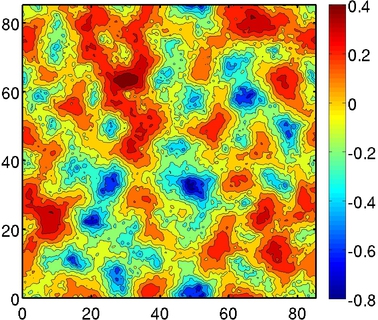}
    \\
    \centerline{$x/D$ }
  \end{minipage}
  \caption{%
    Instantaneous, vertically-averaged fluid velocity
    fluctuations normalized with the instantaneous box-average,  
    $\hat{w}_f^\prime(x,y,t)/
    \langle w_f(\mathbf{x},t)\rangle_{\Omega_f}$, where 
    $\hat{w}_f^\prime(x,y,t)=
    \langle w_f(\mathbf{x},t)\rangle_{z}-\langle
    w_f(\mathbf{x},t)\rangle_{\Omega_f}$: 
    (a) case \caseA{}, %
    (b) case \caseB{}. %
    The instants in time correspond to those in
    figure~\ref{fig:part_pos_col_vel}. 
  } 
  \label{fig:z-box-aver-fluid-vel}
\end{figure}

\begin{figure}
  \begin{minipage}{0.45\textwidth}
    \centerline{$(a)$}
    \includegraphics[width=\textwidth]%
    {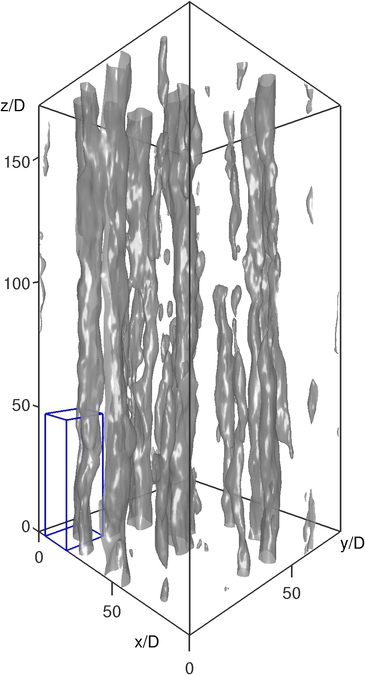}
  \end{minipage}
  \hfill
  \begin{minipage}{0.45\textwidth}
    \centerline{$(b)$}
    \includegraphics[width=\textwidth]%
    {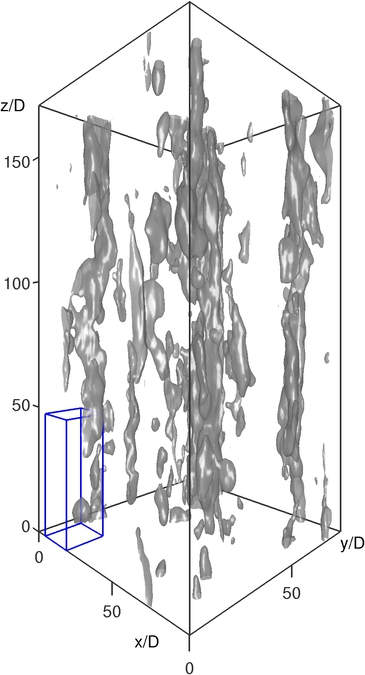}
  \end{minipage}
  \caption{%
    Iso-surfaces of the vertical fluid velocity fluctutations
    $w_f^\prime$ (with respect to the box-averaged value) in case M178
    at one instant in the statistically stationary regime. The
    velocity field has been box-filtered with a filter width of
    $5D$. 
    The graph in $(a)$ shows a value of $-0.4 u_g$, while 
    $(b)$ shows $+0.4u_g$; these values are equivalent to $\pm1.05$
    times the rms value of the unfiltered $w_f^\prime$ field.
    The small cuboid indicated by blue lines corresponds to the
    sub-volume shown in figure~\ref{fig-3d-snapshot-lambda2}. 
  } 
  \label{fig-3d-snapshot}
\end{figure}

\begin{figure}
  \centering
  \begin{minipage}{0.3\textwidth}
    \includegraphics[width=\textwidth]
    {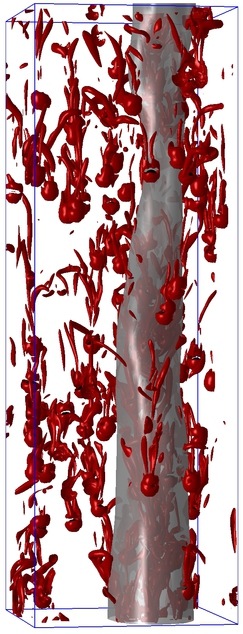}
  \end{minipage}
  \caption{%
    Iso-surfaces of the vortex eduction criterion proposed by
    \cite{jeong:95} at a value of $\lambda_2=-0.25u_g^2/D^2$ (red color), 
    showing a sub-volume of size $16D\times16D\times48D$ in case
    M178. 
    The flow data is taken at the same instant as in
    figure~\ref{fig-3d-snapshot}, where the present sub-volume is
    outlined. 
    The large tube-like structure (indicated in grey color) shows an
    iso-surface of $w_f^\prime=-0.4u_g$, as in
    figure~\ref{fig-3d-snapshot}$(a)$. 
    Gravity acts from top to bottom.
  } 
  \label{fig-3d-snapshot-lambda2}
\end{figure}

\begin{figure}
  \begin{minipage}{2ex}
    $\displaystyle\frac{\tilde{z}}{D}$
  \end{minipage}
  \begin{minipage}{0.3\linewidth}
    \centerline{$(a)$}
    \includegraphics[width=\linewidth,clip=true,
    viewport=335 30 875 860]
    {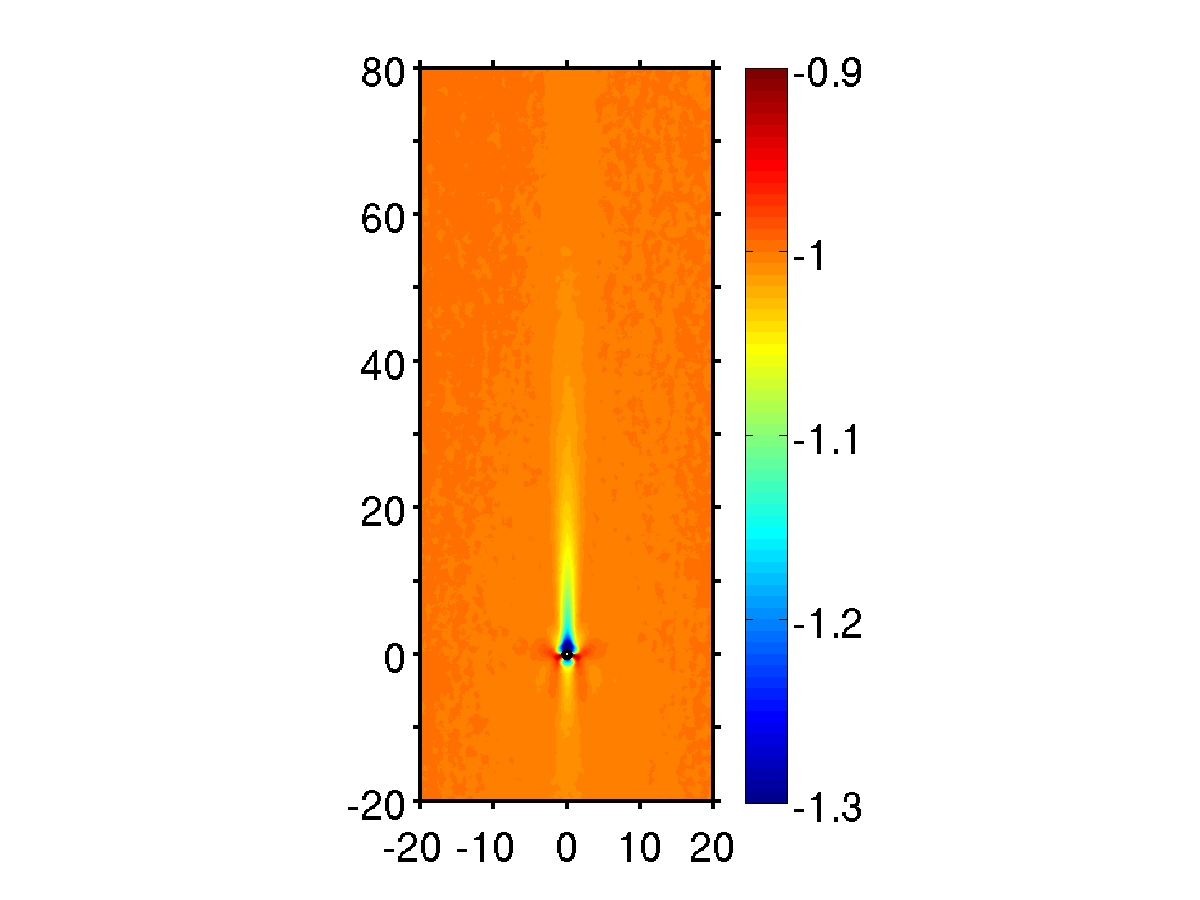}
    \centerline{$\tilde{r}/D$ }
  \end{minipage}
  \raisebox{14ex}{
    \begin{minipage}{4ex}
      $\displaystyle\frac{w_{pr}^{cond}}
      {|\langle w_{pr}\rangle_{p,t}|}$
    \end{minipage}
  }
  \hspace*{.1\linewidth}
  \begin{minipage}{2ex}
    $\displaystyle\frac{\tilde{z}}{D}$
  \end{minipage}
  \begin{minipage}{0.3\linewidth}
    \centerline{$(b)$}
    \includegraphics[width=\linewidth,clip=true,
    viewport=335 30 875 860]
    {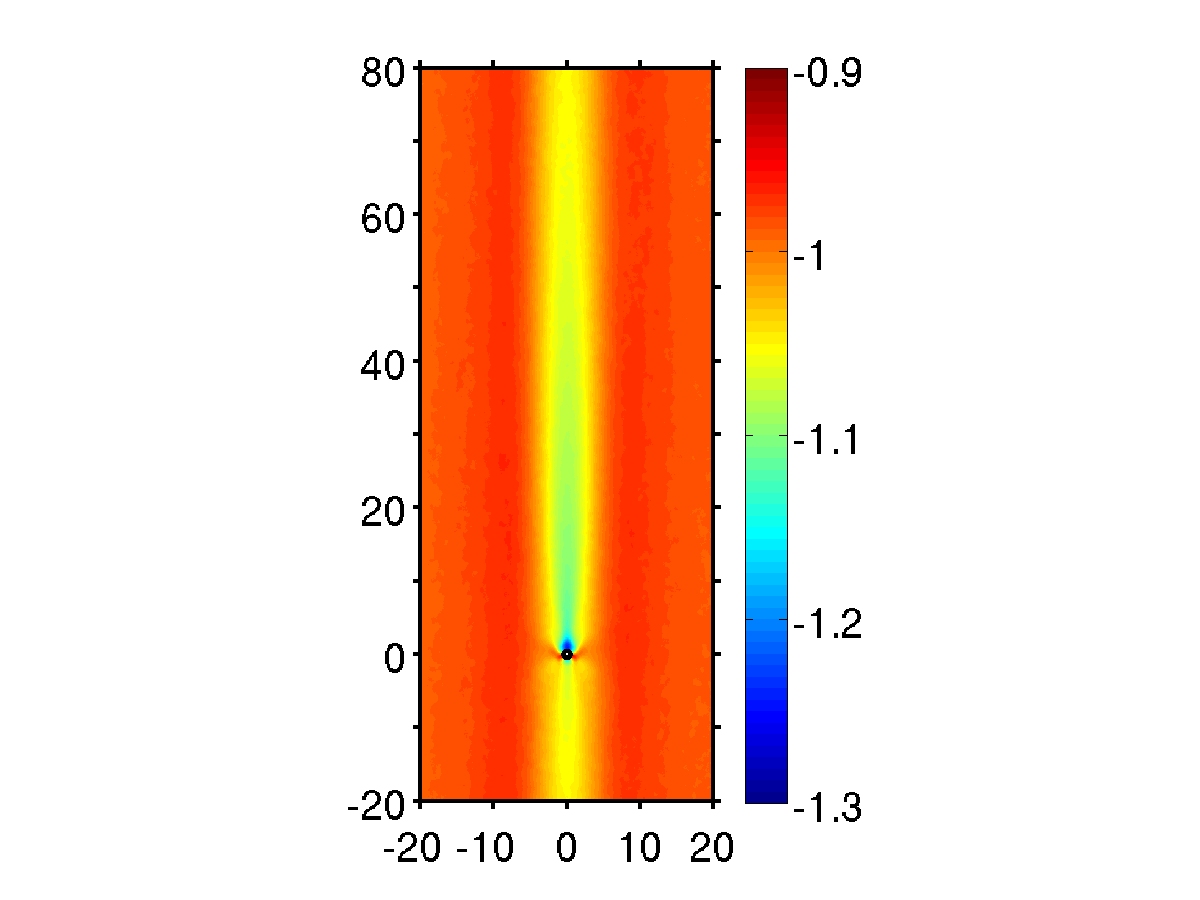}
    \centerline{$\tilde{r}/D$ }
  \end{minipage}
  \raisebox{14ex}{
    \begin{minipage}{4ex}
      $\displaystyle\frac{w_{pr}^{cond}}
      {|\langle w_{pr}\rangle_{p,t}|}$
    \end{minipage}
  }
  \caption{%
    Spatial map of the particle-conditioned average value of the
    settling velocity, $w_{pr}^{cond}$, plotted as a function of
    the position $(\tilde{r},\tilde{z})$
    with respect to a test particle (in a vertical plane): 
    (a) case \caseA{},  
    (b) case \caseB{}.  
  } 
  \label{fig:PartVelPDF2D}
\end{figure}

\begin{figure}
  \begin{minipage}{9ex}
    $\displaystyle\frac{w_{pr}^{cond}}
    {|\langle w_{pr}\rangle_{p,t}|}$
  \end{minipage}
  \begin{minipage}{0.39\linewidth}
    \centerline{$(a)$}
    \includegraphics[width=\linewidth]
    {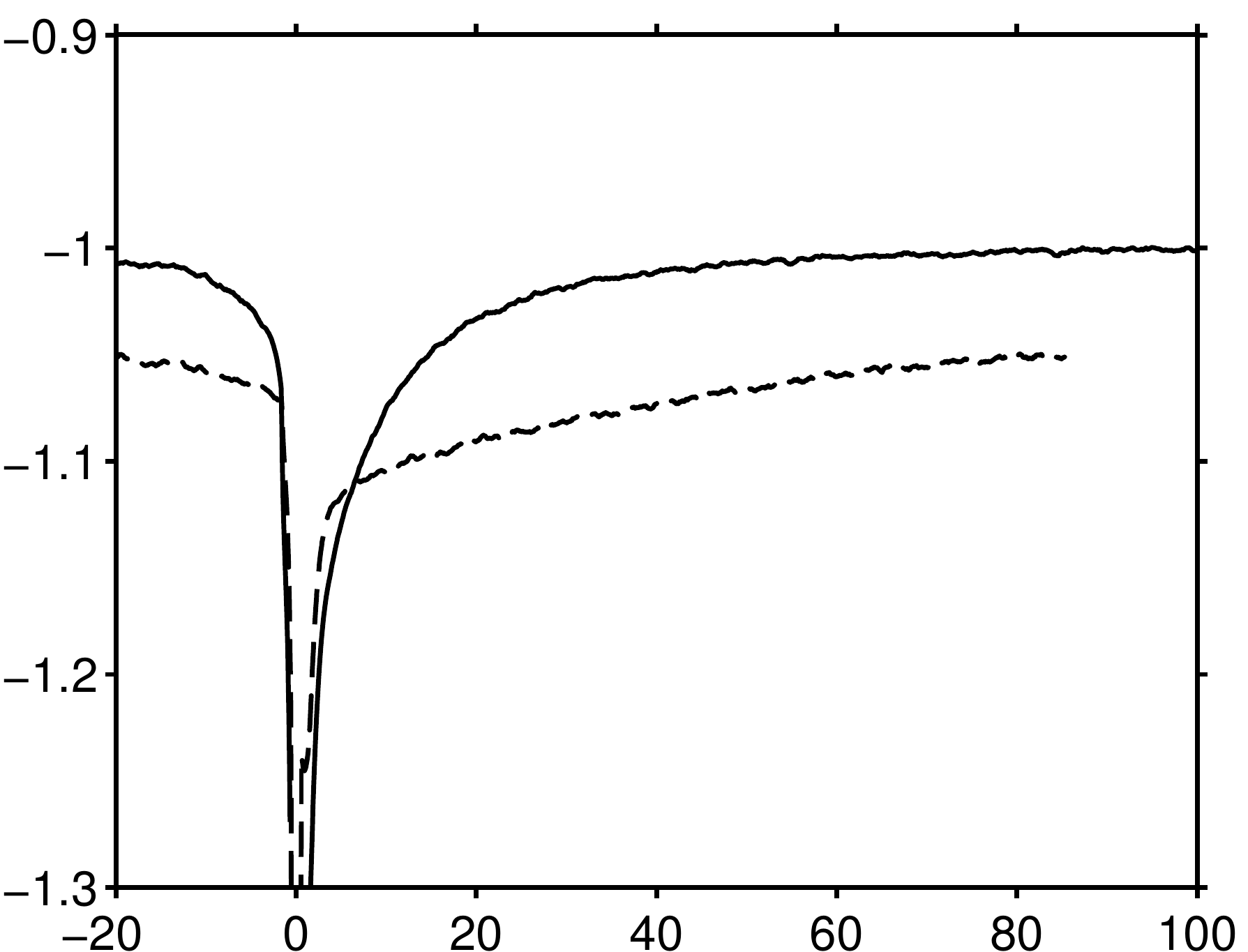}
    \\
    \centerline{ $\tilde{z}/D$ }
  \end{minipage}
  \hfill
  \begin{minipage}{9ex}
    $\displaystyle\frac{w_{pr}^{cond}}
    {|\langle w_{pr}\rangle_{p,t}|}$
  \end{minipage}
  \begin{minipage}{0.39\linewidth}
    \centerline{$(b)$}
    \includegraphics[width=\linewidth]
    {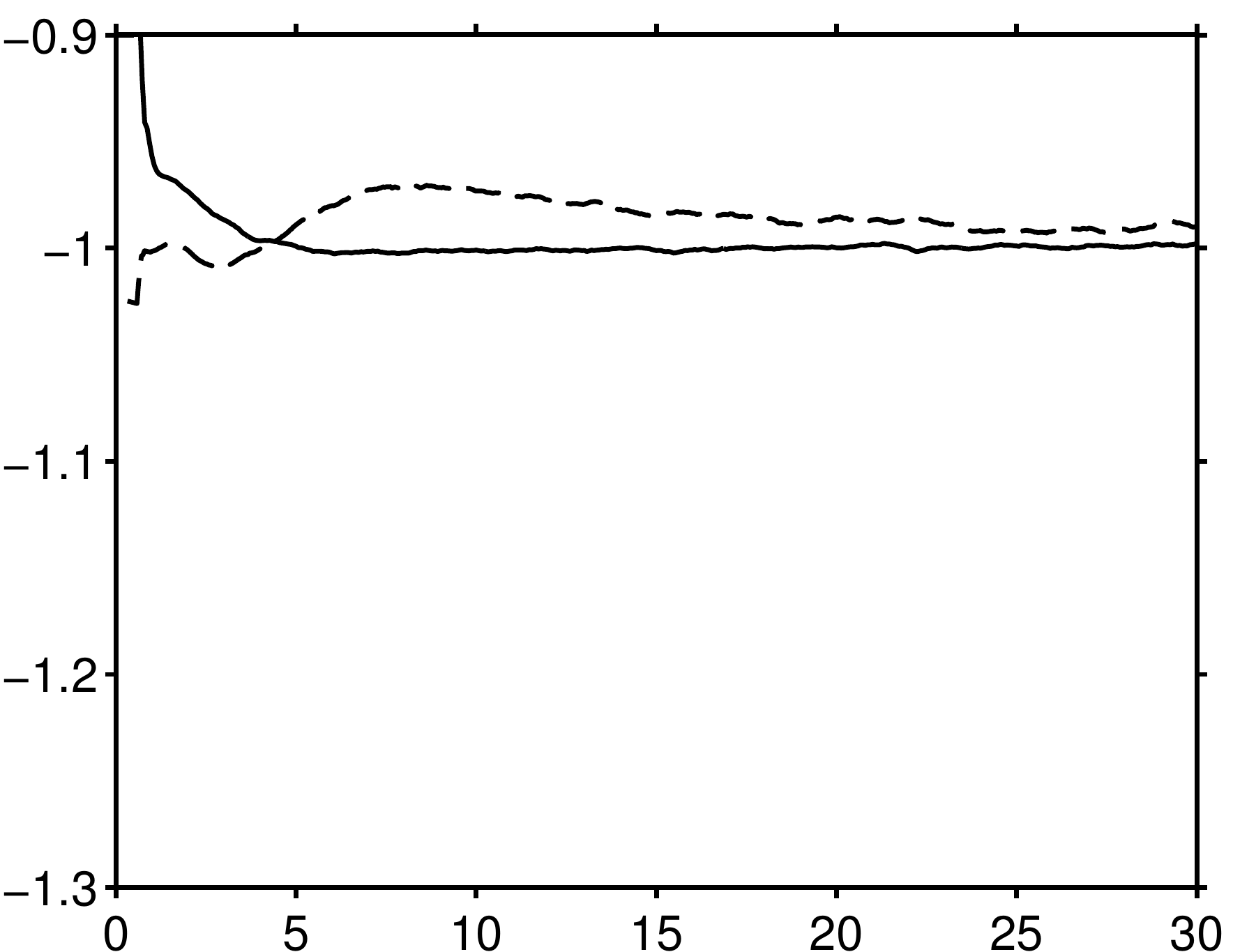}
    \\
    \centerline{ $\tilde{r}/D$ }
  \end{minipage}
  \caption{%
    Variation of the particle-conditioned average value of the
    settling velocity (cf.\ figure~\ref{fig:PartVelPDF2D}) along
    two axes passing through the test particle: 
    $(a)$ vertical axis (with $\tilde{r}=0$), 
    $(b)$ horizontal axis ($\tilde{z}=0$). 
    The line style indicates the flow cases: 
    case M121 (solid line), case M178 (dashed). 
  }
  \label{fig:PartVelPDF2D-lines}
\end{figure}

As the previous analysis shows, in the clustering case M178 it makes
a large difference whether the relative velocity is computed with
respect to the global average of the fluid velocity or whether it is
evaluated locally. This also implies that in the respective flow case
the spatial variability of the settling velocity is large. 
Next, we would like to establish a direct link between the vertical
particle velocity and the (instantaneous) regions where the particle 
concentration is higher/lower than the average. 
For this purpose we have computed the joint p.d.f.\ of the
fluctuations of the particle settling velocity (defined with respect
to the box-averaged fluid velocity, cf.\
equation~\ref{equ-res-trasient-def-uprel}) and the fluctuation of the
\Vor cell volume associated to the respective particle at the
respective time, as shown in
figure~\ref{fig-particles-joint-pdf-settling-velocity-voronoi-cell-volume}.  
It can be seen that the two quantities
are not significantly correlated in case M121 (the correlation
coefficient measures $0.002$). In the
clustering case M178, on the other hand, one can clearly observe a
positive correlation with a correlation coefficient of $0.347$, i.e.\
small \Vor cells tend to correspond to smaller-than-average relative
vertical velocities (equivalent to faster settling particles) and
vice versa. 
This result confirms that in case M178 the fast settling particles
are indeed -- statistically speaking -- located in agglomeration
regions, while the slowly settling particles are found with a higher
probability in void regions. 
  Note that the average deviation of the settling velocity from the
  global mean is not linearly related to the Vorono\"i cell volumes,
  as can be seen from the conditional mean $\langle
  w_{pr}^\prime|V_i^\prime\rangle$ included in
  figure~\ref{fig-particles-joint-pdf-settling-velocity-voronoi-cell-volume}$(b)$. 
  An illustration is provided in
  figure~\ref{fig:part_pos_col_vel} which shows the particle positions
  at one instant (viewed from the top) colored according to their
  settling velocity value $w_{pr}(i,t)$. 
The correlation between agglomerations
(voids) and smaller (larger) than average vertical relative velocity
values is evident in case M178. 
In case M121, on the other hand, such a correlation is not clearly
visible; instead, fast and slowly settling particles appear to be much
less clearly segregated in the horizontal direction.  
In order to demonstrate the spatial correspondence between fast
settling clusters and downward fluid motion (with respect to the
box-average), we compute the instantaneous vertical average of the
vertical component of the fluid velocity (cf.\
equation~\ref{equ-app-avg-def-vertical-fluid-average}); its
fluctuation is defined as follows:   
$\hat{w}_f^\prime(x,y,t)=
\langle w_f(\mathbf{x},t)\rangle_{z}-\langle
w_f(\mathbf{x},t)\rangle_{\Omega_f}$. 
Figure~\ref{fig:z-box-aver-fluid-vel} shows this quantity (normalized
by the box-averaged value) at the same instant as the particle data
shown in figure~\ref{fig:part_pos_col_vel}. 
In case M178 we indeed observe regions of significantly
smaller-than-average fluid velocities at those locations where notable
particle agglomerations occur and vice versa for voids. 
In case M121, on the contrary, the vertically-averaged fluid velocity
$\hat{w}_f^\prime$ is of significantly smaller amplitude and it is
more evenly distributed in space. 
A representative illustration of the flow field is rendered in
figure~\ref{fig-3d-snapshot} where iso-surfaces of the vertical fluid
velocity fluctuations $w_f^\prime$ in case M178 are shown at values 
of $\pm0.4u_g$. Both negative and positive values correspond to
a number of tube-like vertical structures traversing essentially the
whole domain. In case M121, on the contrary, where the amplitude of
the fluctuations is much smaller (cf.\
figure~\ref{fig:fluid-part-vel-rms}$a$), these values are practically
not encountered (figure omitted). 
The shape of the vortical structures in the near-field of the
particles in case M178 can be seen in
figure~\ref{fig-3d-snapshot-lambda2}, where again the criterion of
\cite{jeong:95} has been used on an instantaneous flow field. The
figure shows a sub-volume of the domain which includes part of one
strong downward current (indicated by the large tube-like structure
included therein). In the cluster regions, the particle wakes feature
complex tangles of vortices extending from particle to particle and
beyond. In more quiescent regions, particle wakes are found to be much
closer to the double-threaded state of an isolated sphere at the same
Galileo number (cf.\ regime II in
figure~\ref{fig-wake-regimes-isolated}).  
The corresponding visualization in case M121 (presently omitted) shows
attached toroidal vortex rings for most particles, similar to those
found in regime I of an isolated particle (cf.\
figure~\ref{fig-wake-regimes-isolated}).  
Next we wish to analyze the spatial distribution of the settling
velocity. 
For this purpose we perform particle-conditioned averaging of this
quantity, which is analogous to computing the particle-conditioned
solid volume fraction discussed in
\S~\ref{sec-results-spatial-struct-partconc} (cf.\
figures~\ref{fig:part-2D-num-den-ver-M121} and
\ref{fig:part-2D-num-den-ver-M178}). 
The result of applying the particle-conditioned averaging operator 
(\ref{eq:cond-aver})
to the settling velocity $w_{pr}(i,t)$  
will henceforth be denoted as $w_{pr}^{cond}(\tilde{r},\tilde{z})$. 
A map of this quantity is shown in figure~\ref{fig:PartVelPDF2D}. 
Note that the quantity $w_{pr}(i,t)$ is not invariant with respect
to an interchange of the indices of the two particles making up any
given pair. 
As a consequence, the map of the particle-conditioned average
settling velocity $w_{pr}^{cond}(\tilde{r},\tilde{z})$ 
does not possess a symmetry with respect to the
horizontal axis (i.e.\ the average settling velocity is not
necessarily the same for a leading and for a trailing particle); due
to horizontal homogeneity; it is, however, still symmetric with
respect to the vertical axis passing through the test particle.
The map in figure~\ref{fig:PartVelPDF2D} shows that in case M121
trailing particles are indeed settling faster (i.e.\ their settling
velocity is lower) than the average. However, this effect is limited
to a compact region in the wake of the test particle. 
In case M178, on the other hand, the particle-conditioned settling
velocity is significantly below the average value (i.e.\ the settling
rate is increased) over long distances in the wake of the test
particle, while it is slightly larger than the average (decreased
settling rate) at lateral distances $\tilde{r}$ of approximately 10
particle diameters from the test particle (roughly independent of the
vertical distance $\tilde{z}$).  
A direct comparison of the conditionally-averaged settling velocity in
the two simulations is shown in figure~\ref{fig:PartVelPDF2D-lines},
where the variation along the vertical axis and along a horizontal axis
passing through the test particle is plotted. 
At small positive vertical distances $\tilde{z}$ from the test
particle (i.e.\ focusing on closely trailing particles in
figure~\ref{fig:PartVelPDF2D-lines}$a$), the spatial
variation of the conditioned settling velocity is roughly similar in
both flow cases, featuring a steep increase of the settling velocity with
$\tilde{z}$. Note that the minimum close to the rear of the test
particle measures approximately $1.3$ times the globally-averaged
settling velocity. For larger distances on the vertical axis
($\tilde{z}\gtrsim4D$), however, the conditional settling velocity in
case M178 increases at a much smaller rate than in case M121 
(in both cases the asymptotic value -1 is approached exponentially for
$\tilde{z}\gtrsim15D$). 
Regarding leading particles (i.e.\ negative values of $\tilde{z}$ in
figure~\ref{fig:PartVelPDF2D-lines}$a$), it is found that they are
likewise settling faster (on average) than the global average at any
distance in case M178. In case M121, vertically-aligned leading
particles are also settling somewhat faster, but only within a
distance of a few diameters from the test particle. 
Turning to horizontal distances
(figure~\ref{fig:PartVelPDF2D-lines}$b$), it is found that particles
in case M121 which are located within approximately $5D$ from the test
particle settle on average at a smaller speed than the global
average. 
On the contrary, in case M178 slower settling (i.e.\ 
${w_{pr}^{cond}}/{|\langle w_{pr}\rangle_{p,t}|}>-1$) 
is found only for horizontal distances larger than approximately $5D$,
with a mild peak of ${w_{pr}^{cond}}$ at $\tilde{r}\approx8D$.
In conclusion our analysis of the particle-conditioned
settling velocity has confirmed that clustering in the higher-Galileo
case M178 leads to a patterning of the vertical particle motion which
is distinct from the non-clustering case M121, the main feature in the
former case being a long-range effect in the vertical
direction. The length scales which can be extracted from the
particle-conditioned analysis of the settling velocity provide one way
of characterizing the geometry of the clusters themselves. 
\subsubsection{Fluctuating quantities} 
\label{sec-part-motion-fluct}
\begin{figure}
  \begin{minipage}{2.5ex}
    \rotatebox{90}{$\langle
        u_{f,\alpha}^\prime u_{f,\alpha}^\prime\rangle_{\Omega_f}^{1/2}/
        w_{ref}$}
  \end{minipage}
  \begin{minipage}{0.45\textwidth}
    \centerline{$(a)$}
    \includegraphics[width=\textwidth]
    {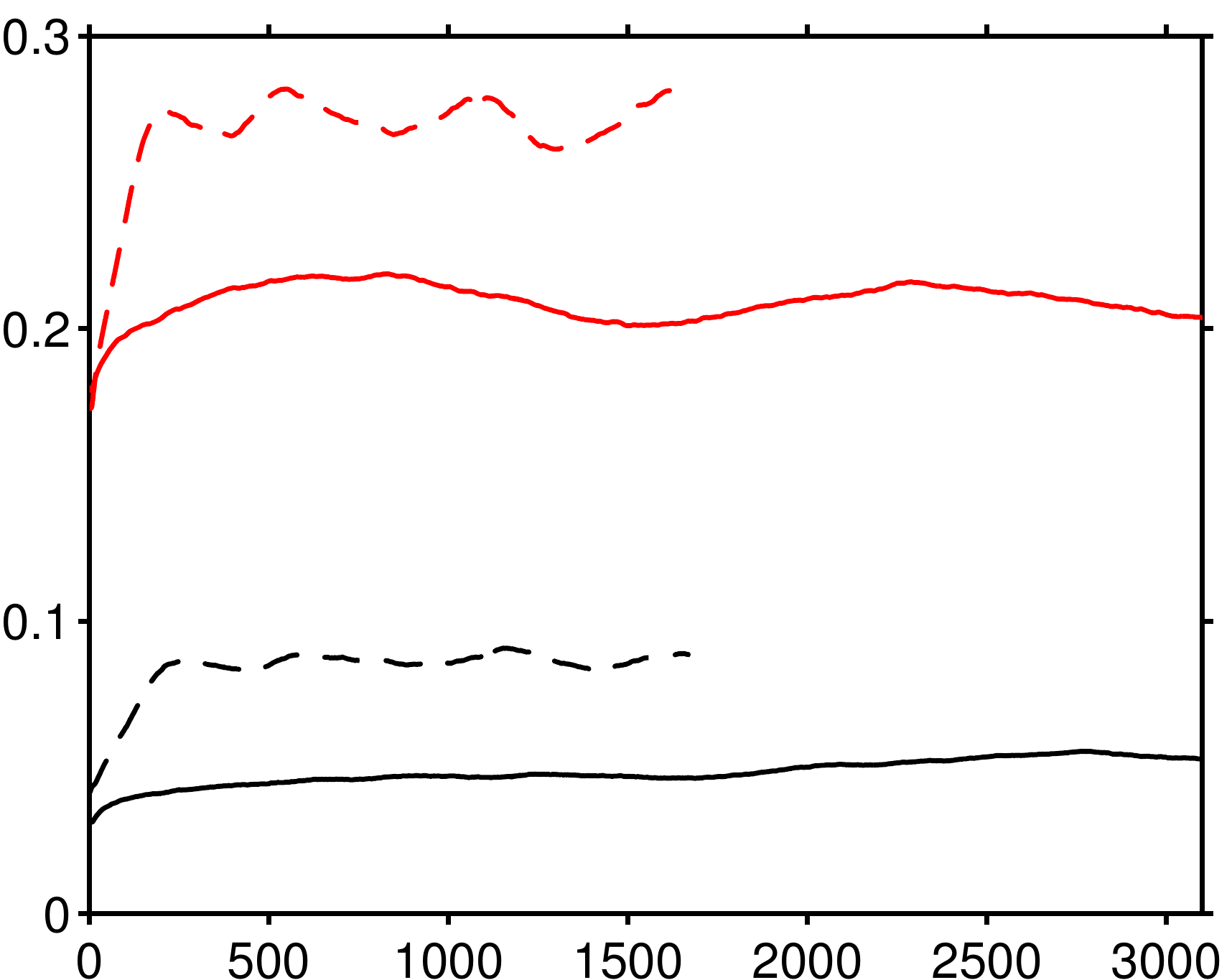}
    \hspace*{-.6\linewidth}\raisebox{.62\linewidth}{vertical ($\alpha=3$)}
    \hspace*{-15ex}\raisebox{.2\linewidth}{horizontal ($\alpha=1,2$)}
    \\
    \centerline{$t/\tau_g$}
  \end{minipage}
  \hfill
  \begin{minipage}{2.5ex}
    \rotatebox{90}{$\langle
        u_{p,\alpha}^\prime u_{p,\alpha}^\prime\rangle_p^{1/2}/
        w_{ref}$ }
  \end{minipage}
  \begin{minipage}{0.45\textwidth}
    \centerline{$(b)$}
    \includegraphics[width=\textwidth]
    {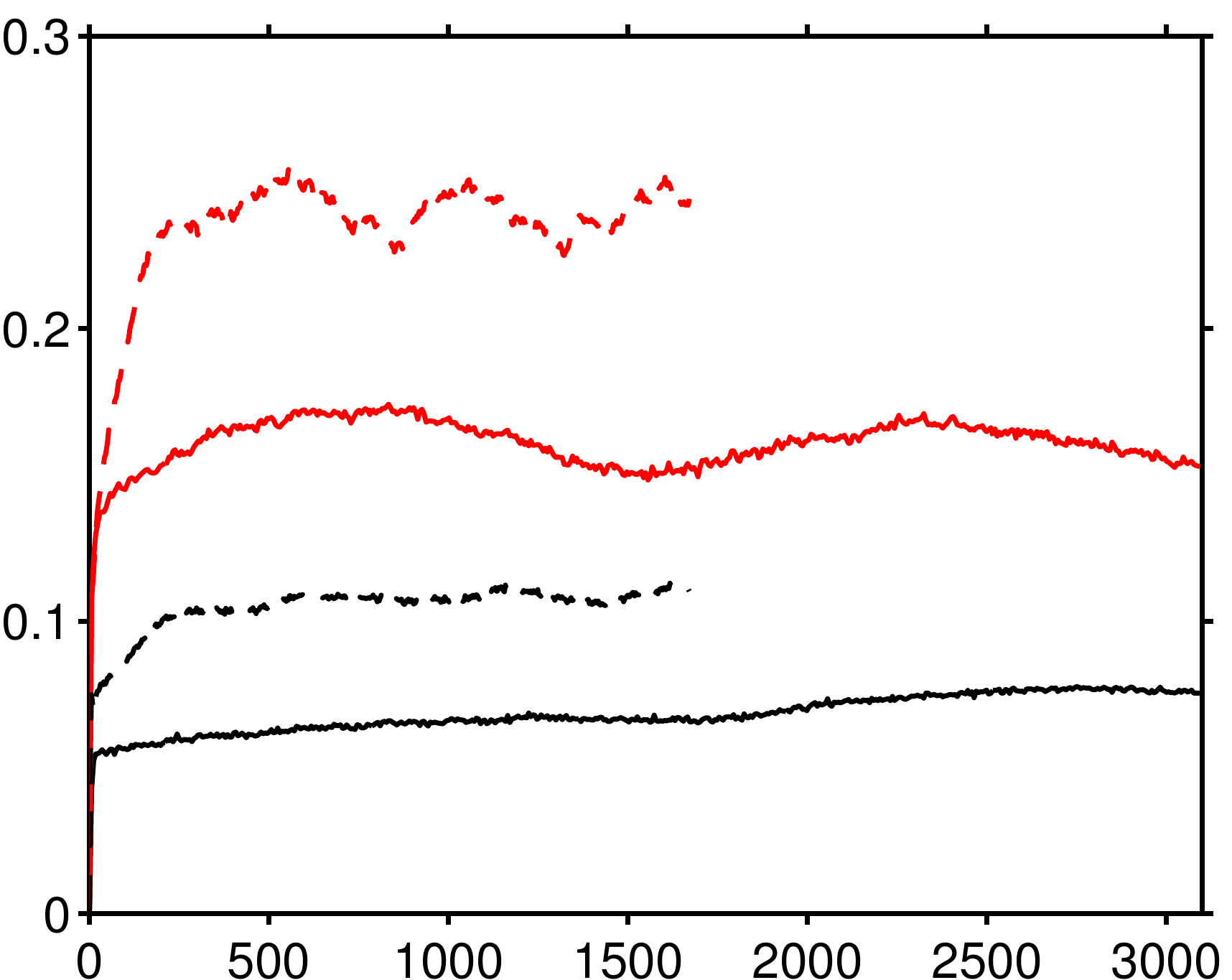}
    \hspace*{-.43\linewidth}\raisebox{.54\linewidth}{$\alpha=3$}
    \hspace*{-7ex}\raisebox{.27\linewidth}{$\alpha=1,2$}
    \\
    \centerline{$t/\tau_g$}
  \end{minipage}
  \caption{%
    Intensity of the velocity fluctuations of: 
    (a) the fluid phase, and 
    (b) the particle phase. 
    Results for case M121 (solid lines) and case M178 (dashed lines)
    are shown. 
    As reference velocity $w_{ref}$ for the purpose of
    normalization, the absolute value of the particle settling
    velocity $|w_{s}|$ of the 
    corresponding single-particle cases  (\caseAA{} and
    \caseBB{}, respectively) has been used. 
  } 
  \label{fig:fluid-part-vel-rms}
\end{figure}

Let us return to the fluctuations of the translational particle
velocity. The rms value of the fluctuations, 
$\langle u_{p,\alpha}^\prime u_{p,\alpha}^\prime\rangle_p^{1/2}$
(with $u_{p,\alpha}^\prime$ defined with respect to the instantaneous
average over all particles, 
cf.\ definition in \S~\ref{sec-numa-notation}), is shown in
figure~\ref{fig:fluid-part-vel-rms}b, which has already been briefly
discussed in \S~\ref{sec-results-transient}. 
Recall that asymptotically an isolated particle in case S121 (same
density ratio and Galileo number as case M121) settles with a constant
velocity along a vertical path, while an isolated particle in case
S178 (corresponding to M178 in terms of $\rho_p/\rho_f$, $Ga$) settles
steadily along an oblique path in a random horizontal direction. 
Therefore, ensemble averaging over many realizations of isolated
particles in case S121 would produce fluctuating particle
velocity rms values equal to zero; in case S178 (due to the randomness
of the horizontal orientation) one would obtain an rms value in any
fixed horizontal direction which is different from zero. In
particular, one would obtain (e.g.\ considering the $x$-direction):
\begin{equation}\label{equ-res-part-motion-uph-single-random-dir-ensemble-rms}
  \frac{\langle u_{p}^\prime u_{p}^\prime\rangle_p^{1/2}}{w_s}
  =
  \frac{u_{pH}}{u_{pV}}\,\frac{1}{\sqrt{2}}
  \,,
\end{equation}
where $u_{pH}$ and $u_{pV}$ are the absolute values of the particle
velocity in the horizontal plane and along the vertical, respectively,
obtained from the simulation of an isolated particle. The value of the
quantity given in
(\ref{equ-res-part-motion-uph-single-random-dir-ensemble-rms}) for an
ensemble of cases S178 is $0.065$ \citep{uhlmann:13a}.  
In the present multi-particle case M178, the corresponding 
time-averaged value over the observation interval, i.e.\ 
$\langle u_{p}^\prime u_{p}^\prime\rangle_{p,t}^{1/2}/w_s$, 
is much larger, measuring $0.11$. 
This comparison shows that the velocity fluctuation amplitude of the
particle ensemble in case M178 cannot be satisfactorily
accounted for by means of a simple superposition of isolated particles
settling in an undisturbed manner. This is the more true when
considering that such superposition yields a zero 
fluctuation amplitude of the vertical particle velocity component.
Next we will decompose the particle velocity in order to separate from
the 
fluctuations those contributions which are due to
the motion of the fluid surrounding the particles. 

\begin{figure}
  \begin{minipage}{3ex}
    \rotatebox{90}{%
      $\langle
      w_{p}^\prime w_{p}^\prime\rangle_p/
      w_{ref}^2$, 
      contributions 
    }
  \end{minipage}
  \begin{minipage}{0.45\textwidth}
    \centerline{$(a)$}
    \includegraphics[width=\linewidth]
    {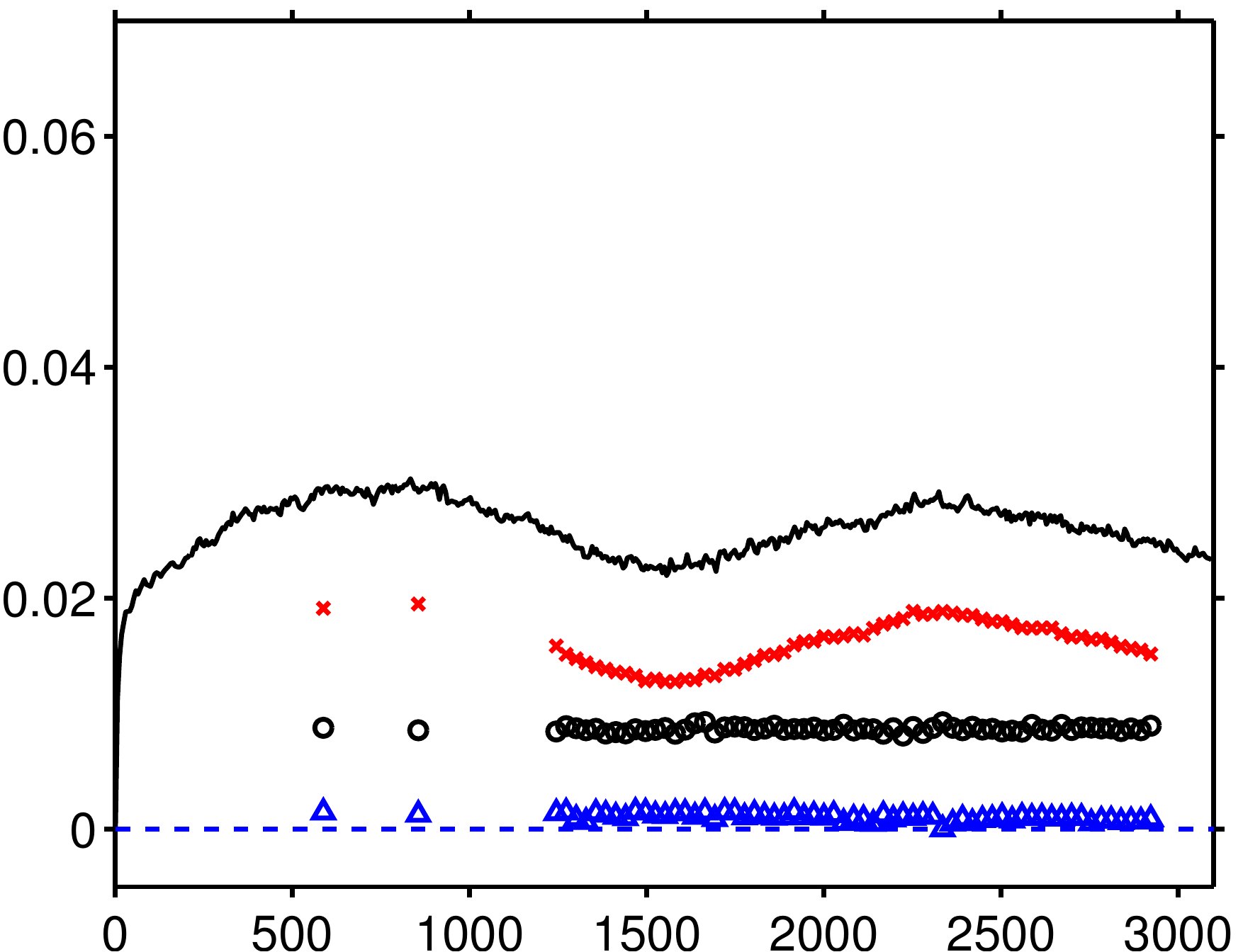}
    \\
    \centerline{$t/\tau_g$}
  \end{minipage}
  \hfill
  \begin{minipage}{3ex}
    \rotatebox{90}{%
      $\langle
      w_{p}^\prime w_{p}^\prime\rangle_p/
      w_{ref}^2$, 
      contributions 
    }
  \end{minipage}
  \begin{minipage}{0.45\textwidth}
    \centerline{$(b)$}
    \includegraphics[width=\linewidth]
    {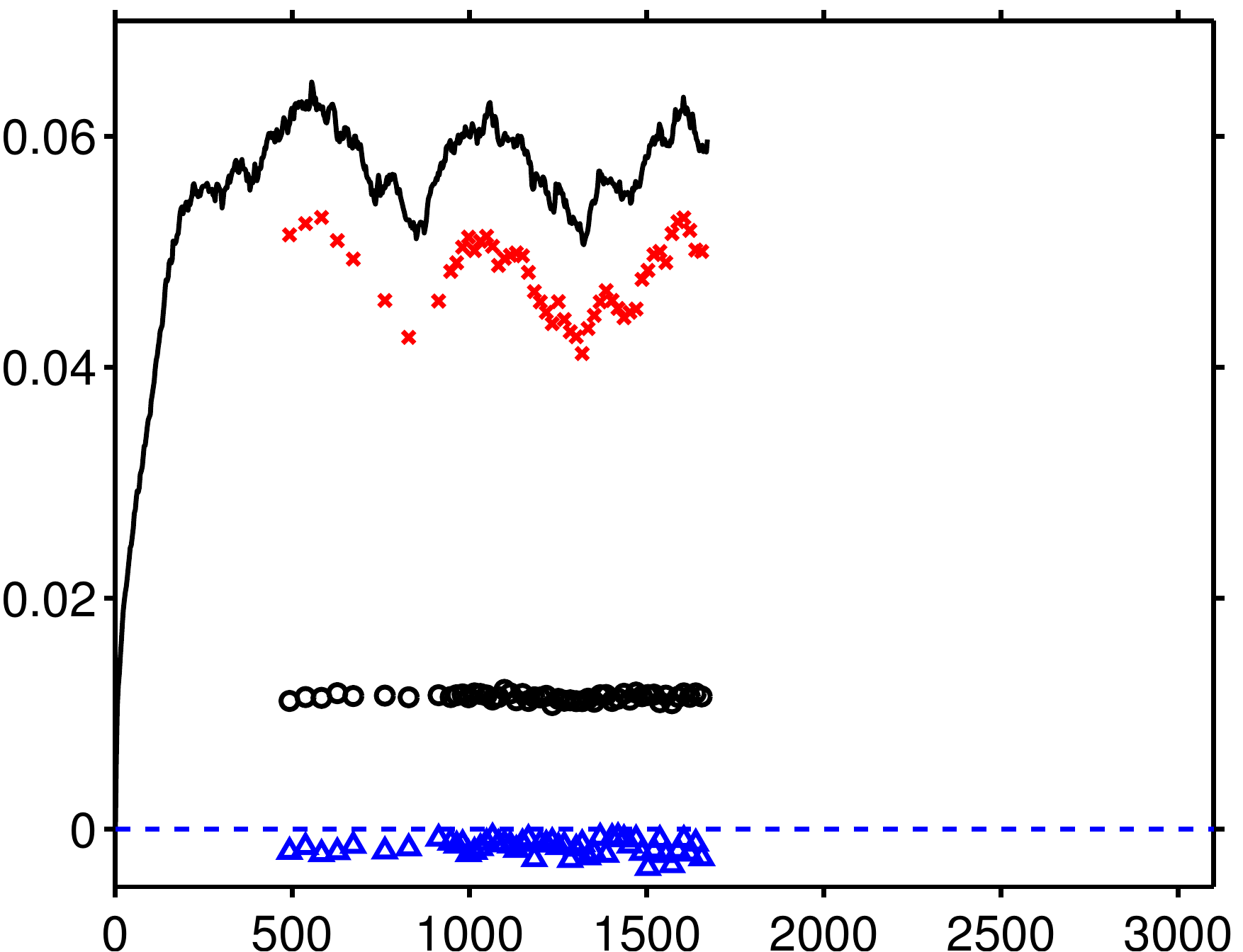}
    \\
    \centerline{$t/\tau_g$}
  \end{minipage}
  \caption{%
    Variance of the vertical velocity of the particle phase, 
    $\langle w_{p}^\prime w_{p}^\prime\rangle_p$ (solid line),
    and its three contributions (cf.\
    equation~\ref{equ-decomp-upart-shell-rms}):   
    $\langle w_{f}^{{\cal S}\,\prime}\,
    w_{f}^{{\cal S}\,\prime}
    \rangle_p$ ({\color{red}$\times$}),
    $\langle w_{pr}^{{\cal S}\,\prime}\,
    w_{pr}^{{\cal S}\,\prime}
    \rangle_p$ ({\color{black}$\circ$}),
    $2\langle w_{pr}^{{\cal S}\,\prime}\,
    w_{f}^{{\cal S}\,\prime}
    \rangle_p$ ({\color{blue}$\vartriangle$}).
    Graph $(a)$ shows case M121, 
    $(b)$ shows case M178.
  }
  \label{fig-particles-wrms-shell-budget}
\end{figure}

\begin{figure}
  \begin{minipage}{2ex}
    \rotatebox{90}{pdf}
  \end{minipage}
  \begin{minipage}{0.45\textwidth}
    \centerline{$(a)$}
    \includegraphics[width=1.0\textwidth]
    {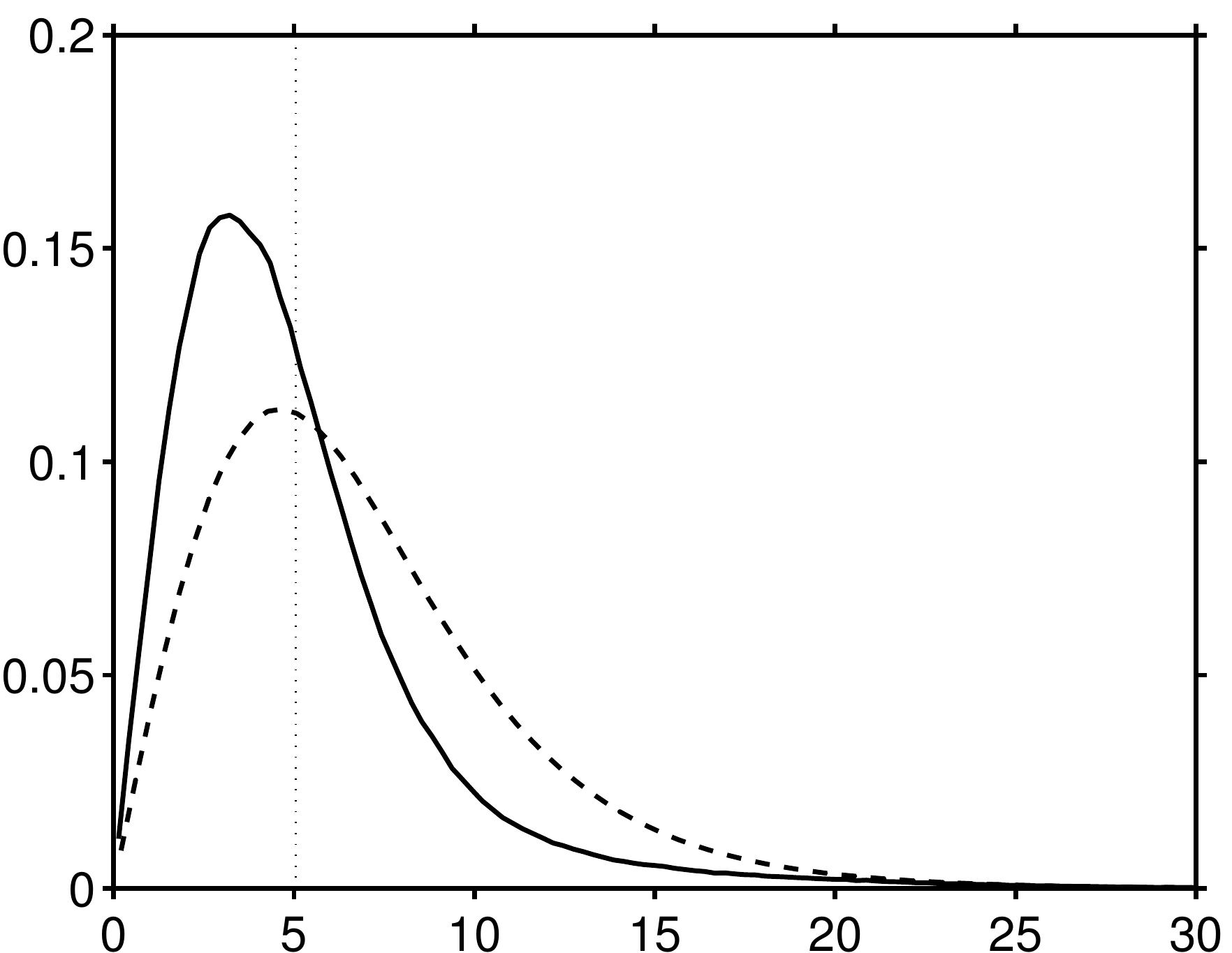}
    \\
    \centerline{$\alpha$}
  \end{minipage}
  \hfill
  \begin{minipage}{2ex}
    \rotatebox{90}{pdf}
  \end{minipage}
  \begin{minipage}{0.45\textwidth}
    \centerline{$(b)$}
    \includegraphics[width=1.0\textwidth]
    {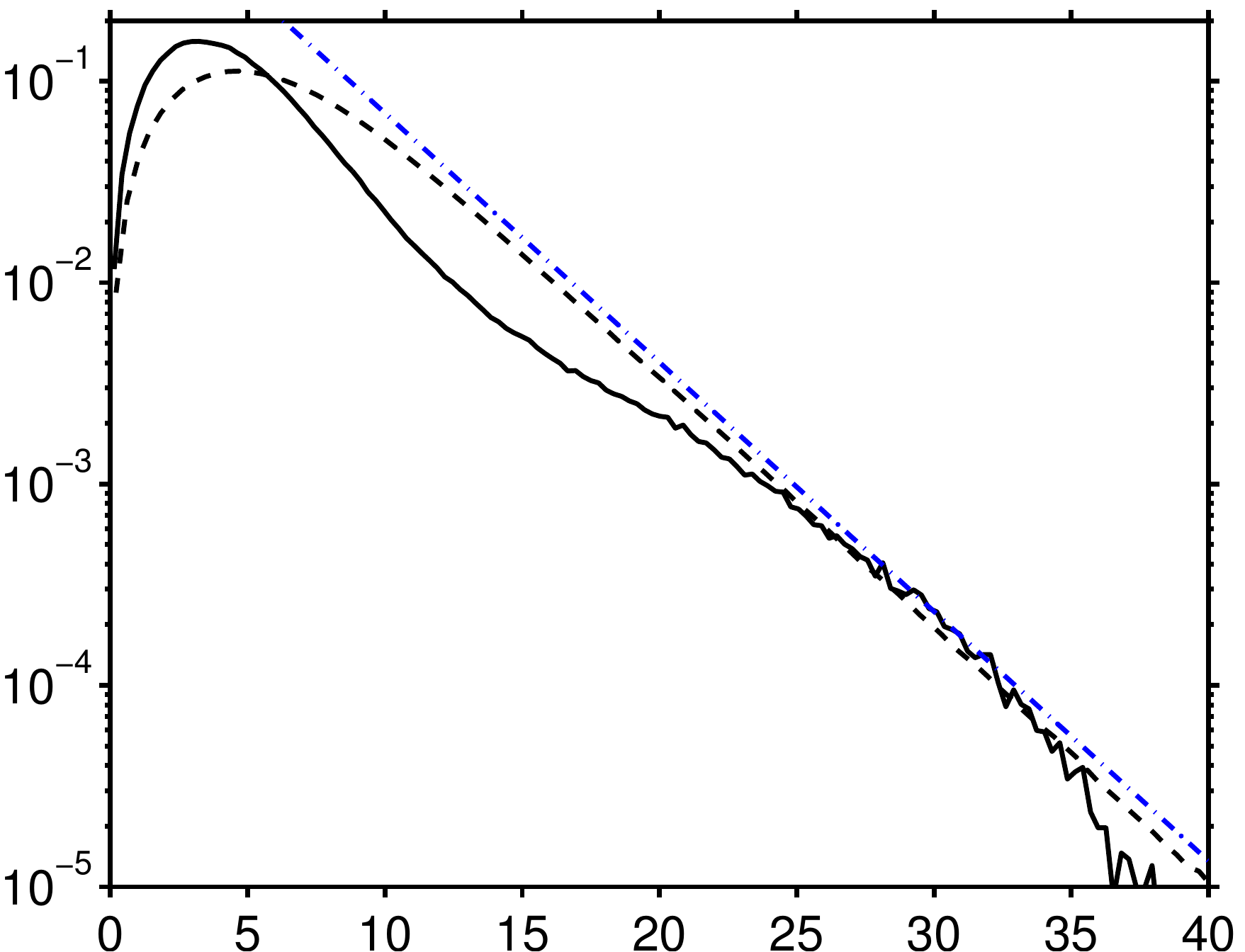}
    \\
    \centerline{$\alpha$}
  \end{minipage}
  \caption{%
    Probability density function of the angle $\alpha$ between the
    relative particle velocity vector and the vertical axis, defined
    by $\tan\alpha=(u_{pr}^2+v_{pr}^2)^{1/2}/|w_{pr}|$, measured in
    degrees, 
    for case \caseA{} (solid line) and case \caseB{} (dashed line). 
    The vertical dotted line denotes the value for 
    an isolated sphere at the same density ratio and Galileo
    number as case M178.
    The graph in $(b)$ shows the same data in semi-logarithmic
    scale. The chain-dotted line therein is proportional to
    $\exp(-a\alpha)$ with $a=0.285$. 
  }
  \label{fig:drift-angle}
\end{figure}

\begin{figure}
  \begin{minipage}{2.5ex}
    \rotatebox{90}{
      $\langle\omega_{p,\alpha}^\prime\omega_{p,\alpha}^\prime\rangle_p^{1/2}
      D/w_{ref}$
    }
  \end{minipage}
  \begin{minipage}{0.45\textwidth}
    \centerline{$(a)$}
    \centerline{\rule{0ex}{3ex}}
    \includegraphics[width=1.0\textwidth]
    {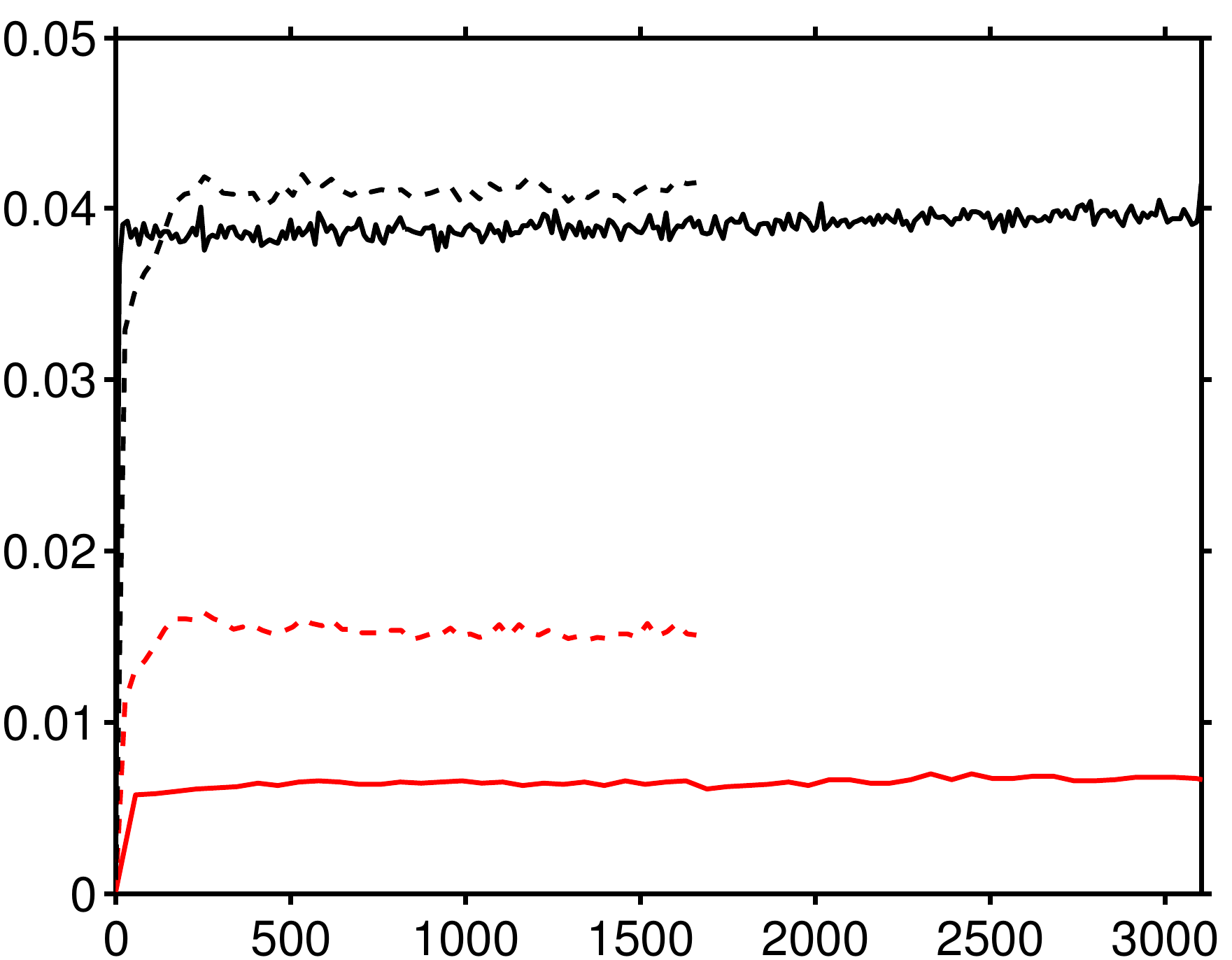}
    \hspace*{-.6\linewidth}\raisebox{.68\linewidth}{horizontal ($\alpha=1,2$)}
    \hspace*{-15ex}\raisebox{.2\linewidth}{vertical ($\alpha=3$)}
    \\
    \centerline{$t/\tau_g$}
  \end{minipage}
  \hfill
  \begin{minipage}{2.5ex}
    \rotatebox{90}{
      $\langle\omega_{p,\alpha}^\prime\omega_{p,\alpha}^\prime\,|\,V_i^\prime\rangle_{p,t}^{1/2}/
      \langle\omega_{p,\alpha}^\prime\omega_{p,\alpha}^\prime\rangle_{p,t}^{1/2}$
    }
  \end{minipage}
  \begin{minipage}{0.45\textwidth}
    \centerline{$(b)$}
    \centerline{\rule{0ex}{3ex}$V_i/(D^3\pi/6)$}
    \includegraphics[width=1.0\textwidth]
    {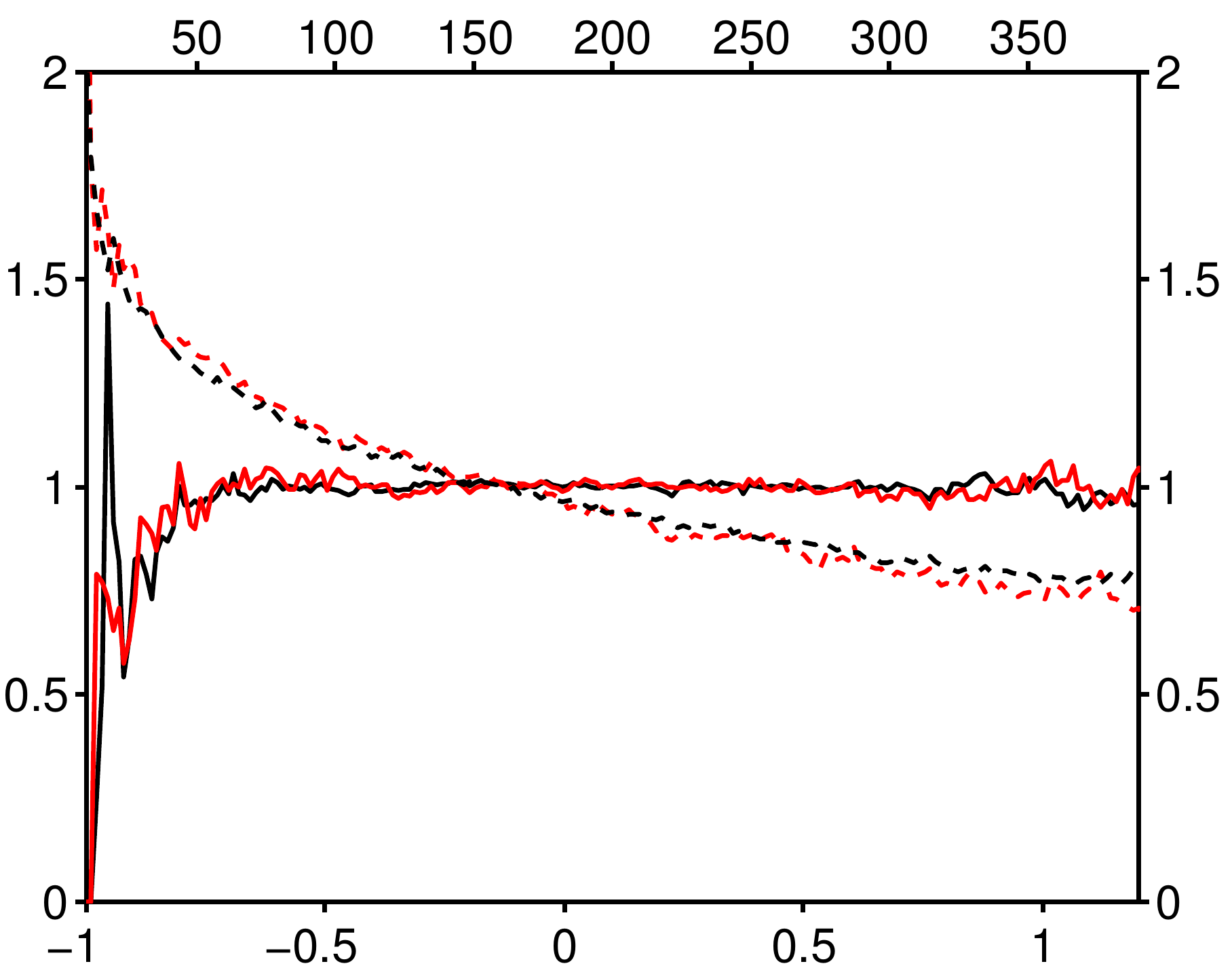}
    \hspace*{-.9\linewidth}\raisebox{.62\linewidth}{case \caseB}
    \hspace*{+11ex}\raisebox{.43\linewidth}{case \caseA}
    \\
    \centerline{$V_i^\prime/\langle V\rangle_{p,t}$}
  \end{minipage}
 \caption{%
     $(a)$ Temporal evolution of the standard deviation of the angular velocity
     of the particle phase.  
     $(b)$ The conditional standard deviation of the angular velocity,
     conditional upon the \Vor cell volume, normalized by the
     unconditional value. In the lower abscissa the fluctuation of the
     \Vor cell volume, normalized by the mean value is used; in the
     upper abscissa, the \Vor cell volume, normalized by the particle
     volume is used. Note that use of both scalings in a single graph is
     possible, since the mean value is the same in both flow cases. 
     In both graphs the results for case M121 (solid lines) and case
     M178 (dashed lines) are shown; the components are indicated by
     black color ($\alpha=1,2$) and red color ($\alpha=3$). 
  }
  \label{fig:part-ang-vel-rms}
\end{figure}

Using our definition of the fluid velocity seen by a particle,
$\mathbf{u}_{f}^{{\cal S}}$ (cf.\
\S~\ref{sec-part-uf-seen-by-partices}), and of the relative velocity
between the particle and the surrounding fluid,
$\mathbf{u}_{pr}^{{\cal S}}$ (defined in equation
\ref{equ-res-part-motion-def-uprel-shell}), the variance of the
particle velocity component in the direction $x_\alpha$ can be
decomposed as follows:
\begin{equation}\label{equ-decomp-upart-shell-rms}
  \langle 
  u_{p,\alpha}^{\prime}\,
  u_{p,\alpha}^{\prime}
  \rangle_p
  =
  \langle 
  u_{pr,\alpha}^{{\cal S}\,\prime}\,
  u_{pr,\alpha}^{{\cal S}\,\prime}
  \rangle_p
  +
  \langle 
  u_{f,\alpha}^{{\cal S}\,\prime}\,
  u_{f,\alpha}^{{\cal S}\,\prime}
  \rangle_p
  +
  2\langle 
  u_{pr,\alpha}^{{\cal S}\,\prime}\,
  u_{f,\alpha}^{{\cal S}\,\prime}
  \rangle_p
  \,,
  \qquad
  \forall\quad\alpha=1,2,3\,,
\end{equation}
where all fluctuations are defined with respect to the instantaneous
average over the particle ensemble, viz.
\begin{equation}\label{equ-upart-various-fluct-def}
  u_{pr,\alpha}^{{\cal S}\,\prime}
  =
  u_{pr,\alpha}^{{\cal S}}-
  \langle 
  u_{pr,\alpha}^{{\cal S}}
  \rangle_p
  \,,\quad
  u_{f,\alpha}^{{\cal S}\,\prime}
  =
  u_{f,\alpha}^{{\cal S}}-
  \langle 
  u_{f,\alpha}^{{\cal S}}
  \rangle_p
  \,.
\end{equation}
The relation (\ref{equ-decomp-upart-shell-rms}) shows that the
energy of the particle velocity fluctuations can be decomposed into
one contribution stemming from the particles' relative motion with
respect to the surrounding fluid (first term on the right-hand-side in
\ref{equ-decomp-upart-shell-rms}),
one contribution due to the fluctuations of the fluid velocity in the
vicinity of the particles (second term on the r.h.s.\ of
\ref{equ-decomp-upart-shell-rms}), and one contribution due to the
correlation between the former two (last term in
\ref{equ-decomp-upart-shell-rms}). 
Figure~\ref{fig-particles-wrms-shell-budget} 
shows the three contributions on the right-hand-side of
(\ref{equ-decomp-upart-shell-rms}) for the vertical direction
($\alpha=3$) along with the total fluctuation energy (note that this
is the square of the quantity depicted in
figure~\ref{fig:fluid-part-vel-rms}$b$).  
It can be seen that in both flow cases the contribution from the
cross-correlation 
$\langle u_{pr,\alpha}^{{\cal
    S}\,\prime}\,u_{f,\alpha}^{{\cal S}\,\prime}\rangle_p$ 
is much smaller than the remaining two terms, indicating that the
fluctuations of the particle velocity are not strongly coupled to the
velocity of the surrounding fluid. 
The variance of the fluid velocity
seen by the particles, 
$\langle w_{f}^{{\cal S}\,\prime}\,w_{f}^{{\cal
    S}\,\prime}\rangle_p$, 
on the other hand, is the largest contribution in both flow cases. Its
time evolution exhibits a similar shape as the total particle velocity
fluctuation energy. In case M178 the time average value $\langle
w_{f}^{{\cal S}\,\prime}\,w_{f}^{{\cal
    S}\,\prime}\rangle_{p,t}/w_{ref}^2$ 
is three times as large as the corresponding value in case M121,
making it the dominant term. 
Since the energy of fluctuations of the particle velocity
relative to the surrounding fluid, 
$\langle w_{pr}^{{\cal S}\,\prime}\,w_{pr}^{{\cal
    S}\,\prime}\rangle_p/w_{ref}^2$, 
is approximately constant in time and of similar value in both flow
cases (its time-average is 30\% larger in case M178), it follows that
the larger overall fluctuation energy in case M178 is mostly due to an 
increase of the fluctuation energy of the fluid velocity
seen by the particles. 
The fact that the (vertical component of the) fluctuating particle
velocity is -- when taken relative to the surrounding fluid -- of
similar intensity in both cases indicates that the present difference
in Galileo number does not significantly affect this quantity. 
However, the occurrence of particle clustering in case M178 does 
lead to a significant increase in the particle velocity fluctuation 
amplitude (as compared to the non-clustering case M121) by way of a
difference in the surrounding fluid velocity which the particles
experience (expressed through the term $\langle w_{f}^{{\cal
    S}\,\prime}\,w_{f}^{{\cal S}\,\prime}\rangle_p$). 
Therefore, the effect of the Galileo number on the variance of the
vertical particle velocity is of indirect nature. 
The instantaneous angle $\alpha$ (with respect to the vertical axis)
of a particle's trajectory relative to the average fluid velocity is
given by the following relation: 
\begin{equation}\label{equ-res-part-motion-def-angle-alpha}
  \tan\alpha=\sqrt{u_{pr}^2+v_{pr}^2}/|w_{pr}|
  \,.
\end{equation}
The probability density functions of $\alpha$ are shown in
figure~\ref{fig:drift-angle}. 
In both flow cases they roughly resemble a Gamma distribution. 
However, closer inspection shows that the right tails for sufficiently
large $\alpha$ are purely exponential (i.e.\
$\mbox{pdf}(\alpha)\sim\exp(-a\alpha)$ with   
positive $a$). 
As can be seen from the semi-logarithmic graph in
figure~\ref{fig:drift-angle}$(b)$, this exponential behavior is
clearly exhibited by the data from case M178 for $\alpha\gtrsim15$
degrees, whereas in case M121 the pure exponential only matches a
small range of angles around $\alpha\approx30$ degrees.  
The figure also shows that in case M178 the angle defined in
(\ref{equ-res-part-motion-def-angle-alpha}) has a peak 
at approximately the value which is observed for an isolated sphere
settling at the same density ratio and Galileo number (i.e.\
$\alpha=5.2$ degrees), but in the present case M178 it also exhibits a
rather large standard deviation (equal to $4.3$ degrees).   
In case M121, the peak of the corresponding pdf is found at a
significantly smaller value of the angle ($\alpha=3.2$ degrees) and it
has a comparatively smaller standard deviation ($3.7$ degrees).  
The most probable trajectory in case M178 is, therefore, inclined with
respect to the vertical axis by a similar angle as in the case of an
isolated sphere. 
In case M121 the most probable trajectory is relatively close to a
vertical path (as in the corresponding single-sphere case). 
These results show that at the current solid volume fraction (and
despite the agglomerations found in case M178) the particle motion is
on average not too far removed from the state of motion of a
corresponding isolated sphere. 
However, the collective effect manifests itself clearly in the form of
broad probability distributions, i.e.\ frequent deviations from the
most probable state. 
  The intensity of the angular 
  particle velocity fluctuations is reported in
  figure~\ref{fig:part-ang-vel-rms}$(a)$. It can be observed that in both
  cases rotational motion around a horizontal axis is much more
  intense than the vertical component (by a factor of $6.0$ and $2.7$  
  in cases M121 and M178, respectively). This strong anisotropy is
  consistent with the corresponding difference in fluid velocity
  fluctuation intensities (cf.\
  figure~\ref{fig:fluid-part-vel-rms}$a$): gradients of the vertical
  fluid velocity drive the horizontal component of the angular
  particle motion and vice versa.  
  The large-scale fluid velocity fluctuations due to clustering in
  case M178, however, appear to have a small influence upon the
  overall intensity of the angular particle motion, presumably due to
  the fact that the cluster scale is significantly larger than the
  particle diameter. This is evident from the near-absence in
  figure~\ref{fig:part-ang-vel-rms}$(a)$ of temporal fluctuations of
  the same scales as in figure~\ref{fig:fluid-part-vel-rms}. Likewise,
  the increase in the horizontal angular velocity from case M121 to
  case M178 measures only 5\%. 
  \%
  We have also analyzed the 
  statistical correlation between angular particle velocity and of the
  fluctuations of the \Vor cell volumes.
  The graph in figure~\ref{fig:part-ang-vel-rms}$(b)$ shows
  that smaller cell volumes
  (i.e.\ larger local concentration values) lead to a significant
  increase in the standard deviation of the angular velocity in the
  clustering case \caseB, both for the horizontal and the vertical
  components in very similar fashion. 
  In the non-clustering case \caseA, on the contrary, the standard
  deviation of all components of the angular particle velocity is
  approximately independent of the \Vor cell volume. 
  The exception is a non-monotonous decrease observed at very small
  cell volumes (below $50$ times the particle volume). 
  Please note that the number of samples decreases considerably at
  small values of the \Vor cell volume in case \caseA\ which leads to
  significant oscillations of the curve.  
  It is noteworthy from the analysis of
  figure~\ref{fig:part-ang-vel-rms}$(b)$ that the \Vor cell volume 
  (i.e.\ the local particle concentration) does not by itself
  constitute a good indicator of an increase of the amplitude of
  angular particle motion. It does so only in conjunction with the
  knowledge of the occurrence of particle clustering, which can be
  deduced e.g.\ from the standard deviation of \Vor cell volumes (cf.\ 
  figure~\ref{fig:voronoi-VOL-STD}). 
\section{Summary and conclusions}
\label{sec-conclusion}
We have simulated the gravity-induced motion of randomly distributed,
finite-size, heavy particles in otherwise quiescent fluid in
triply-periodic domains. Focusing on a solid-to-fluid density ratio of
$1.5$ and on a solid volume fraction of $0.005$, we have considered
two values of the Galileo number, measuring 121 and 178. Isolated
spheres at the same density ratio and Galileo numbers settle steadily
on a vertical and on an oblique path, respectively.
In the multi-particle case, strong clustering of particles
in the form of vertically-elongated columnar regions is observed at
the higher Galileo number value. 
This result is consistent with the finding of \cite{kajishima:04b} at
$Ga\approx210$. 
At the smaller Galileo number the particle arrangement found in the
present study can actually be characterized as slightly more ordered
than a distribution through a random Poisson process. 
The observed clusters in the present work have horizontal extensions
equivalent to some 10 particle diameters; in the vertical direction
they essentially extend throughout the entire domain length of 170
particle diameters.  

In the clustering case, \Vor tesselation analysis further reveals that
approximately 30\% of the particles reside in cluster regions at a
given instant in time, and that their residence time both in clusters
and in void regions is long compared to the gravitational time scale.   

Through comparison between the two flow cases the current
configuration allows us to investigate the impact of clustering upon
the particle motion in detail. It is found that clustering leads to a
significant increase of the magnitude of the average settling velocity
measured relative to the volume-averaged fluid velocity (by 12\% as
compared to the value obtained for an isolated particle). 
Using our definition of the fluid velocity in the vicinity of the
(finite-size) particles we deduce that the particles in the clustering 
case sample the fluid field preferentially in regions of downward
motion (when compared to the box-averaged fluid velocity value). 
As a consequence, the relative velocity between the phases -- when
based upon the local fluid velocity ``seen'' by the particles -- is
similar to the one of a corresponding isolated particle at both
investigated Galileo numbers. 
  This finding suggests that it may be possible to approximate the
  particle settling velocity as the sum of two contributions, adding
  the characteristic meso-scale velocity of the surrounding fluid to
  the settling velocity value of an isolated sphere. This idea was
  suggested by an anonymous referee in analogy to the model proposed
  by \cite{aliseda:02} for the enhancement of the settling velocity of
  sub-Kolmogorov-size droplets due to clustering in grid-generated
  turbulence.  
  In their case it was possible to model the additional term as a
  function of the local particle concentration by identifying the
  particle clusters with macro-particles of a linear dimension
  equivalent to the experimentally determined cluster size; 
  invoking Stokes flow and using a concentration-dependent mixture
  density the model was successful in reproducing the experimental
  trends. 
  In the present case, however, the assumption of Stokes flow is not
  appropriate, and the dependency of the settling
  velocity upon the local concentration is not linear, as seen in
  figure~\ref{fig-particles-joint-pdf-settling-velocity-voronoi-cell-volume}. 
  Furthermore, our study features only a single case with significant
  particle clustering. Therefore, additional data (at different solid
  volume fractions and Galileo numbers) is required in order to
  further pursue this avenue.  

Concerning the fluctuations of the particle motion, it is found that
simple superposition of isolated sphere data cannot account for the
observed amplitudes of the multi-particle system. A decomposition of
the variance of the particle velocity is proposed which separates the
contribution due to the particle motion with respect to the
surrounding fluid from the contribution due to the fluctuations of the
fluid velocity in the vicinity of the particles. 
In the clustering case the time-averaged variance of the vertical
particle velocity is more than twice as large as in the non-clustering
case. The decomposition shows that this increase is mainly due to a
three-fold increase in the latter contribution (i.e.\ the variance of
the fluid velocity surrounding the particles). 
Therefore, the observed increase in the particle velocity fluctuation
amplitude in that case is due to a substantial amount of
particles being located in clusters or void regions which constitute
large scale fluid velocity perturbations with respect to the
box-averaged value.  

In the present work we have also analyzed the micro-structure of the
particulate phase in detail. 
The particle-conditioned solid volume fraction has revealed that
clustering not only manifests itself through a very slowly decaying
excess of particles vertically above a test particle, but that it also
leads to a significantly increased probability of finding particles at
small horizontal distances of several diameters from the test
particle.  
For the clustering case, particle-conditioned averaging also shows
that the settling velocity of particles located inside a roughly
cylindrical region with diameter equivalent to approximately $10D$
surrounding a test particle are settling at a higher speed than the
global average. Since a large number of particles resides in
regions of increased particle concentration in this case, the pattern
of increased settling speed is consistent with the geometry of the
clusters themselves. 

A further analysis of the fluid flow induced by the settling
particles (such as a characterization of the pseudo-turbulence,
particle-conditioned averaging of the near field, etc.) has not been
performed in the current paper. This will be the subject of a future
publication. 

One question of interest obviously concerns the mechanism of particle
cluster formation. 
  Wake attraction in the sense of ``drafting-kissing-tumbling''
  \citep[cf.][]{fortes:87,wu:98} has already been identified as the
  key ingredient in the process \citep{kajishima:04b}.  
  However, the details of cluster formation still await further
  clarification before modelling of the phenomenon becomes possible.  
In particular, with respect to the present results it is not clear how 
(for otherwise identical parameters) a critical value of
the Galileo number for the appearance of clustering arises. 
In the present work we have not attempted to obtain a more precise
value of the critical Galileo number due to the high computational
cost associated to the additional simulations which would be
required. 
However, it is now established that for a value of $Ga=121$ (whence an 
isolated sphere settles on a straight path) no clustering takes place,
while at $Ga=178$ (corresponding to an oblique path for an isolated
sphere) clustering is observed. 
  The lateral motion of isolated particles (with random orientation)
  may promote the tendency to cluster in dilute multi-particle systems
  by increasing the probability of close particle encounters, which
  then -- through said ``drafting'' effect -- lead to statistically
  significant accumulations.  
From this observation it appears tempting to relate the onset of
clustering to the bifurcation point (from axi-symmetric to
planar-oblique) of the wake of an isolated sphere -- at least under
similarly dilute conditions as in the present case. 
This conjecture would mark the critical point at a Galileo value of
approximately $155$. 

As a future perspective, we believe it will be worthwhile to determine
the critical Galileo number for the onset of clustering more
precisely. 
  Near that threshold it should then be possible to 
  determine whether the decisive factor triggering cluster formation
  is indeed related to the tendency of particles to follow
  non-vertical paths or possibly due to an alternative mechanism.
\vspace*{1ex}
Thanks is due to J.\ Du\v{s}ek for fruitful discussions. 
This work was supported by the German Research Foundation (DFG) under
projects UH~242/1-1 and UH~242/1-2. 
The simulations were partially performed at J\"ulich Supercomputing
Center (grant hka09), at LRZ M\"unchen (grant pr58cu) and at SCC Karlsruhe.  
The computer resources, technical expertise and assistance provided by  
these centers are gratefully acknowledged. 
\appendix
\section{Averaging procedures} \label{sec:app-averaging} 
The current appendix provides detailed definitions of the various
averaging operators used in the present work.  
\subsection{Phase averaging}\label{sec:app-phase-averaging} 
Let us first define indicator functions for both the fluid and the
dispersed phase.  
An indicator function for the fluid phase $\Phi_f(\mathbf{x},t)$
specifies whether a given point $\mathbf{x}$ at time $t$ is
located inside the region $\Omega_f(t)$ occupied by the fluid at time
$t$, viz.  
\begin{equation}
  \Phi_f(\mathbf{x},t) = \left\{
    \begin{array}{l l}
      1 \qquad \mbox{if} \quad \mathbf{x} \in \Omega_f(t) \\
      0 \qquad \mbox{else} \\
    \end{array} \right. 
  \,.
  \label{eq:Indicator_f}
\end{equation}
The corresponding indicator function for the dispersed phase
$\Phi_p(\mathbf{x},t)$ can be readily obtained as:  
\begin{equation}
	\Phi_p(\mathbf{x},t) = 1 -
        \Phi_f(\mathbf{x},t). 
        \label{eq:Indicator_p} 
\end{equation}
Based on the definitions in equations (\ref{eq:Indicator_f}) and
(\ref{eq:Indicator_p}) a discrete counter of sample points occupied by
the carrier phase, $n_f$, and by the dispersed phase, $n_p$, at 
time $t^m$ can be defined as follows (with subscript $\alpha=f$ or
$\alpha=p$): 
\bea
n_\alpha (t^m) &=& \sum_{i=1}^{N_x} \sum_{j=1}^{N_y} \sum_{k=1}^{N_z}
{\Phi_\alpha(\mathbf{x}_{ijk},t^m)} \qquad \: \: \forall \: \: m \in
[1,N^{(\alpha)}_t],
\eea
where $N_x$, $N_y$ and $N_z$ are the number of grid nodes in the
corresponding coordinate directions,
$\mathbf{x}_{ijk}$ 
is the spatial location of a grid node with indices $i$, $j$, $k$, 
and $N_{t}^{(f)}$ ($N_{t}^{(p)}$) is the
number of available instantaneous flow fields (particle states). 

The fluid-phase averaging for any fluid-related quantity
$\psi_f(\mathbf{x}_{ijk},t^m)$ at time $t^m$ can be defined as: 
\ben
\label{equ-app-avg-def-fluid-space-avg}
	\langle \psi_f \rangle_{\Omega_f} (t^m) =
        \frac{1}{n_f(t^m)}\sum_{i=1}^{N_x}\sum_{j=1}^{N_y}\sum_{k=1}^{N_z}
        {\Phi_f(\mathbf{x}_{ijk},t^m)
          \psi_f(\mathbf{x}_{ijk},t^m)}, \qquad \: \: \forall \:
        \: m \in [1,N_{t}^{(f)}] . 
\een
Similarly, by defining a counter of fluid phase samples along the
vertical direction, $n_f^{(z)}$, viz.
\begin{equation}\label{equ-app-avg-def-vertical-fluid-sample-counter}
  n_f^{(z)} (x_i,y_j,t^m) = \sum_{k=1}^{N_z}
  {\Phi_f(\mathbf{x}_{ijk},t^m)} \qquad \forall \,m \in
  [1,N^{(f)}_t],
  \,i\in[1,N_x],
  \,j\in[1,N_y]
  \,,
\end{equation}
the spatial average of a fluid-related quantity
$\psi_f(\mathbf{x}_{ijk},t^m)$ over the vertical direction is defined
as follows:  
\begin{eqnarray}\label{equ-app-avg-def-vertical-fluid-average}
  \langle \psi_f\rangle_z (x_i,y_j,t^m) &=& \frac{1}{n_f^{(z)}(x_i,y_j,t^m)}
  \sum_{k=1}^{N_z} {\Phi_f(\mathbf{x}_{ijk},t^m)
    \psi_f(\mathbf{x}_{ijk},t^m)}, \\\nonumber
  &&\forall\,
  m \in[1,N_{t}^{(f)}],
  \,i\in[1,N_x],
  \,j\in[1,N_y]
  \,.
\end{eqnarray}
The dispersed-phase averaging operator $\langle \cdot
\rangle_p$ for a given quantity $\psi_p^i(t^m)$ related to the $i$th
particle at time $t^m$ can be defined as: 
\ben
\label{equ-app-avg-def-particle-avg}
  \langle \psi_p\rangle_p (t^m)  = \frac{1}{N_p}
  \sum_{i=1}^{N_p}{\psi_p^i(t^m)}, \qquad \: \: \forall \: \: m \in
  [1,N_{t}^{(p)}]. 
\een
A time-averaging operator for any quantity $\psi(t^m)$ related to either the
dispersed phase or the carrier phase can be defined as follows: 
\bea
\label{equ-app-avg-def-time-avg}
	\langle \psi\rangle_{t} =
        \frac{1}{N_{t}^{(\alpha)}}\sum_{m=1}^{N_{t}^{(\alpha)}}{\psi(t^m)}, 
\eea
where $\alpha=p$ or $\alpha=f$, respectively. 

Combining the operators defined in
(\ref{equ-app-avg-def-fluid-space-avg}) and 
(\ref{equ-app-avg-def-time-avg}), the time and 
space average of a fluid quantity $\langle \psi_f
\rangle_{\Omega_f,t}$ is obtained. 
Analogously, the average of a particle-related quantity over all
available particle samples, $\langle \psi_p \rangle_{p,t}$, is defined
from the combination of operators (\ref{equ-app-avg-def-particle-avg})
and (\ref{equ-app-avg-def-time-avg}).  
\subsection{Particle-conditioned averaging}	
\label{sec:app-cond-averaging}
Let us first define a three-dimensional averaging domain $\Omega_A$
with grid node locations $\mathbf{y}_{ijk}$.
The averaging domain $\Omega_A$
is discretized in the same manner as the simulations, i.e.\ with an
isotropic mesh with constant grid spacing in all spatial
directions. The averaging domain is first shifted to the position
of one of the particles.  
Then the instantaneous coordinate of a grid node $\mathbf{y}_{ijk}
\in \Omega_A$ relative to the position of the $l$-th particle is
defined as: 
\ben
	\tilde{\mathbf{x}}_{ijk}^l(t^m) = \mathbf{y}_{ijk} -
        \mathbf{x}_{p}^l(t^m), \qquad \: \: \forall \: \: m \in
        [1,N_{p,t}], \qquad l=1 \ldots N_p. 
\een
We can now define a discrete counter field
$\tilde{n}_{ijk}^p(\mathbf{y}_{ijk})$ which holds the number of
samples obtained through dispersed-phase averaging at a given grid
node in the averaging domain $\Omega_A$ at time $t^m$: 
\ben
  \tilde{n}_{ijk}^p (\tilde{\mathbf{x}}_{ijk}) =
  \sum_{m=1}^{N_{p,t}}\sum_{l=1}^{N_p}{\Phi_p
    (\tilde{\mathbf{x}}_{ijk}^l (t^m), t^m
    )}\,, \label{eq:loc-fluid-aver-vert} 
\een
where $\Phi_p$ is defined in (\ref{eq:Indicator_p} ). 
Under proper normalization, the quantity
$\tilde{n}^p_{ijk}(\tilde{\mathbf{x}}_{ijk})$ defines the
particle-conditioned local solid volume fraction $\phi_s^{cond}$,
viz. 
\ben
  \phi_s^{cond} (\tilde{\mathbf{x}}_{ijk})= \frac{1}{N_{p,t} N_p}
  \tilde{n}^p_{ijk}(\tilde{\mathbf{x}}_{ijk}) 
  \,.
\een

The particle-conditioned space-time averaging of any particle-related
quantity $\psi_p^l(t^m)$ can then be defined as follows:  
\ben
  \psi_p^{cond} (\tilde{\mathbf{x}}_{ijk}) =
  \frac{1}{\tilde{n}_{ijk}^p (\tilde{\mathbf{x}}_{ijk})}
  \sum_{m=1}^{N_{p,t}}\sum_{l=1}^{N_p}{\Phi_p
    (\tilde{\mathbf{x}}_{ijk}^l (t^m), t^m )
    \psi_p^l(t^m)}. 
  \label{eq:cond-aver} 
\een
\bibliographystyle{model2-names}
\addcontentsline{toc}{section}{References}
\clearpage
\begin{table}
  \setlength{\tabcolsep}{4pt}
  \centering
  \begin{tabular}{lccccccc}
    case & $\Phi_s$ & $\rho_p/\rho_f$ & $Ga$ & $Re$ & $L_x \times L_y
    \times L_z$ & $N_p$ & $T_{obs}/\tau_g$\\[.5ex]  
    \caseA & $5\cdot10^{-3}$ & $1.5$ & $121.24$&$141.6$
    & $68 D \times 68 D \times 341 D$ &
    $15190$&$2967$\\ 
    \caseB & $5\cdot10^{-3}$ & $1.5$ & $178.46$ &$260.6$
    & $85 D \times 85 D \times 170 D$ & $11867$&$1472$\\ 
    \caseAA & $5.3\cdot10^{-5}$ & $1.5$ & $121.24$ &$141.1$
    & $8.5 D \times 8.5 D \times 136 D$ & $1$ & --\\ 
    \caseBB & $2.7\cdot10^{-5}$ & $1.5$ & $178.46$ &$233.1$
    & $10.6D \times 10.6D \times 170 D$ & $1$ & --\\ 
  \end{tabular}
  \caption{Physical parameters in the present simulations: 
    global solid volume fraction $\Phi_s$, 
    density ratio $\rho_p/\rho_f$, 
    Galileo number $Ga=u_gD/\nu$, 
    Reynolds number $Re= |\langle w_s\rangle_t| D/\nu$ 
    (based on the average settling velocity 
    $\langle w_s\rangle_t$ defined in
    equation~\ref{equ-res-trasient-def-w-settling}, 
    particle diameter $D$ and fluid viscosity $\nu$), 
    size of the computational domain $L_\alpha$ (in the
    coordinate direction $x_\alpha$) 
    and the number of particles $N_p$.
    $T_{obs}$ is the duration of the observation
    interval over which statistics were accumulated. 
  }
  \label{tab:phys_params}
\end{table}
\begin{table}
  \setlength{\tabcolsep}{4pt}
  \centering
  \begin{tabular}{lcc}
    case & $D/\Delta x$ & $N_x \times N_y \times N_z$ \\[.5ex] 
    \caseA   & $15$   & $1024 \times 1024 \times 5120$ \\
    \caseB   & $24$   & $2048 \times 2048 \times 4096$ \\
    \caseAA   & $15$   & $128 \times 128 \times 2048$ \\
    \caseBB   & $24$   & $256 \times 256 \times 4096$ \\
  \end{tabular}
  \caption{Numerical parameters employed in the present
    simulations. Particle resolution $D/\Delta x$, number of grid
    nodes $N_i$ in the $i$-th coordinate direction.} 
  \label{tab:num_params}
\end{table}
\begin{table}
  \centering
  \begin{tabular}{lccc}
    case&
    $(\langle w_f^{\cal S}\rangle_{p,t}
    -\langle w_f\rangle_{\Omega_f,t})/
    w_{ref}$
    &
    $\sigma_{p,t}(w_f^{\cal S})/w_{ref}$
    &
    $\sigma_{\Omega_f,t}(w_f)/w_{ref}$
    \\
    M121&
    $-0.0315$&
    $0.1262$&
    $0.2081$
    \\
    M178&
    $-0.1155$&
    $0.1972$&
    $0.2407$
  \end{tabular}
  \caption{%
    The difference (in the mean and in the standard deviation)
    between the particle-conditioned vertical fluid velocity
    $w_f^{\cal S}(i,t)$ ($\forall\,i=1\ldots N_p$) and the
    unconditioned velocity field $w_f(\mathbf{x}\in\Omega_f,t)$.  
    The operator $\sigma_{p,t}$ refers to the standard deviation with
    respect to the ensemble of particles and temporal snapshots; 
    $\sigma_{\Omega_f,t}$ denotes the standard deviation with respect
    to all available samples in the fluid phase and the available
    temporal records. 
    All values are normalized with the absolute value of the
    conventional time-average settling velocity, i.e.\
    $w_{ref}=|\langle w_s(t)\rangle_{t}|$ (cf.\
    equation~\ref{equ-res-trasient-def-w-settling}).  
  }
  \label{tab-res-w-shell-moments}
\end{table}
\clearpage
\newpage
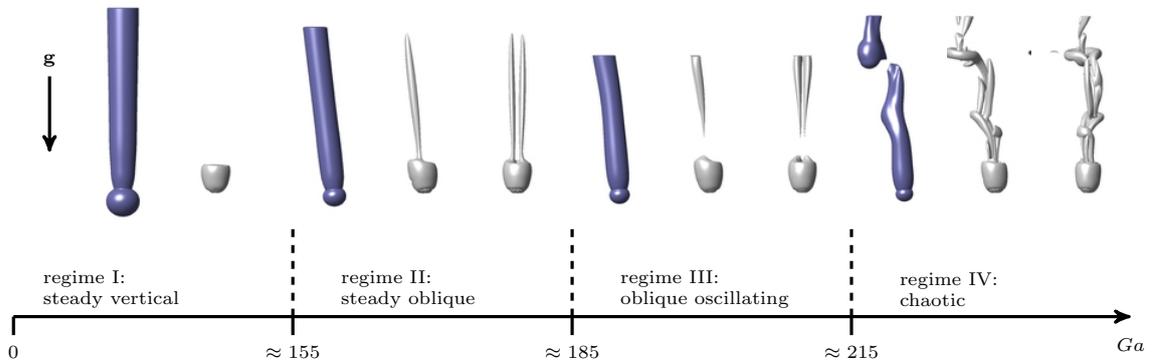
\begin{figure}
  \begin{minipage}{.245\linewidth}
    \begin{center}
      \raisebox{10ex}{
        \begin{minipage}{2ex}
          \begin{tikzpicture}[scale=1,font=\scriptsize]
            \draw [stealth'-,very thick] (0,0) -- (0,+1) node [above=0ex] (B) {$\mathbf{g}$};
          \end{tikzpicture}
        \end{minipage}
      }
      \includegraphics[width=.3\linewidth]%
      {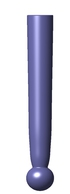}
      \includegraphics[width=.3\linewidth]%
      {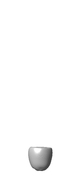}
    \end{center}
  \end{minipage}
  \begin{minipage}{.245\linewidth}
    \includegraphics[width=.3\linewidth]%
    {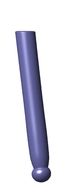}
    \includegraphics[width=.3\linewidth]%
    {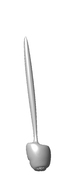}
    \includegraphics[width=.3\linewidth]%
    {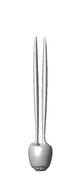}
  \end{minipage}
  \begin{minipage}{.245\linewidth}
    \includegraphics[width=.3\linewidth]%
    {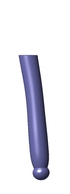}
    \includegraphics[width=.3\linewidth]%
    {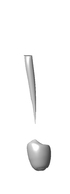}
    \includegraphics[width=.3\linewidth]%
    {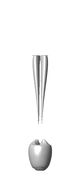}
  \end{minipage}
  \begin{minipage}{.245\linewidth}
    \includegraphics[width=.3\linewidth]%
    {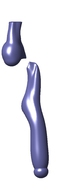}
    \includegraphics[width=.3\linewidth]%
    {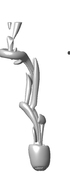}
    \includegraphics[width=.3\linewidth]%
    {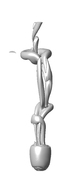}
  \end{minipage}
  \\[0ex]
  \begin{minipage}{.99\linewidth}
    \begin{tikzpicture}[scale=2.45,font=\scriptsize]
      \coordinate [] (A) at (-.5,0);
      \draw [-stealth',very thick] (A) -- ++ (6,0) node [below=1ex] (B) {$Ga$};
      \coordinate [] (c1) at (1,0);
      \coordinate [] (c2) at (2.5,0);
      \coordinate [] (c3) at (4,0);
      \draw [very thick] (A) -- ++ (0,-.1) node [below]
      {$0$};
      \draw [very thick] (c1) -- ++ (0,-.1) node [below]
      {$\approx155$};
      \draw [very thick,dashed] (c1) -- ++ (0,.5);
      \draw [very thick] (c2) -- ++ (0,-.1) node [below]
      {$\approx185$};
      \draw [very thick,dashed] (c2) -- ++ (0,.5);
      \draw [very thick] (c3) -- ++ (0,-.1) node (a220) [below]
      {$\approx215$};
      \draw [very thick,dashed] (c3) -- ++ (0,.5);
      \node [black,align=left] at (0.2,.15) {\parbox{25ex}{regime I:\\steady vertical}} ;
      \node [black] at (1.8,.15) {\parbox{25ex}{regime II:\\steady oblique}} ;
      \node [black] at (3.3,.15) {\parbox{25ex}{regime III:\\oblique oscillating}} ;
      \node [black] at (4.8,.15) {\parbox{25ex}{regime IV:\\chaotic}} ;
    \end{tikzpicture}
  \end{minipage}
  \caption{%
    The four different regimes of particle motion encountered as a
    function of the Galileo number for a density ratio
    $\rho_p/\rho_f=1.5$. 
    In the upper part of the figure, the typical wake structure at one 
    value of the Galileo number is shown for each regime (i.e.\ 
    from left to right $Ga=\{144,178,190,250\}$). 
    In each case, the blue surfaces indicate the locations where the fluid
    velocity relative to the particle in the direction given by its 
    instantaneous velocity (i.e.\ the axial component) measures
    $1.2u_g$. 
    The grey-colored surfaces are visualizations of the vortical
    structures by means of the $\lambda_2$ criterion of
    \cite{jeong:95}. 
    In regimes II-IV the vortical structures are additionally shown
    from a second viewpoint, rotated around the particle by an angle
    of $\pi/2$ in the horizontal plane with respect to the first one. 
    The regime boundaries ($Ga=155$, $185$, $215$)
    correspond to particle Reynolds numbers of approximately 
    $203$, $253$ and $304$, respectively.
    The flow data is from the study of \cite{uhlmann:13a}. 
  }
  \label{fig-wake-regimes-isolated}
\end{figure}
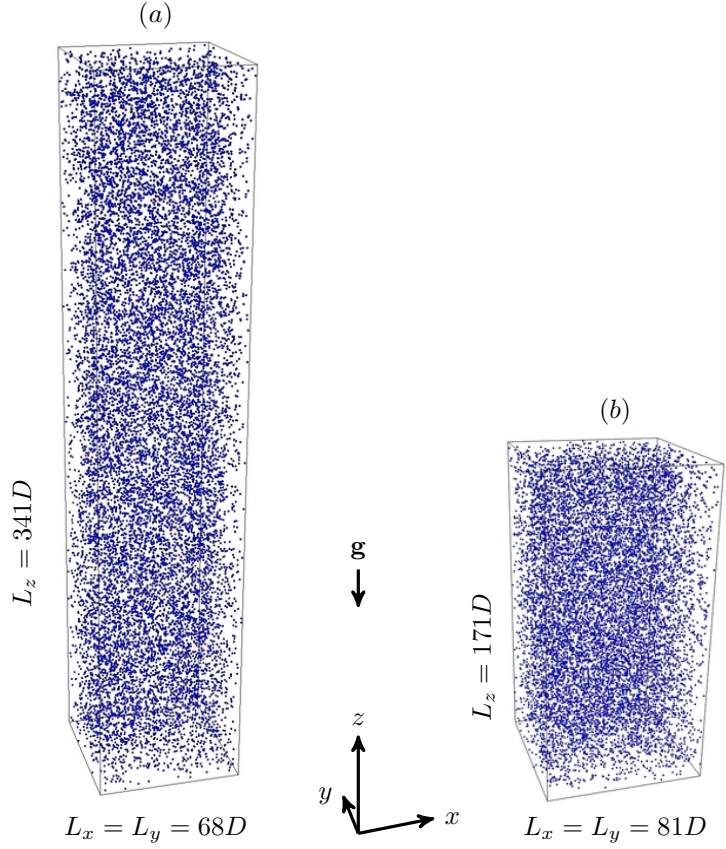
\begin{figure}
  \centering
  \raisebox{25ex}{
    \begin{minipage}[c]{1ex}
      \rotatebox{90}{$L_z=341D$}
    \end{minipage}
  }
  \begin{minipage}[b]{0.2\linewidth}
    \centerline{$(a)$}
    \includegraphics[width=\linewidth]
    {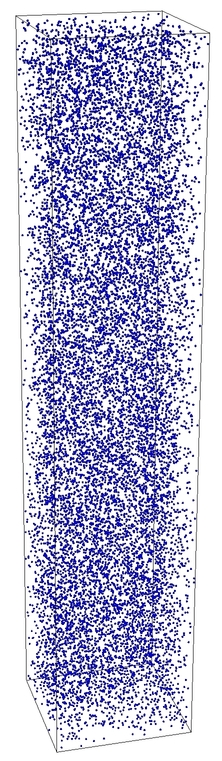}
    \\
    \centerline{$L_x=L_y=68D$}
  \end{minipage}
  \hspace*{.02\linewidth}
  \begin{minipage}[b]{0.1\linewidth}
    \begin{tikzpicture}[scale=1,font=\normalsize]
      \draw [stealth'-,very thick] 
      (0,3) -- (0,3.5) 
      node [above=0ex] (B) {$\mathbf{g}$};
      \draw [-stealth',very thick] 
      (0,0) -- (1,.2) 
      node [right=0ex] (X) {$x$};
      \draw [-stealth',very thick] 
      (0,0) -- (-.2,.5) 
      node [left=0ex] (Y) {$y$};
      \draw [-stealth',very thick] 
      (0,0) -- (0,1.3) 
      node [above=0ex] (Z) {$z$};
    \end{tikzpicture}
  \end{minipage}
  \hspace*{.02\linewidth}
  \raisebox{15ex}{
    \begin{minipage}[c]{1ex}
      \rotatebox{90}{$L_z=171D$}
    \end{minipage}
  }
  \begin{minipage}[b]{0.2\linewidth}
    \centerline{$(b)$}
    \includegraphics[width=\linewidth]
    {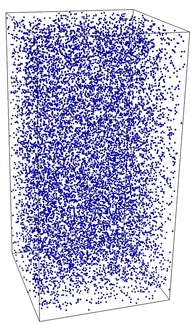}
    \\
    \centerline{$L_x=L_y=81D$}
  \end{minipage}
  \caption{Dimensions of the computational domain with the
    initial particle distribution for (a) case \caseA{} and (b)
    case $M178$. $D$ denotes the particle diameter; $\mathbf{g}$ is
    the vector of gravitational acceleration. 
    Periodic boundary conditions are applied in all three spatial
    directions.}  
  \label{fig:box_part_pos}
\end{figure}

\clearpage
\newpage

\clearpage
\newpage

\clearpage
\end{document}